\documentstyle[12pt]{article}
\newlength{\pubnumber} 

\catcode`\@=11
\@addtoreset{equation}{section}

\def\section{\@startsection{section}{1}{\z@}{3.5ex plus 1ex minus .2ex}
 {2.3ex plus .2ex}{\large\bf}}
\def\subsection{\@startsection{subsection}{2}{\z@}{2.3ex plus .2ex}
 {2.3ex plus .2ex}{\bf}}

\begin{document}

\begin{titlepage}
\samepage{
\setcounter{page}{1}
\rightline{\tt hep-th/0412185}
\rightline{CPTH RR 065.1104}
\rightline{LPT-ORSAY 04-123}
\rightline{UMN-TH-2333/04}
\rightline{December 2004}
\vfill
\begin{center}
   {\Large \bf A Calculable Toy Model of the Landscape\\}
\vfill
   {\large
      Keith R. Dienes$^1$\footnote{
     E-mail address:  dienes@physics.arizona.edu}, $\,$ Emilian Dudas$^{2,3}$\footnote{
     E-mail address:  emilian.dudas@cpht.polytechnique.fr}, $\,$ Tony Gherghetta$^4$\footnote{
     E-mail address:  tgher@physics.umn.edu} \\}
\vspace{.20in}
 {\it  $^1$  Department of Physics, University of Arizona, Tucson, AZ  85721  USA\\
       $^2$  Centre de Physique Th\'eorique, Ecole Polytechnique, \\
                   F-91128, Palaiseau Cedex, France\\
       $^3$  LPT, B\^at.~210, Univ.\ Paris-Sud, F-91405, Orsay Cedex, France\\
       $^4$  School of Physics and Astronomy, University of Minnesota, \\
                 Minneapolis, MN 55455 USA\\}
\end{center}
\vfill
\begin{abstract}
  {\rm 
    Motivated by recent discussions of the string-theory landscape,
    we propose field-theoretic realizations of models with large numbers
    of vacua.  These models contain multiple $U(1)$ gauge groups, and
    can be interpreted as deconstructed versions of higher-dimensional gauge 
    theory models with fluxes in the compact space.  We find that the vacuum structure of 
    these models is very rich, defined by parameter-space regions with different 
    classes of stable vacua separated by boundaries.
    This allows us to explicitly calculate physical quantities such as 
    the supersymmetry-breaking scale, the presence or absence of $R$-symmetries,
    and probabilities of stable versus unstable vacua. 
    Furthermore, we find that this landscape picture evolves with 
    energy, allowing vacua to undergo phase transitions as they cross
    the boundaries between different regions in the landscape.
    We also demonstrate that supergravity 
    effects are crucial in order to stabilize most of these vacua, and in 
    order to allow the possibility of cancelling the cosmological constant. 
   }
\end{abstract}
\vfill
\smallskip}
\end{titlepage}

\setcounter{footnote}{0}

% ========================= DEFINITIONS ===================================
\def\beq{\begin{equation}}
\def\eeq{\end{equation}}
\def\beqn{\begin{eqnarray}}
\def\eeqn{\end{eqnarray}}
\def\half{{\textstyle{1\over 2}}}
\def\quart{{\textstyle{1\over 4}}}

\def\vac#1{{\bf \{{#1}\}}}

\def\calO{{\cal O}}
\def\calE{{\cal E}}
\def\calT{{\cal T}}
\def\calM{{\cal M}}
\def\calF{{\cal F}}
\def\calN{{\cal N}}
\def\calY{{\cal Y}}
\def\ie{{\it i.e.}\/}
\def\eg{{\it e.g.}\/}

\newcommand{\newc}{\newcommand}
\newc{\gsim}{\lower.7ex\hbox{$\;\stackrel{\textstyle>}{\sim}\;$}}
\newc{\lsim}{\lower.7ex\hbox{$\;\stackrel{\textstyle<}{\sim}\;$}}

%==============================================================================
\hyphenation{su-per-sym-met-ric non-su-per-sym-met-ric}
\hyphenation{space-time-super-sym-met-ric}
%==============================================================================

%================== BLACKBOARD BOLD CHARACTERS ==============================

\def\inbar{\,\vrule height1.5ex width.4pt depth0pt}

\def\IC{\relax\hbox{$\inbar\kern-.3em{\rm C}$}}
\def\IQ{\relax\hbox{$\inbar\kern-.3em{\rm Q}$}}
\def\IR{\relax{\rm I\kern-.18em R}}
 \font\cmss=cmss10 \font\cmsss=cmss10 at 7pt
\def\IZ{\relax\ifmmode\mathchoice
 {\hbox{\cmss Z\kern-.4em Z}}{\hbox{\cmss Z\kern-.4em Z}}
 {\lower.9pt\hbox{\cmsss Z\kern-.4em Z}}
 {\lower1.2pt\hbox{\cmsss Z\kern-.4em Z}}\else{\cmss Z\kern-.4em Z}\fi}

%========================================================================
%          MACROS FOR REFERENCES
%========================================================================
\def\NPB#1#2#3{{\it Nucl.\ Phys.}\/ {\bf B#1} (#2) #3}
\def\PLB#1#2#3{{\it Phys.\ Lett.}\/ {\bf B#1} (#2) #3}
\def\PRD#1#2#3{{\it Phys.\ Rev.}\/ {\bf D#1} (#2) #3}
\def\PRL#1#2#3{{\it Phys.\ Rev.\ Lett.}\/ {\bf #1} (#2) #3}
\def\PRT#1#2#3{{\it Phys.\ Rep.}\/ {\bf#1} (#2) #3}
\def\CMP#1#2#3{{\it Commun.\ Math.\ Phys.}\/ {\bf#1} (#2) #3}
\def\MODA#1#2#3{{\it Mod.\ Phys.\ Lett.}\/ {\bf A#1} (#2) #3}
\def\IJMP#1#2#3{{\it Int.\ J.\ Mod.\ Phys.}\/ {\bf A#1} (#2) #3}
\def\NUVC#1#2#3{{\it Nuovo Cimento}\/ {\bf #1A} (#2) #3}
\def\etal{{\it et al.\/}}

% Redefine caption to put text and formulas in smaller font
\long\def\@caption#1[#2]#3{\par\addcontentsline{\csname
  ext@#1\endcsname}{#1}{\protect\numberline{\csname
  the#1\endcsname}{\ignorespaces #2}}\begingroup
    \small
    \@parboxrestore
    \@makecaption{\csname fnum@#1\endcsname}{\ignorespaces #3}\par
  \endgroup}
\catcode`@=12

\input epsf
%============================== TEXT BEGINS HERE ============================

%=============================================================================
\setcounter{footnote}{0}
\section{Introduction}

Recent developments in the study of string-theory compactifications 
suggest that there exist huge numbers of string vacua, with
different cosmological constants and different 
low-energy phenomenological properties~\cite{bp}. 
The resulting picture, dubbed the ``landscape''~\cite{landscape}, 
has stimulated a statistical analysis of the number 
of string vacua~\cite{statistics},
the supersymmetry-breaking scale, and other phenomenological
features~\cite{dgt}.  Anthropic principles have even been advanced
to resolve difficult issues such as the cosmological constant
problem~\cite{anthropic}.  

One of the problems facing
this landscape picture of string theory is that of
calculating physical parameters.  This is, to a large extent,
due to the limited technology for performing string calculations 
in the relevant flux vacua~\cite{fluxes}. 
It is therefore useful to present field-theoretic counterparts of such 
constructions, \ie, field-theoretic models which naturally give rise to very large
numbers of vacua,  and to be able to quantitatively determine
statistical distributions of  relevant physical quantities such as     
the cosmological constant, the scale of supersymmetry breaking, the 
Higgs mass, gauge and Yukawa couplings, and
the like.

The purpose of this paper is to present such field-theoretic examples based on 
multiple Abelian gauge groups and multiple 
charged scalar fields. 
As we shall see, such models naturally lead to large numbers of vacua,
and allow us to quantitatively address
many of the pressing questions that such pictures raise.

In Sect.~2, as a brief warm-up, we discuss the standard Fayet-Iliopoulos model,
now reconsidered from our perspective. 
In Sect.~3, we then  describe, in some detail, the construction
of a relatively simple toy model 
which nevertheless gives rise to many of the landscape features that we
wish to study.
Specifically, we consider a supersymmetric
model with two $U(1)$'s with Fayet-Iliopoulos terms,
and three charged  chiral superfields. 
We show that this model contains seven distinct classes of vacua which populate
the parameter space in different pie-slice regions, giving rise to vastly different
physical properties such as supersymmetry-breaking scale and numbers of
unbroken gauge and global symmetries.  We also show that renormalization-group 
evolution can bring the theory from one vacuum to another,
while the boundaries between different regions correspond to parameter-space surfaces 
on which a single scalar particle becomes massless.

In Sect.~4, we consider a generalization of this toy model to more complicated
supersymmetric models with large numbers of $U(1)$ gauge groups.   Such models
easily give rise to huge numbers of distinct classes of vacua.
The simplest example is provided by $n$ copies of the Fayet-Iliopoulos
model,  thereby giving rise to $3^n$ distinct classes of vacua. 
We concentrate most of our efforts, however, on a four-dimensional model 
which we believe is the ``deconstruction''~\cite{deconstruction} 
of a six-dimensional model with internal
magnetic fluxes.  Our interest in this respect is to provide
a purely four-dimensional description of these higher-dimensional flux models
which are among the simplest models that generate chirality in four dimensions.
This flux-compactification framework is
also probably the simplest and most successful framework 
for the construction of semi-realistic string models
with intersecting branes~\cite{intersecting}.  We also comment on the role of mixed anomalies and 
their cancellation 
in the vacua we consider, and show 
that almost all of these vacua are unstable,
in analogy with the Nielsen-Olesen instabilities of internal magnetic-flux compactifications. 

In Sect.~5, we demonstrate that
supergravity effects can lead to large soft mass terms which stabilize most of these 
unstable vacua in certain regions of the parameter space.
   We also show that qualitative arguments based on cosmological
constant cancellation generically require a large gravitino mass which,
in a gravity-mediated supersymmetry breaking scenario, is consistent with the
large soft terms that stabilize our unstable vacua.
  Finally, we conclude with comments about possible applications and
generalizations of these constructions, and prospects for future research.

%=============================================================================
\setcounter{footnote}{0}
\section{Starting small:  The Fayet-Iliopoulos model}

\begin{figure}[ht]
\centerline{ 
   \epsfxsize 5.8  truein \epsfbox {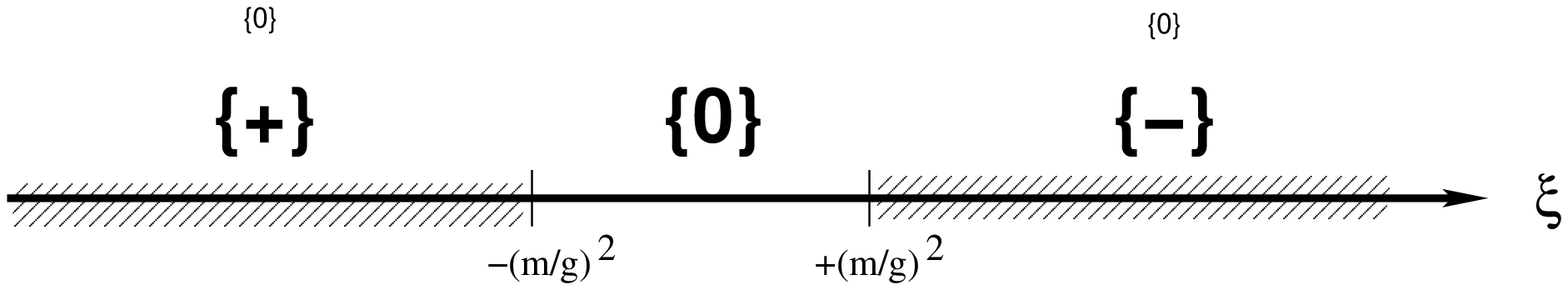}
    }
\caption{The vacuum structure of the Fayet-Iliopoulos~(FI) model as a function
of the FI coefficient $\xi$.  The different classes of vacua
outlined in the text (and here denoted \vac{$\emptyset$}, 
\vac{$+$}, and \vac{$-$} respectively) 
occupy non-overlapping regions along the $\xi$-axis.  In the regions
corresponding to \vac{$+$} and \vac{$-$} vacua, we have 
indicated the existence of an unstable \vac{$\emptyset$} extremum 
in parenthesis.  We have also shaded the regions corresponding
to vacua with at least one non-zero vacuum expectation value,
corresponding to vacua exhibiting $R$-symmetry breaking.} 
\label{prefig}
\centerline{ 
   \epsfxsize 5.0  truein \epsfbox {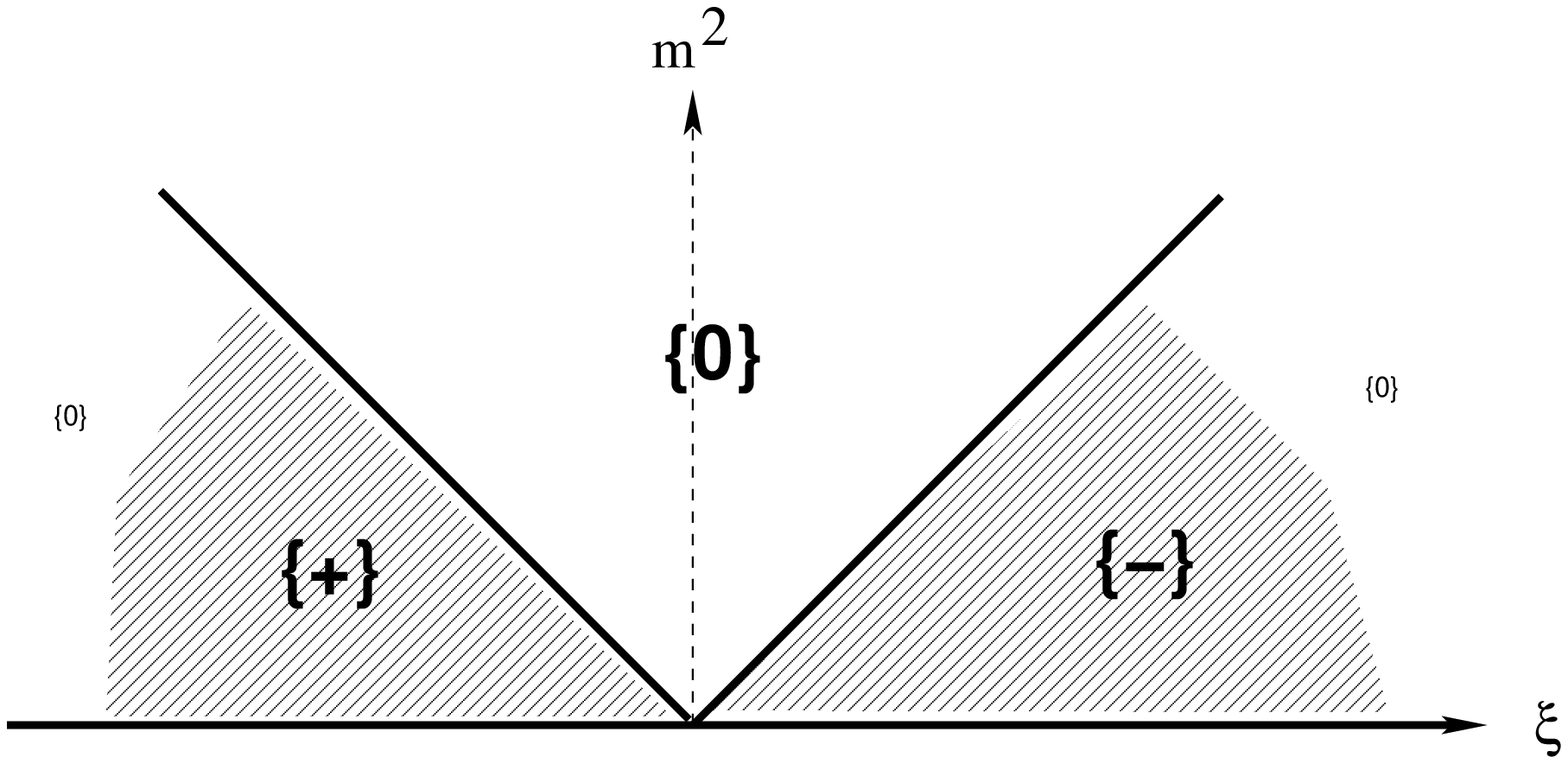}
    }
\caption{The vacuum structure of
the FI model, now sketched as a function of $(\xi,m)$.
Each region now has the shape of a pie-slice.
The slopes of the diagonal lines are $\pm g^2$.} 
\label{prefigtwo}
\end{figure}

The simplest example of  a model exhibiting a
non-trivial, multiple-vacuum structure that
we have in mind is the Fayet-Iliopoulos~(FI) model of supersymmetry 
breaking~\cite{fi}.
Recall that this model has a single $U(1)$ gauge group with gauge coupling $g$ and
FI coefficient $\xi$, 
two charged chiral superfields $\Phi^{(\pm)}$ of charges $\pm 1$,
and a superpotential $W = m \Phi^{(+)} \Phi^{(-)}$. 
It turns out that the scalar potential for this model has three classes
of extrema:
\begin{itemize}
\item  Extrema with $v^{(\pm)}\equiv \langle \phi^{(\pm)}\rangle =0$.
       Such extremum solutions exist for all $\xi$, but represent
       stable vacua only for $m^2 \pm g^2 \xi \ge 0$.
\item  Extrema with $v^{(+)}\not=0$, $v^{(-)}=0$.   These solutions
       exist if $m^2 + g^2 \xi <  0$, and are stable whenever they exist.
\item  Extrema with $v^{(-)}\not=0$, $v^{(+)}=0$.   These solutions
       exist if $m^2 - g^2 \xi <  0$, and are stable whenever they exist.
\end{itemize}
Note that in the first class of solutions, both the gauge symmetry and 
$R$-symmetry of the model are unbroken, whereas in the second two classes
of solutions they are both broken.\footnote{$D$-term breaking always preserves 
    $R$-symmetries, independently of their detailed definition.  By contrast,
    $F$-type breaking only breaks $R$-symmetries if the auxiliary field $F$ 
    is charged under the $R$-symmetry.
    In all of the examples to be considered in this paper, 
    we find that $F$-terms are charged under the $R$-symmetry  
    if we define $R$-symmetry by requiring the superpotential to 
    be invariant and assume
    that all matter fields have the same $R$-charge. 
    Thus, $F$-breaking is equivalent 
    to $R$-symmetry breaking. Indeed, for the FI model, if the 
     superpotential charge is equal to 
    $2$, then the matter superfield $R$-charges are $1$ and therefore the 
    $F$-terms have charges $-1$.}
This vacuum structure is illustrated in Fig.~\ref{prefig}, which indicates
the appropriate vacuum as a function of $\xi$.
 
%================== FIGURE ============================================
%================== END OF INSERTED FIGURE ============================

Of course, the FI model is actually defined by more than the single parameter
$\xi$;  we can just as easily consider $m$ as an additional defining parameter.
We would then sketch the vacuum structure of this model as in 
Fig.~\ref{prefigtwo}, which shows the vacuum structure as a function
of both $\xi$ and $m$.
The slopes of the diagonal lines are $\pm g^2$.

Finally, we may recognize that $g$ also represents 
an additional degree of freedom defining the model.
In this case, it is relatively easy to visualize the resulting 
vacuum structure as a function of location in a three-dimensional 
space.

Suppose we were to ask a simple question:  given the vacuum structure
of this model, what is the net probability of randomly finding a vacuum
that preserves $R$-symmetry?
In this simple example, 
the answer clearly depends on the variables over which we average.
If we consider $\xi$ to be the only variable as in Fig.~\ref{prefig} 
and perform our averaging 
with measure $d\xi$, we would claim that we have a virtual impossibility
of realizing such a vacuum for any given (fixed) value of $(m/g)^2$.
This is because the \vac{$\emptyset$} region in Fig.~\ref{prefig} 
ultimately occupies only a region of finite size (set of measure zero)
within the infinitely long $\xi$-axis.\footnote{In practice, of course,
   we must have $\xi \lsim M_P^2$ where $M_P$ is the Planck scale, 
    so our parameter space is not truly infinite in extent.  
   Also, the FI coefficients $\xi$ are typically quantized. 
   Throughout this paper, however, we shall  consider the continuous limit
    and imagine that our parameter spaces are infinite in extent.} 
On the other hand, if we consider both $\xi$ and $m$ to be our relevant
variables as in Fig.~\ref{prefigtwo} and calculate our average with measure 
$d\xi dm^2 = 2 m d\xi dm$,
we would find that this probability is given by
\beq
   \hbox{Prob (R~{\rm symm})} ~=~ 1 - {2\over \pi}\, \tan^{-1} g^2 ~.
\label{FImodelprobR}
\eeq
Indeed, this is nothing but the ratio of the angles of the pie-slices
in Fig.~\ref{prefigtwo}.
Of course, this result would change if we were to take our measure to
be, for example, $d\xi dm$ rather than $d\xi d m^2$.
Finally, if we wish to consider $g$ as a third defining degree of 
freedom, we need to
integrate the result in Eq.~(\ref{FImodelprobR}) over possible values of
$g$.  Clearly, if we use the measure $dg$, we would obtain a vanishing
probability because of the infinite relative weighting given to vacua with 
large $g$. 
However, motivated by the pie-slice diagram in Fig.~\ref{prefigtwo},
we can define an angular variable $\theta_g\equiv \tan^{-1} g^2$
and take our measure to be
\beq
         d \theta_g ~=~ {1\over 1+g^4} \,dg^2~.
\label{ameasure}
\eeq
This naturally suppresses the contributions from large-$g$ vacua,
and yields the integrated result
\beq
        P~\equiv~ {2\over \pi} \int_0^{\pi/2} d\theta_g  
             \left( 1-{2\over \pi}\theta_g\right)~=~
                {1\over 2}~.
\eeq
Of course, in practice we should perhaps only be integrating up to the
limits of our perturbative treatment, restricting to $g\lsim \sqrt{4\pi}$,
but this would introduce a possible strong dependence on the cutoff employed. 
Likewise, the measure in Eq.~(\ref{ameasure}) is only one 
of many possible measures we could use.
In any case, this discussion dramatically illustrates the importance
of clearly defining both our relevant variables, our 
integration measure, and all associated cutoffs  when averaging over multiple vacua.

It may seem strange to consider a coupling constant as a coordinate
parametrizing a space of vacua.  However, we recall that in string theory,
the coupling constant is related to the vacuum expectation value (vev) of 
the dilaton, and  such a vacuum expectation value is traditionally taken as 
one coordinate of an ensuing moduli space of vacua.

The issue of determining the appropriate landscape coordinates is 
clearly important, but unfortunately model-dependent. In the popular
compactifications with fluxes~\cite{fluxes}, the gauge couplings on D3-branes
depend on the dilaton which, in turn, depends on the fluxes. 
However, since we ultimately have to restrict our attention to gauge couplings 
at high scales 
which are compatible with the experimentally measured couplings at low scales,
these couplings should be $\sim {\cal O}(1)$.
A pragmatic attitude
would then be to focus directly on that subset of fluxes leading
to realistic gauge couplings.
By contrast, the FI terms depend on moduli fields and charge assignments; 
they are certainly landscape-dependent and are therefore natural landscape
coordinates.  The supersymmetric mass in the FI
model can also be generated by suitable fluxes, in analogy with
the $\mu$-term of MSSM~\cite{ibanez}.  
Likewise, the supergravity-induced soft mass
terms that we shall introduce in subsequent sections are also
flux-dependent.
Thus, we see that essentially all
mass-like parameters are natural landscape coordinates.  

An important constraint is that the FI terms and supersymmetric and soft
mass terms are bounded in any string compactifications.  Indeed, FI
terms are proportional to vev's of various moduli fields and to the sum of
$U(1)$ charges, and are always smaller than the string and the Planck
scale.  Likewise, the various fluxes must generically satisfy Ramond-Ramond
tadpole constraints;  in most cases, these constraints place
upper limits on their possible values.
Thus, in Figs.~\ref{prefig} and \ref{prefigtwo} (as well as all 
subsequent parameter-space diagrams in this paper), 
we should implicitly impose a cutoff
on  the FI terms and various mass parameters.
This cutoff is model-dependent, but 
should roughly be of size $\sim {\cal O}(M_P)$ where
$M_P$ is the Planck mass.

An interesting result of our analysis of the FI model in
Fig.~\ref{prefigtwo}  is that the vacuum structure is 
energy-dependent.  Indeed, if $Z_m(E)$ denotes
the wavefunction renormalization factor for the matter fields
(which we take to be equal for the two superfields)
and $Z(E)$ denotes
the wavefunction renormalization factor for 
the gauge superfield,
then the boundaries between the 
\vac{$+$} and \vac{$\emptyset$} regions 
and the \vac{$\emptyset$} and \vac{$-$} regions
will scale according to the energy-dependent 
expressions $\pm g^2 Z_m (E)/ Z (E)$. 
These boundaries therefore rotate
under the renormalization group.  This opens up the possibility of having
a vacuum which passes from one region to another under RG flow, 
a phenomenon which
can generate phase transitions with potentially interesting cosmological
implications.  We will study this phenomenon in more detail in Sect.~3,
within the framework of a more interesting 
(and also slightly more complicated) example.

%==================================================================================
\section{Constructing a toy model}

It turns out that the FI model is not sufficiently complex for our
purposes.  In this section, we shall therefore turn to the next-simplest model which,
as we shall see,
will exhibit almost all of the relevant features which we shall    
encounter when we proceed to consider more complicated situations
in subsequent sections.  In particular, we shall see that this 
toy model gives rise to
a non-trivial ``landscape'' consisting of multiple 
stable (or metastable) vacua with different low-energy phenomenologies,
unstable extrema, phase transitions, etc.   
Moreover, even though this toy model is relatively simple,
we expect that the resulting landscape is literally a component of the
full string-theory landscape in cases of string models with multiple
$U(1)$ gauge factors. 

Our model consists of two $U(1)$ gauge symmetries, denoted $U(1)_1$ and $U(1)_2$,
and three $\calN=1$ chiral superfields, $\Phi_{i=1,2,3}$.  The charges of these 
chiral superfields under the $U(1)$ gauge symmetries are chosen as in Table~\ref{table1}.
In a string-theory context, such $U(1)$ gauge factors can be imagined as arising from
different D-branes, and the $\Phi_i$ fields can arise as strings stretched
between these branes.
We shall also assume that the $\calN=1$ supersymmetry is broken by Fayet-Iliopoulos
$D$-term coefficients $\xi_1$ and $\xi_2$, and by a renormalizable 
Wilson-line superpotential of the form
\beq
             W ~=~ \lambda\, \Phi_1 \Phi_2 \Phi_3~. 
\label{superpot}
\eeq
Our model is thus defined by three parameters, $\lbrace \xi_1, \xi_2, \lambda\rbrace$,
and our goal will be to study the vacuum structure of this model as a function of these
parameters. 
Of course, the resulting physics is unchanged if $\lambda\to -\lambda$.
We shall therefore restrict ourselves to $\lambda\geq 0$ for simplicity.

\begin{table}[t]
\centerline{
   \begin{tabular}{||c||cc||}
   \hline
   \hline
    $~$ &  $U(1)_1$  & $U(1)_2 $  \\
   \hline
   \hline
    $\Phi_1$  &  $-1$ & 0  \\
    $\Phi_2$  &  $1$ & $-1$  \\
    $\Phi_3$  &  $0$ & $1$  \\
   \hline
   \hline
   \end{tabular}
 }
\caption{$U(1)$ charge assignments for chiral superfields in our toy model.}
\label{table1}
\end{table}

It is straightforward to analyze the vacuum states of this theory.
As usual, the scalar potential is given by
\beq
                  V ~=~ \half \sum_a g_a^2 D_a^2 ~+~ \sum_i |F_i|^2
\label{potential}
\eeq
where the $D$- and $F$-terms are given by
\beq 
            D_a ~=~ \sum_i \, q_i^{(a)} |\phi_i|^2 ~+~ \xi_a~,~~~~~~~
            F_i~=~ {\partial W\over \partial \phi_i}~.
\label{potentialtwo}
\eeq
Here $a=1,2$ is the index of the gauge group $U(1)$ factor,
$g_a$ is the gauge coupling corresponding to the $U(1)_a$ factor,
and $i=1,2,3$ is the index of the chiral superfield.  Thus, 
$q_i^{(a)}$ denotes the $U(1)_a$ charge of $\Phi_i$.   In most
of our considerations the gauge couplings $g_a$ will not be important,
so we will henceforth consider $g_1=g_2 = 1$ for simplicity.
We will reinstall gauge couplings
whenever relevant for our analysis.
Our task is to determine the extrema of $V$ by seeking solutions to the simultaneous
equations
\beq
                      {\partial V\over \partial \phi_i}~=~
                      {\partial V\over \partial \phi^\ast_i}~=~0~,
\label{extrema}
\eeq
and then to determine whether these extrema represent stable (or metastable) vacua by
calculating the eigenvalues of the corresponding mass matrix
\beq
  {\cal M}^2 ~\equiv~ \pmatrix{\displaystyle
   {\partial^2 V \over \partial \phi^\ast_i  \partial \phi_j}  &
   \displaystyle  {\partial^2 V \over \partial \phi^\ast_i  \partial \phi^\ast_j} \cr
   \displaystyle {\partial^2 V \over \partial \phi_i  \partial \phi_j}  &
   \displaystyle {\partial^2 V \over \partial \phi_i  \partial \phi_j^\ast} \cr}
\label{massmatrix}
\eeq
evaluated at the extrema.
Note that in general, there will be a zero eigenvalue for each spontaneously 
broken $U(1)$;  these eigenvalues correspond to the resulting Nambu-Goldstone bosons.
The extrema defined by Eq.~(\ref{extrema}) represent
stable (or metastable) vacua only if each of the remaining eigenvalues is
positive.

Just as with the FI model in Sect.~2, it will prove convenient to group the extrema of $V$ 
into classes depending on which combinations of chiral superfields receive non-zero
vacuum expectation values.  This classification will help in determining such features
as whether the sources of SUSY-breaking are primarily $D$-terms or $F$-terms,
and whether they are likely to lead to $R$-symmetry breaking when incorporated into a supergravity framework.  
Since there are three chiral superfields in this example, there are
correspondingly $8=2^3$ classes of extrema of $V$.  
Denoting $\langle \Phi_i \rangle \equiv v_i$,
we shall define our classes of extrema according to their values of $v_i$,
using the notation \vac{ijk...} to indicate the class of vacua
in which $v_i$, $v_j$, $v_k$, ... are all non-zero (with
\vac{$\emptyset$} denoting the vacua in which all $v_i$ vanish).
Note that in this toy model, we can choose all $v_i$ to be real
without loss of generality.

%===========================================================================================
\subsection{$\lambda=1$}

Let us first consider the vacuum structure of this model when $\lambda=1$.
Our results are then as follows.
We find that \vac{$\emptyset$} extrema exist for all values of $(\xi_1,\xi_2)$, but are always unstable.
This result, of course, is intuitively obvious, since \vac{$\emptyset$} vacua 
are essentially at the unstable center of the SUSY-breaking ``Mexican hat''. 
By contrast, \vac{1} extrema exist only for $\xi_1>0$.  However, these extrema are 
stable only within the subregion defined by $|\xi_2|< \xi_1$. 
Likewise, \vac{2} extrema exist
for $\xi_2>\xi_1$ and are stable in the region $\xi_1<0$, $\xi_2>0$,
while \vac{3} extrema exist only for $\xi_2 <0$ and are stable 
within the subregion $|\xi_1| < |\xi_2|$. 
Similarly, \vac{12} extrema exist only for $0<\xi_1<\xi_2$ and are stable
everywhere within this region, while \vac{23} extrema exist only for
$\xi_1<\xi_2<0$ and are also stable everywhere within this region.
Finally, for $\lambda=1$, 
extrema in Classes~\vac{13} and \vac{123} do not exist 
for any values of $(\xi_1,\xi_2)$.

\begin{figure}[th]
\centerline{ 
   \epsfxsize 3.8  truein \epsfbox {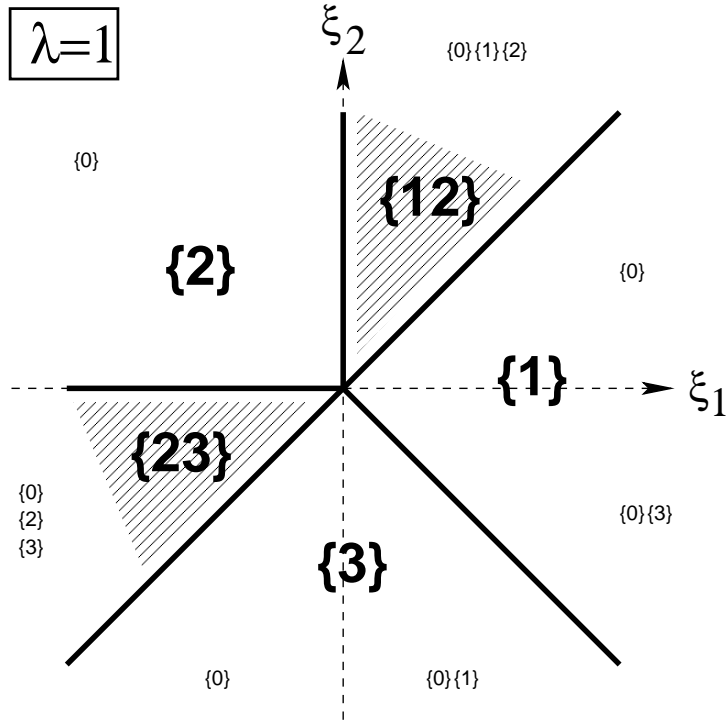}
    }
\caption{The ``landscape'' of our toy model with two $U(1)$ gauge fields
and three chiral superfields, sketched for $\lambda=1$.
The different classes of vacua (labelled according to which chiral
fields receive non-zero vacuum expectation values) 
occupy the
non-overlapping regions indicated above in bold type, while unstable extrema
in each region are indicated in smaller type.  We have shaded the regions 
corresponding to vacua with two non-zero vacuum expectation values 
(Classes~\vac{12} and \vac{23});  these are regions in which $R$-symmetry
is potentially broken.}
\label{fig1}
\end{figure}

It is clear from these results that for $\lambda=1$, 
there are no regions in $(\xi_1,\xi_2)$
parameter space in which there are multiple stable vacua.
Instead, the different classes of vacua occupy non-overlapping regions
which completely fill out this space.
This situation is sketched in Fig.~\ref{fig1}.

%================== FIGURE ============================================
%================== END OF INSERTED FIGURE ============================

It is important to interpret the diagram in Fig.~\ref{fig1} correctly.
Each point in the $(\xi_1,\xi_2)$ plane represents a unique theory,
\ie, a unique model parametrized by its defining parameters $\xi_1$ and
$\xi_2$.  Thus, this plane is nothing but a field-theoretic ``landscape''
for this two-$U(1)$ theory with three chiral superfields.
For each point in the landscape, there is a 
unique vacuum whose energy can be calculated from Eq.~(\ref{potential}).
For example, for models within the region 
specified by $\xi_2> -\xi_1$ and $\xi_2<0$ 
(\ie, the lower half of the \vac{1} region in Fig.~\ref{fig1}), the 
energy of the stable \vac{1} extremum is 
given by $V_\vac{1}=\xi_2^2/2$.
This is to be contrasted with the {\it unstable}\/ \vac{$\emptyset$} and \vac{3}
extrema which also exist for models in this region, 
with energies given by $V_\vac{\emptyset}= (\xi_1^2+\xi_2^2)/2$ and
$V_\vac{3}= \xi_1^2/2$ respectively.
Note that indeed $V_\vac{1}<V_{\vac{\emptyset},\vac{3}}$.

At this stage, we can already reach an interesting conclusion:
for the $\lambda=1$ landscape, the average ratio of stable extrema relative to all
extrema is only
\beqn
 \hbox{Probability that extremum is stable} 
             &=& {1\over 8}\left( {1\over 4} + {1\over 2} + {1\over 3} + {1\over 3} + {1\over 2}
             + {1\over 4} + {1\over 2} + {1\over 2} \right)  \nonumber\\
                ~&=& {19\over 48}~\approx~ 0.396~.
\label{prob1}
\eeqn
Note that in calculating this probability, we are averaging over the landscape
with a uniform measure (averaging the stability probabilities in each of the 
eight octants of the plane), 
imagining that all values of our landscape coordinates
$(\xi_1,\xi_2)$ are equally likely.
This result is significantly larger than the value $(1/2)^6$ which we 
might have na\"\i vely
expected from the fact that our mass matrix in Eq.~(\ref{massmatrix})
has six eigenvalues,
any one of which randomly could have been negative.
Even if we allow for potentially as many as two zero eigenvalues
corresponding to Nambu-Goldstone bosons for the two $U(1)$ gauge factors
in this toy model, we still have a much higher probability of finding
stable extrema than expected.

In a similar vein, we can also see from Fig.~\ref{fig1} that 
a randomly chosen model in the landscape has exactly a 25\% chance
of breaking $R$-symmetry when $\lambda=1$.  Those regions populated by $R$-symmetry
breaking vacua are shaded in Fig.~\ref{fig1}.  

In this landscape, there do not exist any regions
with overlapping stable vacua.  In other words, there are no 
values of $(\xi_1,\xi_2)$ giving rise to
multiple stable vacua.  Thus, each vacuum is absolutely
stable with respect to dynamics within that model.  However,
these different vacua fill out a landscape in theory space,
as sketched in Fig.~\ref{fig1}, and these vacua ultimately 
have different energies which depend on the 
landscape coordinates $(\xi_1,\xi_2)$.
For $\lambda=1$, the different classes of extrema have energies:
\beqn
         V_\vac{\emptyset} &=&  \half\, (\xi_1^2+\xi_2^2)~=~V_\vac{1}+V_\vac{3} \nonumber\\
         V_\vac{1} &=&  \half\, \xi_2^2\nonumber\\
         V_\vac{2} &=&  \quart\,( \xi_1 + \xi_2)^2 \nonumber\\
         V_\vac{3} &=&  \half\, \xi_1^2\nonumber\\
         V_\vac{12} &=&  \quart\,( -\xi_1^2 + 2\xi_1\xi_2 + \xi_2^2)~=~ 
                                                 V_\vac{2} -V_\vac{3}\nonumber\\
         V_\vac{23} &=&  \quart\,( \xi_1^2 + 2\xi_1\xi_2 - \xi_2^2)~=~ V_\vac{2}-V_\vac{1}~.
\label{vacenergies}
\eeqn

Moreover, note from Eq.~(\ref{vacenergies}) that Class~\vac{1} vacua with $\xi_2=0$,
Class~\vac{2} vacua with $\xi_2= -\xi_1$, and 
Class~\vac{3} vacua with $\xi_1=0$ are all SUSY-preserving, 
with $V=0$.
Even though these SUSY-preserving vacua only form
a one-dimensional space of vacua within the larger two-dimensional
landscapes we have plotted (thereby essentially forming a set of measure zero
within the full  landscape),  
these vacua represent the lowest-energy vacua within their
respective regions.  
However, unless
transitions between different vacua in the landscape are somehow possible,
the existence of such supersymmetric vacua at one landscape location
does not destabilize the non-supersymmetric vacua
that may exist at other landscape locations.

We see, then, that this ``landscape'' is indeed a very
realistic, and ultimately completely calculable,
model of the string-theoretic landscape. 
Indeed, we have a space of models, 
with each parametrized by the model-defining
parameters $(\xi_1,\xi_2)$, 
and each with a different vacuum
and vacuum energy.

%===========================================================================================
\subsection{$\lambda<1$}

\begin{figure}[ht]
\centerline{ 
   \epsfxsize 3.1 truein \epsfbox {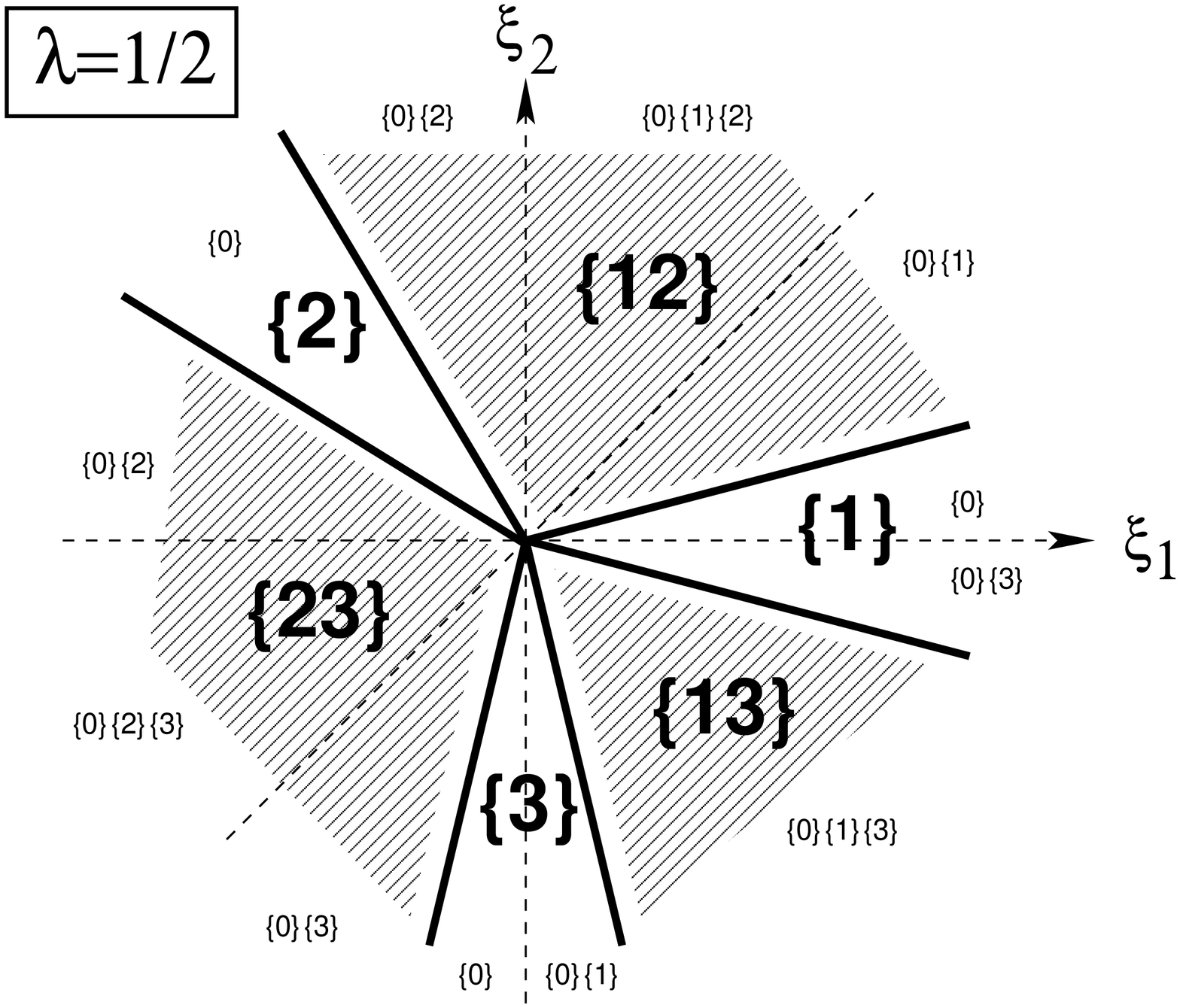}
   \epsfxsize 3.1 truein \epsfbox {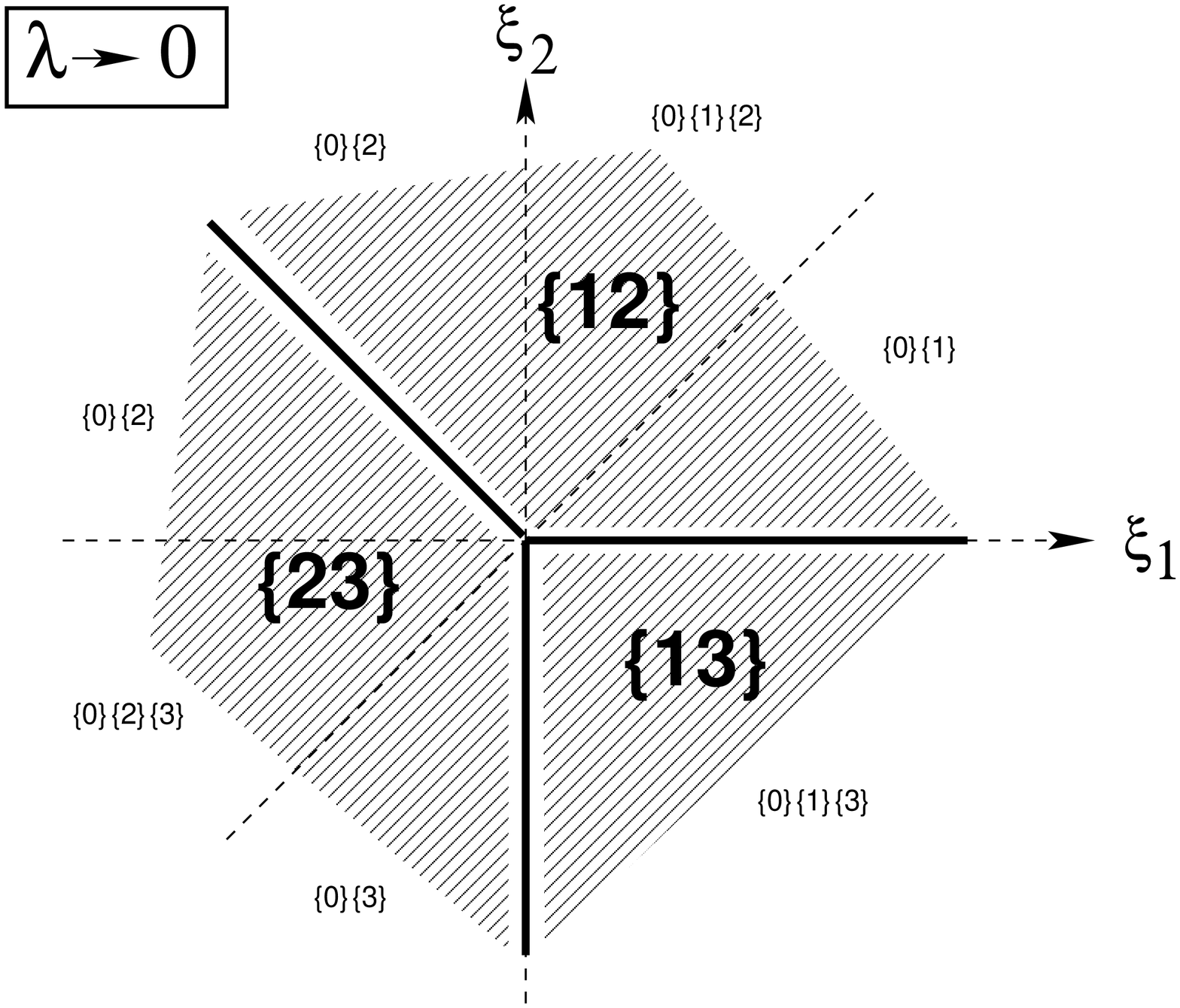}
    }
\caption{(a) {\it Left figure:} Same landscape as in Fig.~\protect\ref{fig1},
now sketched for $\lambda=1/2$.
The \vac{1}, \vac{2}, and \vac{3} regions have each become smaller, remaining equal in size, 
while the \vac{12} and \vac{23} regions
(as well as the new \vac{13} region) have expanded to fill in the
gaps.  Just as for $\lambda=1$, 
the different classes of vacua 
occupy non-overlapping regions in this landscape, and we have indicated the
unstable extrema in parentheses.  (b) {\it Right figure:} As $\lambda\to 0$, the
\vac{1}, \vac{2}, and \vac{3} regions
disappear entirely, leaving a landscape entirely dominated by 
non-overlapping $\vac{12}$, \vac{23}, and \vac{13} regions.  These regions
ultimately represent flat directions in the $\lambda=0$ limit.}
\label{fig2}
\end{figure}

It is important to consider how this landscape picture
is modified for $\lambda\not=1$. 
Considering first the case with $\lambda<1$,
we find that
Class~\vac{$\emptyset$} extrema continue to exist for all $(\xi_1,\xi_2)$
and continue to be unstable everywhere.
Class~\vac{1} extrema, by contrast, continue to exist for all $\xi_1>0$,
but are now stable only within the smaller region $|\xi_2|< \lambda^2 \xi_1$. 
Similarly, 
Class~\vac{2} extrema continue to exist for all $\xi_2>\xi_1$ and 
are stable only within the region defined by 
\beq
     (1-\lambda^2)\, \xi_2~ <~-(\lambda^2+1)\, \xi_1~,~~~~~
     (\lambda^2+1)\,\xi_2~>~ (\lambda^2-1)\, \xi_1~,  
\label{pieslice}
\eeq
while 
Class~\vac{3} extrema continue to exist for all $\xi_2<0$
but are stable only within the smaller region $|\xi_1| <\lambda^2 |\xi_2|$.
All of these results reduce to the previous case as $\lambda\to 1$.
In general, these results indicate that the \vac{1}, \vac{2}, and \vac{3} regions
become smaller
as $\lambda\to 0$, occupying narrower and narrower ``pie-slices''
in the $(\xi_1,\xi_2)$ landscape plane.
Specifically, each of these ``pie-slices'' has 
total angle $\theta_{\vac{1},\vac{2},\vac{3}}=2\theta_\lambda$,
where
\beq
           \theta_\lambda ~\equiv~  \tan^{-1} \lambda^2~,
\label{thetadef}
\eeq
and differ only in their orientations in the
$(\xi_1,\xi_2)$ plane, with  the Class~\vac{1} pie-slice centered around 
the positive $\xi_1$-axis,
the Class~\vac{2} pie-slice centered around the $\xi_2 = -\xi_1 >0$ diagonal axis,
and the Class~\vac{3} pie-slice centered around the negative $\xi_2$-axis.
Thus, as $\lambda\to 0$, the \vac{1}, \vac{2}, and \vac{3} regions disappear
entirely.
A sketch of these results for $\lambda=1/2$ 
is shown in Fig.~\ref{fig2}(a).

%================== FIGURE ============================================
%================== END OF INSERTED FIGURE ============================

Just as in the $\lambda=1$ case, extrema in Classes~\vac{12} and \vac{23} continue 
to exist in the gaps between the \vac{1} and \vac{2} regions,
and \vac{2} and \vac{3} regions, respectively.
As evident from Fig.~\ref{fig2}(a), each of these regions has 
angle $\theta_{\vac{12},\vac{23}}=3\pi/4-2\theta_\lambda$.
Moreover, these extrema are stable everywhere within these regions.
Thus, as $\lambda\to 0$, these regions become larger rather than
smaller.  The new feature for $\lambda<1$, however, is the emergence
of new Class~\vac{13} extrema which populate the gap that has opened
up between the \vac{1} and \vac{3} regions, with angle 
$\theta_\vac{13}=\pi/2-2\theta_\lambda$.
Just as with the extrema in Classes~\vac{12} and \vac{23},
extrema in Class~\vac{13} are stable wherever they exist.
There continue to be no extrema, stable or unstable, in
Class~\vac{123} for $\lambda>0$.

We see, then, that just as for $\lambda=1$, 
the different classes of vacua 
continue to occupy non-overlapping regions for $\lambda<1$.
Indeed, as $\lambda\to 0$, the \vac{1}, \vac{2}, and \vac{3} regions
disappear entirely, leaving a landscape entirely dominated by 
non-overlapping \vac{12}, \vac{13}, and \vac{23} regions, 
with angles 
$\theta_{\vac{12},\vac{23}}= 3\pi/4$ and $\theta_\vac{13}=\pi/2$ respectively.
This situation is illustrated in Fig.~\ref{fig2}(b).

The $\lambda=0$ limit is somewhat special.
In this limit, of course,
we find that the three regions in Fig.~\ref{fig2}(b) continue to exist.
However, rather than representing stable vacua, each of these regions
now represents a class of {\it flat directions}.
This makes sense, since SUSY is unbroken for these extrema in the $\lambda=0$ limit, even for
arbitrary values of the FI coefficients $\xi_i$.
Likewise, $R$-symmetry is preserved in these regions.
Of course, the unstable extrema listed in Fig.~\ref{fig2}(b) continue
to represent solutions with broken SUSY and positive vacuum energies,
even when $\lambda=0$.

It is easy to interpret this flat direction in terms of the 
vacuum expectation values $v_i$.
Recall that for any $\lambda>0$, there are no extrema in the \vac{123} class.
However, for $\lambda=0$, a new \vac{123} extremum appears, corresponding to the
general solution
\beq
          |v_1|^2 ~= ~ |v_2|^2 + \xi_1~,~~~~~~~
          |v_3|^2 ~= ~ |v_2|^2 - \xi_2~.
\label{gensoln}
\eeq
These equations have solutions for $(v_1,v_2,v_3)$ for all $(\xi_1,\xi_2)$,
and thus can be considered to exist throughout the entire $(\xi_1,\xi_2)$ landscape.
They are also SUSY-preserving, maintaining $V=0$ for all choices of $|v_2|^2$ in
Eq.~(\ref{gensoln}), and serve as the 
algebraic representation of the flat direction when $\lambda=0$.
Indeed, depending on the value of $(\xi_1,\xi_2)$, the
corresponding \vac{12}, \vac{13}, or \vac{23} solution 
that appears in Fig.~\ref{fig2}(b) is only a special case 
of the more general \vac{123} solution in Eq.~(\ref{gensoln}).
For example, the \vac{12} region in Fig.~\ref{fig2}(b), which corresponds to
solutions with $|v_1|^2=\xi_1+\xi_2$, $|v_2|^2=\xi_2$, and $|v_3|^2=0$, can be viewed
as a special case of Eq.~(\ref{gensoln}) with $|v_2|^2=\xi_2$.
Thus, when $\lambda=0$, we should more correctly state 
that the entire $(\xi_1,\xi_2)$ landscape
can be described as a single flat direction 
corresponding to the \vac{123} solution
in Eq.~(\ref{gensoln}).

%===========================================================================================
\subsection{$\lambda>1$}

Let us now turn to the situation when $\lambda >1$.
In such cases, the above algebraic conditions for the existence and stability
of the \vac{$\emptyset$}, \vac{1}, \vac{2}, and \vac{3} extrema 
do not change.  
Thus, as $\lambda$ grows beyond $1$, we see that the pie-slices occupied by stable
extrema in these classes actually grow rather than shrink, with angles still
given by $\theta_{\vac{1}, \vac{2}, \vac{3}}=2\theta_\lambda= 2\tan^{-1}\lambda^2$.
Indeed, as $\lambda\to\infty$, each of these stable subregions expands to
fill its corresponding existence region, so that {\it all}\/
extrema in these classes become stable as $\lambda\to \infty$.

Given these results, we see that Regions~\vac{1} and \vac{3} actually overlap
for $\lambda>1$, with overlap angle $\theta_{\vac{1}+\vac{3}}=2\theta_\lambda-\pi/2$.  
Likewise, the \vac{2} region separately overlaps with the \vac{1} region
and with the \vac{3} region for $\lambda> \lambda_\ast$, where 
\beq
        \lambda_\ast~\equiv~ (1+\sqrt{2})^{1/2} ~\approx~ 1.55
     ~~~~~~ \Longleftrightarrow ~~~~~~
         \theta_\ast ~\equiv~ \tan^{-1}\lambda_\ast^2 ~=~ {3\pi\over 8}~.
\label{lambdaastdef}
\eeq
In the latter case, these overlap regions have 
angles $\theta_{\vac{1}+\vac{2}}=\theta_{\vac{2}+\vac{3}}= 
2\theta_\lambda-3\pi/4= 2(\theta_\lambda-\theta_\ast)$.
Thus, we see that we now have cases 
in which a single point in the landscape 
can give rise to {\it multiple}\/ (meta)stable extrema.
Of course, only the extremum with the lowest energy in each case is truly stable within
the specified model,
while the remaining extremum is only metastable.  
In this example, we find from Eq.~(\ref{vacenergies})
that the \vac{1} vacuum is truly stable
for $\xi_2<\xi_1$, while the \vac{3} vacuum is truly stable for
$\xi_1<\xi_2$.  This conclusion holds for all $\lambda >1$.

We stress that this metastability exists within a single point in the landscape
(\ie, within a single model characterized by specific values of $\xi_1$ and $\xi_2$). 
This is quite different from 
instabilities
 {\it across}\/ different models at different landscape locations;  such
theory-to-theory transitions are beyond the framework of quantum field theory,
but have been conjectured to occur in string theory.  
If such transitions can occur in string theory,
then we can expect the full dynamics on such a landscape to involve both
intra-model as well as inter-model metastability.
However, there is no evidence that such inter-model transitions can occur.

\begin{figure}
\centerline{ 
   \epsfxsize 4.0 truein \epsfbox {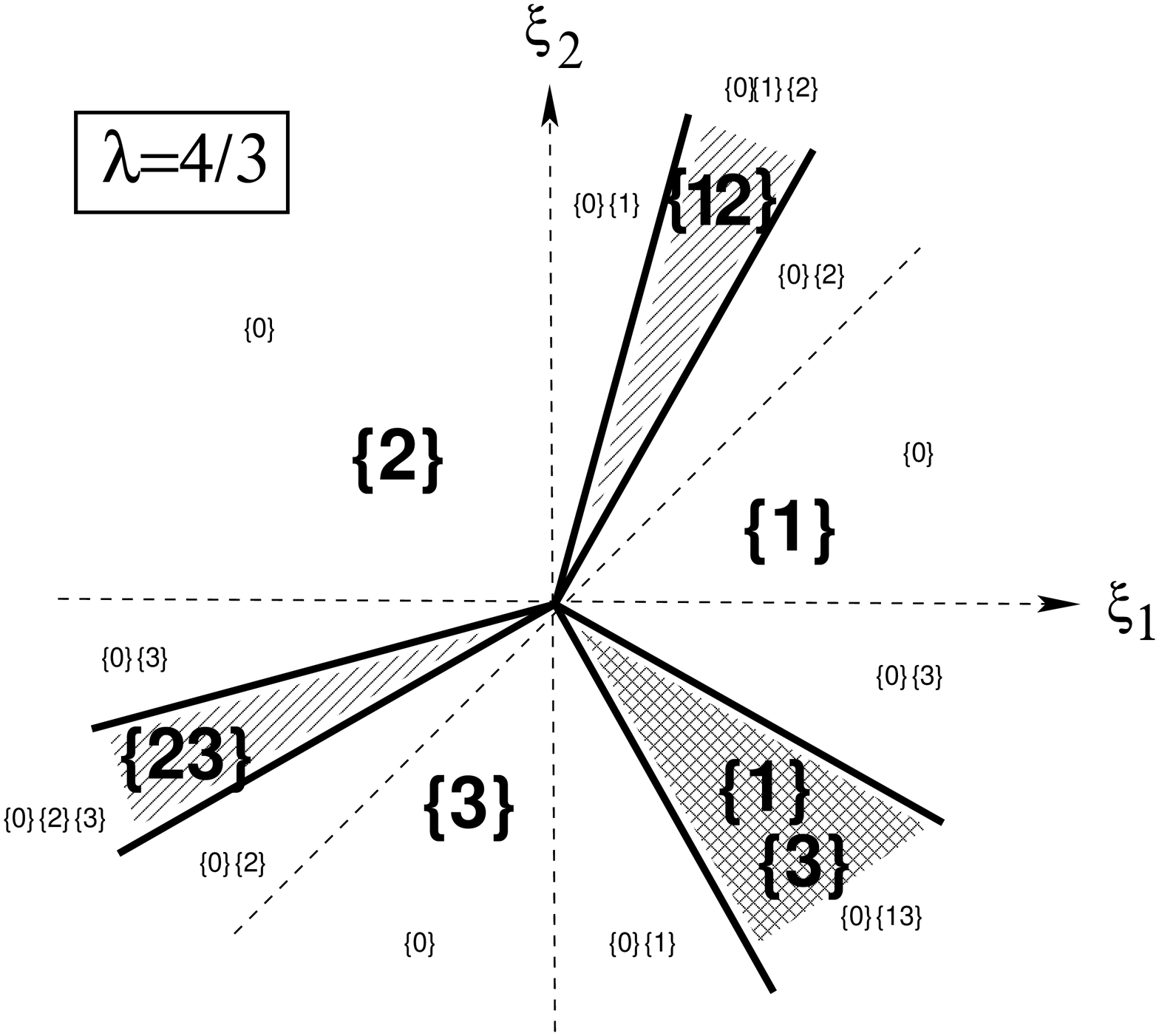}
    }
\caption{Same landscape as in previous figures,
now sketched for $\lambda=4/3$.
Regions~\vac{1}, \vac{2}, and \vac{3} have each become larger, with a new overlap region
(indicated with double cross-hatching) between the \vac{1} and \vac{3} regions.
Class~\vac{13} extrema exist but are now unstable in this overlap region,
while stable Class~\vac{12} and \vac{23} vacua continue to populate 
the gaps between Regions~\vac{1} and \vac{2}, and Regions~\vac{2} and \vac{3}, respectively.
There continue to be no overlaps amongst Regions~\vac{1}/\vac{12}/\vac{2} or 
Regions~\vac{2}/\vac{23}/\vac{3}.} 
\label{fig3}
\centerline{ 
   \epsfxsize 3.4 truein \epsfbox {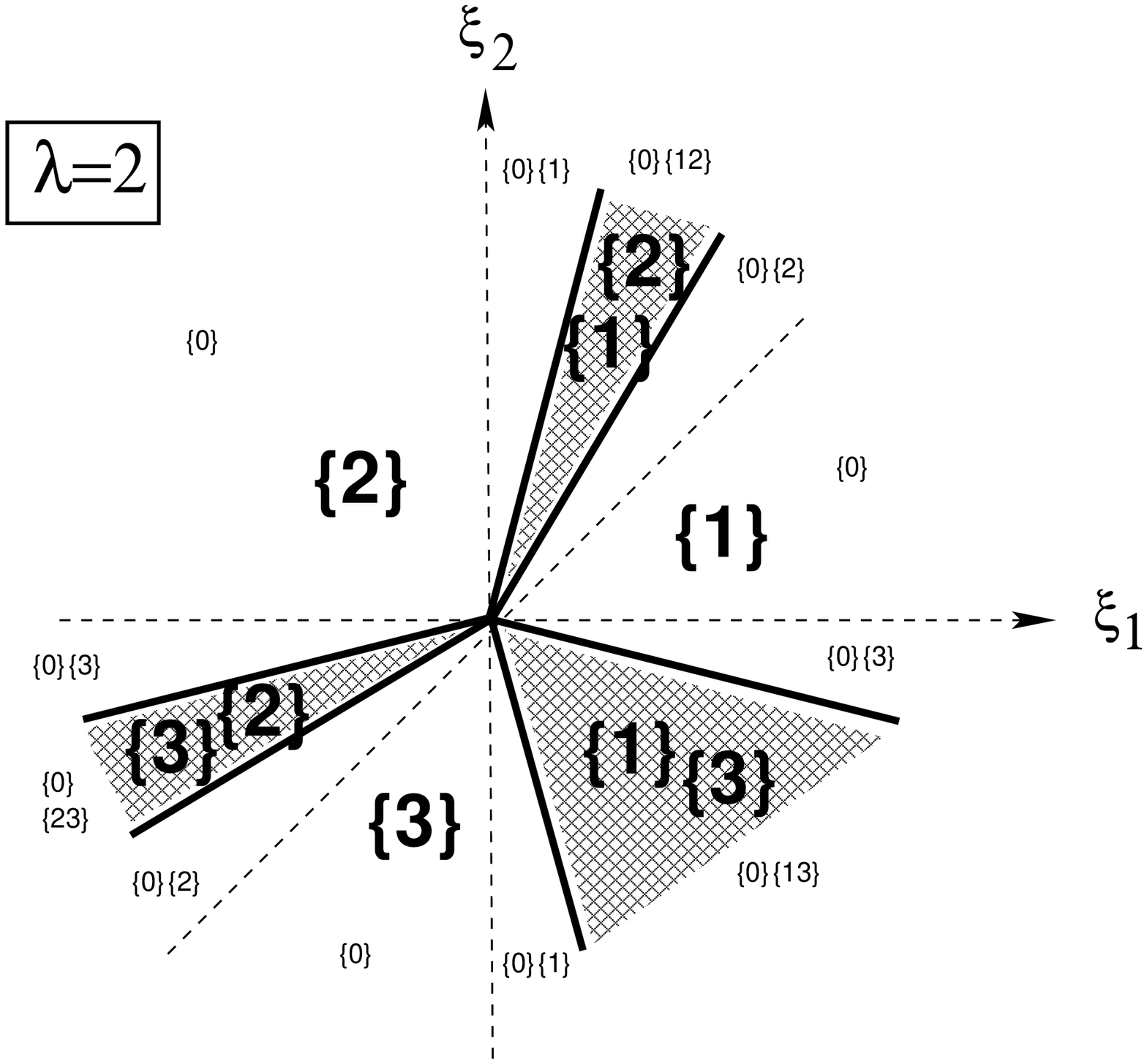}
   \hskip -0.3 truein
   \epsfxsize 3.4 truein \epsfbox {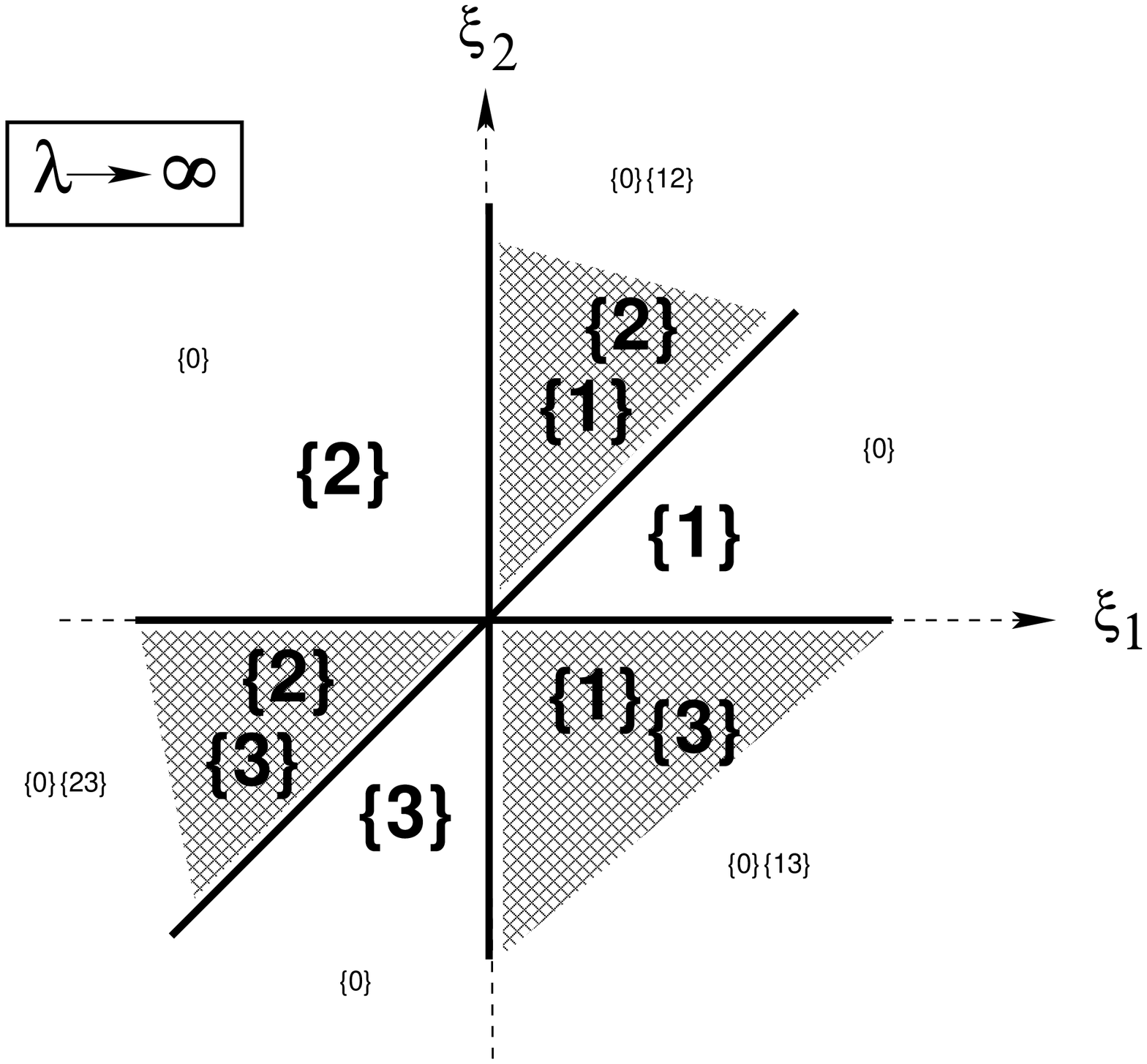}
    }
\caption{Same landscape as in previous figures,
now sketched (a) for $\lambda=2$ (left), and (b) as $\lambda\to\infty$ (right).
Since $\lambda>\lambda_\ast$ in each case, 
the \vac{1}, \vac{2}, and \vac{3} regions all mutually overlap; 
these three overlap regions are 
indicated with double cross-hatching.
Extrema in Classes~\vac{12}, \vac{13}, and \vac{23} exist in these overlap 
regions, but are always unstable for $\lambda>\lambda_\ast$.  } 
\label{fig4}
\end{figure}

An important question is to determine the 
fate of the \vac{12}, \vac{13}, and \vac{23} extrema for $\lambda>1$.
For $1<\lambda<\lambda_\ast$,
extrema in the \vac{12} and \vac{23} classes 
continue to exist within their remaining gap regions
and are always stable, with angles $\theta_{\vac{12},\vac{23}}=3\pi/4-2\theta_\lambda$. 
Moreover, for all $\lambda>1$,
\vac{13} extrema actually re-emerge within the 
Class~\vac{1}$+$\vac{3} {\it overlap}\/ region, but are now {\it unstable}.
As always in this toy model, there are no extrema in Class~\vac{123} except at $\lambda=0$.
We thus obtain a landscape as sketched in Fig.~\ref{fig3}.

%================== FIGURE ============================================
%================== END OF INSERTED FIGURE ============================

The final remaining case corresponds to $\lambda>\lambda_\ast$.
For $\lambda$ in this range,
the \vac{1}, \vac{2}, and \vac{3} regions all mutually overlap in a pairwise fashion.
(Note that these three regions never have a simultaneous common overlap.)
In such cases, we then find that extrema in Classes~\vac{12}, 
\vac{13}, and \vac{23} all
continue to exist in their respective overlap regions, but that 
all of these extrema are now unstable.
We thus obtain a landscape as sketched in Fig.~\ref{fig4}(a).
Note that as $\lambda\to \infty$, 
our landscape reduces to that sketched in Fig.~\ref{fig4}(b).

\vfill
\eject

%===========================================================================================
\subsection{Landscape probabilities and integration measures}

\begin{figure}[hb]
\centerline{ 
   \epsfxsize 3.3  truein \epsfbox {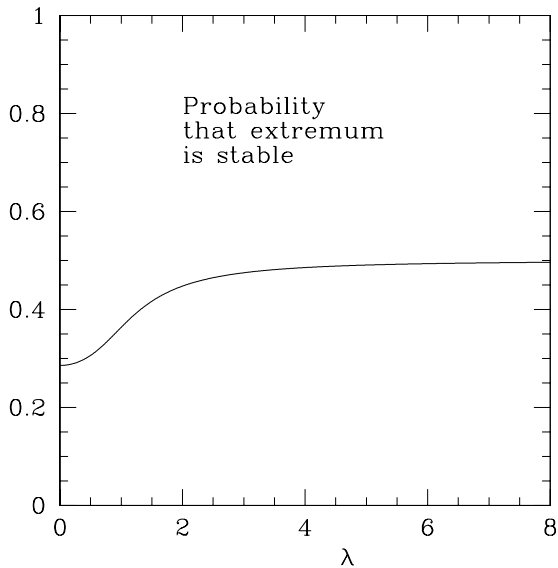}
    }
\caption{The probability that a given extremum in our toy model is stable,
   plotted as a function of the Yukawa coupling $\lambda$ and integrated
   over all $(\xi_1,\xi_2)$.
   This function increases monotonically from $7/24$ when $\lambda=0$,
   and asymptotes to $1/2$ as $\lambda\to\infty$.
   Note that in all cases, this probability is significantly greater
   than we would have na\"\i vely expected based on a random 
   assignment of signs for the eigenvalues of the six-dimensional mass matrix.}
\label{plotfig}
\end{figure}

Just as in the $\lambda=1$ case, we can calculate the fraction of {\it stable}\/ 
extrema relative to all extrema  as a function of $\lambda$;
we merely evaluate this fraction in each 
relevant pie-slice, and then properly weight these pie-slices according to
their opening angles. 
It turns out that as a function of $\lambda$, 
this fraction is given by
\beq
 \hbox{Prob (stable;$\,\lambda$)} ~=~\cases{
      (7\pi + 10 \tan^{-1}\lambda^2 )/ 24\pi &  for $\lambda\leq 1$~,\cr
      (15\pi + 16 \tan^{-1}\lambda^2 )/ 48\pi &  for $1\leq \lambda\leq \lambda_\ast$~,\cr
      (\pi + 2 \tan^{-1}\lambda^2 )/ 4\pi &  for $\lambda\geq \lambda_\ast$~.\cr}
\eeq
%  It is remarkable that this single function applies
%  for all $\lambda$, given that our landscape undergoes relatively
%  dramatic changes at $\lambda=1$ and $\lambda=\lambda_\ast$, with
%  new regions emerging, overlapping regions developing, and stable vacua 
%  becoming unstable.
Note that this probability agrees with Eq.~(\ref{prob1}) when $\lambda=1$,
and generally increases monotonically from $7/24 \approx 0.29$ at $\lambda= 0$
to exactly $1/2$ at $\lambda\to\infty$.
For example, it is straightforward to verify from Fig.~\ref{fig4}(b) 
that this fraction is indeed $1/2$ in the $\lambda\to \infty$ limit.
A plot of this probability as a function of $\lambda$ is shown in Fig.~\ref{plotfig}.
Once again, we stress that 
we are averaging over our $(\xi_1,\xi_2)$ landscape with a uniform measure
when calculating this probability. 
As already noted below Eq.~(\ref{prob1}), we see that
this probability is significantly greater for all values of $\lambda$
than we might have na\"\i vely expected.

%================== FIGURE ============================================
%================== END OF INSERTED FIGURE ============================

Of course, $\lambda$ is actually a landscape coordinate;  
just as with $\xi_1$ and $\xi_2$, the value of the 
parameter $\lambda$ specifies a particular model under study.\footnote{
   In a realistic setting,
   all three of these quantities may change dynamically 
   under renormalization-group flow.
   This issue will be discussed further in Sect.~3.5.
   However, for the purposes of studying the vacuum structure of our toy
   model, we are considering these quantities as primordial fixed parameters 
   defining our specific model.}  
In other words, as discussed at the beginning of this section,
our toy model really has a {\it three}\/-dimensional landscape, 
with these figures showing only particular fixed-$\lambda$ slices through
this landscape.
Thus, in calculating the probability of finding a stable extremum 
from amongst all extrema, we should really be averaging over $\lambda$ 
(our third direction) as well.

However, this illustrates another critical issue for the landscape.
What measure should we use when performing our $\lambda$-integration?
If we were to perform our averaging process using the measure $d\lambda$,
we would find that the resulting
average probability would be close to $1/2$  
because of the overwhelming dominance of the contributions
from large $|\lambda|$.
This suggests that the simple measure $d\lambda$ may not be the most appropriate
measure for the Yukawa coupling $\lambda$, even with an appropriate perturbative cutoff.
As briefly mentioned in Sect.~2, another option prompted by the geometry of 
our landscape 
might be to consider an alternate
measure based on our {\it angular}\/ Yukawa variable $\theta_\lambda\equiv \tan^{-1}\lambda^2$: 
\beq
             d \theta_\lambda ~=~ 
                           {d\lambda^2 \over 1+\lambda^4}~.
\label{angularmeasure}
\eeq
Note that this measure suppresses the contributions from large $|\lambda|$,
as we desired, yielding a finite result.
For example, using this angular measure, we find that the average 
probability for finding stable extrema 
becomes
\beq
            \hbox{Prob$\,$(stable)} ~=~ 
        {2\over \pi} \int_0^{\pi/2} d \theta_\lambda ~ \hbox{Prob$\,$(stable;$\,\lambda$)} ~=~ 
                     {151\over 384}~\approx~ 0.393~.
\eeq 
One could also implement a perturbative cutoff for $\theta_\lambda$,
again yielding cutoff-sensitive results.
However, regardless of the particular measure or cutoffs chosen,
this example illustrates in a graphic and concrete way the need to have
an unambiguous way of determining appropriate measures and cutoffs
in order to meaningfully calculate
any statistical quantities on the landscape.

\begin{figure}[bht]
\centerline{ 
   \epsfxsize 3.3  truein \epsfbox {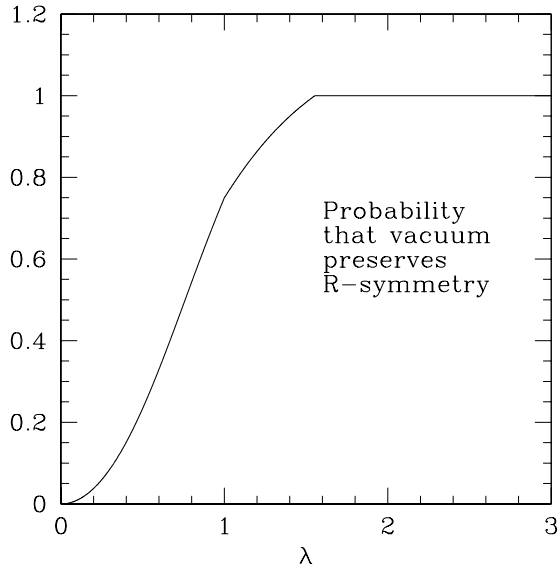}
    }
\caption{The probability that a given vacuum in our toy model preserves
   $R$-symmetry,
   plotted as a function of the Yukawa coupling $\lambda$ and integrated
   over all $(\xi_1,\xi_2)$.
   This function increases monotonically from zero when $\lambda=0$,
   and saturates at $1$ for $\lambda\geq \lambda_\ast$.}
\label{plotfig2}
\end{figure}

Another quantity that we can calculate in our toy model is the probability that 
a given (stable) vacuum state preserves $R$-symmetry.
In our toy model, such $R$-symmetry-preserving vacua
are those in Classes~\vac{1}, \vac{2}, and \vac{3}, 
while the vacua in Classes~\vac{12}, \vac{13}, and \vac{23} 
have the potential to break $R$-symmetry.
Thus, calculating the probability of randomly obtaining an $R$-symmetric vacuum
from amongst all stable vacua as a function of $\lambda$, we obtain 
\beq
   \hbox{Prob$\,$(R~{\rm symm};$\,\lambda$)} ~=~\cases{
       (3/\pi) \tan^{-1} \lambda^2 &  for $\lambda\leq 1$~,\cr 
        (1/ 4) + (2/ \pi) \tan^{-1} \lambda^2 &  for $1\leq \lambda\leq \lambda_\ast$~,\cr
        1 & for $\lambda\geq \lambda_\ast$~.\cr }
\label{ProbR}
\eeq
This probability is plotted in Fig.~\ref{plotfig2} as a function of $\lambda$.

%================== FIGURE ============================================
%================== END OF INSERTED FIGURE ============================

Once again, just as with the probability of 
finding stable extrema, we recognize that we should really average
this result over $\lambda$ (the third landscape coordinate).
If we were to use the measure $\int_0^\infty d\lambda$
or $\int_{-\infty}^\infty d\lambda$,
we would obtain the result that the average probability is one,
since the contributions from large $|\lambda|$
(or from $|\lambda|\geq \lambda_\ast$)
dominate over the contributions from  
smaller $\lambda$.
However, using the ``angular'' measure in Eq.~(\ref{angularmeasure}),
we obtain the exact analytical result
\beq
   \hbox{Prob$\,$(R~{\rm symm})}~=~ 
        {2\over \pi}\, \int_0^{\pi/2}  {d\theta_\lambda}~
   \hbox{Prob$\,$(R~{\rm symm};$\,\lambda$)}~=~ {21\over 32}~\approx~ 0.656~.
\eeq
Clearly, this result depends strongly on the measure employed, and 
possible cutoffs for $\lambda$ that might be imposed.

%===========================================================================================
\subsection{Renormalization-group flow and boundary crossings}

One of the interesting features of the vacuum structure we are seeing
in this toy model is the fact
that all of the boundaries between different vacuum regions 
are actually energy-dependent (or temperature-dependent in the early
universe). Therefore,  it is possible that a
vacuum can cross a boundary between regions as the result of
renormalization group evolution, either because the landscape location
of the vacuum changes, because the boundary changes, or both.
 
In order to understand this potentially interesting observation, 
let us now analyze the evolution of our toy model under renormalization-group
(RG) flow.  
As we have seen, this model actually contains several quantities which  
are potentially renormalized:  these include the FI coefficients $\xi_i$
and the Yukawa coupling $\lambda$.  However, we must remember  
that our toy model also depends on the $U(1)_i$ gauge couplings
$g_i$ that were implicitly dropped from Eq.~(\ref{potential}).
In this model, 
the RGEs for the two gauge couplings $g_1$, $g_2$, and the Yukawa
coupling $\lambda$ are given by
\beqn
   \mu {d\over d\mu} g_i &=& \frac{g_i^3}{16 \pi^2}{\rm Tr}\, Q_i^2~, \nonumber\\
   \mu {d \over d \mu} \lambda  &=& \frac{\lambda}{16\pi^2} \, (3\lambda^2 -
   4 g_1^2 - 4 g_2^2 ) ~,
\label{RGEs}
\eeqn
where ${\rm Tr}~Q_i^2$ is the sum of (squared) charges under $U(1)_i$. 

In order to illustrate the phenomenon of boundary crossing, we will 
assume for simplicity that the entire matter content of our theory
consists of the three chiral superfields $\Phi_1$, $\Phi_2$, and $\Phi_3$,
and that the FI terms $(\xi_1,\xi_2)$ are introduced at tree level.
We therefore find that
${\rm Tr}\,Q_1^2= {\rm Tr}\,Q_2^2=2$.  
Moreover, since 
${\rm Tr}\, Q_1 = {\rm Tr}\, Q_2 =0$ in this case, the one-loop induced FI 
coefficients are zero,  and therefore the tree-level FI coefficients
$(\xi_1,\xi_2)$ are not renormalized.
Thus, we see that in this toy model, the location of an individual vacuum
in the landscape is invariant, and the only changes that can occur are
those which change the topography of the
landscape itself (moving boundaries, growing or shrinking regions, {\it etc}\/.) 
with respect to that fixed location.

Let us consider initial boundary values
$g_i^2 (\Lambda) = g_{i,0}^2$ and $\lambda^2 (\Lambda) = \lambda_0^2$, where
$\Lambda$ is some initial reference ultraviolet (UV) scale.
In order to make this calculation tractable, let us further assume
that our gauge couplings are originally equal at the UV scale: 
$g_{1,0}= g_{2,0}\equiv g_0$.
We therefore find from Eq.~(\ref{RGEs}) that our gauge couplings 
remain equal at all subsequent energy scales:  $g_1(\mu)=g_2(\mu)\equiv g(\mu)$.

The question that we wish to pursue, then, is that of determining the
evolution of the entire landscape structure as a function of the energy
scale $\mu$.
We shall do this in several steps.

First, 
since $g_1(\mu)=g_2(\mu)\equiv g(\mu)$, 
we see that we can easily restore the gauge couplings to 
all of our previous landscape calculations simply by rescaling
$\lambda(\mu)\to \lambda(\mu)/g(\mu)$. 
This holds because the vacuum structure of the potential
$V$ in Eq.~(\ref{potential}) is not affected if we balance
the rescaling of the $D$-terms with a corresponding rescaling 
of the $F$-terms.
Of course, such a universal rescaling of $\lambda$ would not be 
possible if we did not assume that $g_{1,0}=g_{2,0}$.

Next, let us consider the sequence of landscape ``snapshots''
that we have already generated in Figs.~\ref{fig1},
\ref{fig2}(a), \ref{fig2}(b), \ref{fig3}, \ref{fig4}(a),
and \ref{fig4}(b).
Reordering these figures, we see that they describe 
the landscape as a function of the parameter $\lambda(\mu)$, which should now
be interpreted as the rescaled parameter $\lambda(\mu)/g(\mu)$.
Therefore, if we wish to understand the RG flow in the landscape,
we simply need to determine the flow of the single quantity
$\lambda/g$ which parametrizes which landscape sketch is appropriate
at which energy scale.  
Our original location on the landscape doesn't change with energy, 
but we simply have to look at the correct figure corresponding to the
    appropriate value of $\lambda/g$ in order to know
    the physics at this location.  
This observation is important, because it confirms
that the RG flow simply corresponds to moving between landscape pictures
while holding our original landscape location $(\xi_{1,0},\xi_{2,0})$ fixed.
There is therefore no need to perform any RG calculation for the behavior
of each individual boundary region.

Note that this last observation would not hold if we did not assume
that $g_{1,0}=g_{2,0}$.  In the case of unequal gauge couplings, our
landscape pictures in previous sections might not apply, and new
regions of stability/instability might emerge.  However, because of
the universal scaling argument which is possible when
$g_{1,0}=g_{2,0}$, we see
that our previous landscape calculations are guaranteed to apply
regardless of $\mu$.

Thus, in order to determine the evolution between different
landscape pictures, we simply need to determine the
RG equation for our rescaled Yukawa coupling $Y(\mu)\equiv 
\lambda(\mu)/g(\mu)$. 
For example, when $g_1(\mu)=g_2(\mu)\equiv g(\mu)$,
the boundaries demarcating the $\vac{1}+\vac{3}$ overlap
region for $Y(\mu)>1$ can be specified directly in terms of $Y(\mu)$:
\beq
           -Y^2 (\mu) \, \xi_1 ~<~ \xi_2 ~<~ - {1\over Y^2(\mu)} \, \xi_1~.
\eeq
Given Eq.~(\ref{RGEs}), it is relatively straightforward to
determine the RG equation for $Y(\mu)$.
We find
\beq
        \mu {d\over d\mu} Y(\mu) ~=~ {Y\over 16\pi^2} \, (3 Y^2-10) \,
            g^2(\mu)~,
\label{RGEtwo}
\eeq
and substituting the explicit solution for $g^2(\mu)$ from Eq.~(\ref{RGEs}),
we can integrate Eq.~(\ref{RGEtwo}) to obtain the solution
\beq
        3-{10\over Y^2(\mu)} ~=~ 
        \left( 3-{10\over Y_0^2} \right) \, \left[ {g(\mu)\over g_0}\right]^{10}~.
\label{RGEsoln}
\eeq
Thus, we see that quantity $3-10/Y^2$ scales according to
the ratio of gauge couplings $\left[ g(\mu)/ g_0\right]^{10}$.
Since this ratio always decreases as we flow towards the infrared,
we see that $3-10/Y^2$ always evolves towards zero,
either from above or below,
as we flow towards the infrared. 
In other words, regardless of the initial value $Y_0$,
our theory always flows towards an {\it infrared fixed point}\/
\beq
      \overline{Y}~\equiv ~ \sqrt{10\over 3}~\approx~ 1.826~.
\label{fixedpoint}
\eeq

The physical interpretation of this behavior is as
follows.  We begin our evolution with some initial values
$\xi_{i,0}$ (which determine our location in the landscape)
as well as initial values $\lambda_0$ and $g_0$ (which determine
which landscape sketch is appropriate).
Our landscape location is invariant as we flow towards the
infrared, but the underlying landscape picture (boundaries,
 {\it etc.}) will evolve. 
If we find that we are originally in a landscape with 
$Y_0\equiv \lambda_0/g_0<\overline{Y}$,
we flow {\it upwards}\/ through the landscape diagrams towards
the landscape with $Y=\overline{Y}$.
Conversely, if we are originally in a landscape with
$Y_0>\overline{Y}$,
we flow {\it downwards}\/ through the landscape diagrams towards
the landscape with $Y=\overline{Y}$.
In all cases, however, the net result of the RG evolution towards the
infrared is to wind
up in the landscape corresponding to $Y= \overline{Y}$.
This is the universal infrared limit.
 
Given this, it is straightforward to trace the RG dynamics
that a single point in the landscape might experience.
Note that a generic feature of the boundaries
separating two different vacuum stability regions is the presence of a massless scalar in
the spectrum. This is intuitively clear, since boundaries demarcate 
the stability regions of various vacua. 
Thus, crossing a boundary is in some sense equivalent to a phase transition,
with the appearance of long-range order as the new massless state appears
in the spectrum.
The order of the phase transition can then be determined from the behavior
of the vacuum energy $V$ as the boundary is crossed.

Note that in general, there are only two generic classes of boundaries
that appear in our landscape diagrams for this toy model: 
\begin{itemize}
\item  Boundaries separating single-vev regions, such as \vac{1},
    and two-vev regions, such as \vac{12}, in which one of the two vev's
    is the same as that in the single-vev region.
    These sorts of boundaries appear in Fig.~\ref{fig2}(a), for example,
    and tend to dominate at small $Y$.
\item  Boundaries separating single-vev regions, such as \vac{1},
    and overlap regions, such as \vac{1}+\vac{2}, 
    in which one of the overlapping vacua is the same as the vacuum
    in the single-vev region.
    These sorts of boundaries appear in Fig~\ref{fig4}(a),
    for example, and tend to dominate at large $Y$.
    Note that near the boundary, it is the common vacuum 
    that has the lower vacuum energy $V$ in the overlap region,
    while the other vacuum in the overlap region is only metastable.
\end{itemize}
It is easy to verify that the vacuum energy $V$ is continuous across
the first class of boundaries, while for the second class, the stable
and the metastable vacua are not degenerate in energy on the boundary.   
Note that the only other types of boundaries are those that separate
single-vev regions from each other (such as the boundary separating
the \vac{1} and \vac{3} regions in Fig.~\ref{fig1}), but these 
are fine-tuned and occur only at the critical values $Y=1$ and
$Y=Y_\ast\approx 1.55$
[where $Y_\ast$ corresponds to $\lambda_\ast$ in Eq.~(\ref{lambdaastdef})].

Given these classes of boundaries, let us therefore consider the 
kinds of phase transitions which can result as a consequence of 
RG flow in our toy model.

If $Y_0>\overline{Y}$, then we find ourselves in a landscape
containing only single-vev and overlap regions.
Thus, we have only boundaries of the second type.  Moreover,
since RG flow pushes us  
towards landscapes with smaller values of $Y$,  
we can only have situations in which our overlap regions
are getting smaller rather than larger.  Thus, depending
on our original (fixed) landscape location,
the only type of boundary crossing that may occur in this case 
is one in which our location changes from being within an
overlap region to within a single-vev region.
For example, we may be originally be located in the
\vac{1}$+$\vac{3} overlap region, closer to the \vac{1} region
than to the \vac{3} region.  In such a case, as the boundary between
the \vac{1}$+$\vac{3} region and the \vac{1} region
moves towards us and passes our location, 
we ultimately find ourselves in the \vac{1} region.

In this case, the resulting physics depends on our initial 
vacuum state.
Recall that in the overlap region, the true stable vacuum is
determined according to the nearer single-vev region.  For example,
in \vac{1}$+$\vac{3} overlap region, 
the \vac{1} vacuum has lower energy than the (metastable) \vac{3}
vacuum if we are closer to the \vac{1} region than to the \vac{3}
region.
Thus, if we are originally in the \vac{1} state, then there is
no effect as the boundary passes our location.
However, if we are originally in the metastable \vac{3} vacuum,
this vacuum becomes unstable and the \vac{1} 
solution becomes the only stable vacuum
as the boundary passes our location. 
Of course, a key ingredient in making this determination is
the timescale for boundary crossing compared with the timescale
for decaying from the metastable \vac{3} vacuum to the truly stable \vac{1}
vacuum even without a boundary crossing.  
If the latter timescale is shorter than the former,
the decay from the metastable \vac{3} vacuum to the truly stable
\vac{1} vacuum can occur
even before the boundary reaches us.  

The analysis is slightly more complicated for $Y_0<\overline{Y}$. 
If $Y_0 < 1$, then our original landscape has only one-vev and two-vev regions,
with the two-vev regions shrinking as a result of the RG flow 
towards larger $Y$-values.
If we are originally located in a two-vev region of this landscape, 
then we will necessarily eventually experience a second-order phase
transition into a one-vev vacuum as a result of RG flow.
However, if $1<Y_0<Y_\ast$ 
[where $Y_\ast$ corresponds to $\lambda_\ast$ in Eq.~(\ref{lambdaastdef})],
then our landscape consists of a mixture of one-vev, two-vev, and overlap regions.
As a result of RG flow towards greater $Y$-values, 
the two-vev regions are shrinking and the overlap regions are growing.
Thus, two different types of transitions are possible:
either we can be located in a two-vev region and experience
a second-order phase transition into a one-vev region, as described above, 
or we can be originally located in
a single-vev region next to a growing overlap region.
In the latter case, our vacuum state in the single-vev region continues to 
be the truly stable vacuum state in the overlap region, so there is no
phase transition.
Finally, if $Y_\ast< Y_0< \overline{Y}$, 
we find ourselves in a landscape in which there are only single-vev
regions and overlap regions, with the overlap regions growing as a result
of RG flow.  In such a case, as above, no phase transitions
are possible:  if we pass from a single-vev region into an overlap region,
our original vacuum state continues to be the truly stable vacuum state in 
the overlap region, and no phase transition occurs.
 
Clearly, the possibility of such RG-induced phase transitions
in the landscape is of great interest and relevance for cosmology,
and in particular for inflationary models.

One important consequence of the infrared fixed-point behavior towards
$Y=\overline{Y}\approx 1.826$ follows 
directly from Eq.~(\ref{ProbR}):  {\it our theory always flows 
in the infrared to one in which $R$-symmetry 
is preserved}\/.  This observation is true in our toy model regardless of the
original  landscape location or Yukawa/gauge couplings.

Indeed, the emergence of an infrared  fixed point has an even more
significant consequence:  in such cases, the low-energy phenomenology 
becomes {\it insensitive}\/ to the plethora of (ultimately string-theoretic) 
variables that define the ultraviolet landscape physics.
If such infrared fixed points are generic features of the string-theoretic
landscapes, their existence suggests that it may not be necessary to
understand the full ultraviolet string theory in order to extract
physically testable predictions from the landscape.   

Of course, we have made a number of assumptions in performing
our analysis in this section.  In a more realistic situation,
or with $g_{1,0}\not= g_{2,0}$, the true landscape topography
is likely to be much more complicated, with an even richer 
set of possible phase transitions.  
Other important effects that we have ignored include, for example, the possibility of
kinetic mixing~\cite{kn}.
Similarly, the RG evolution of the gauge couplings must clearly stop
below the scale of gauge symmetry breaking;  likewise, the beta-function
coefficients depend on the scale of supersymmetry breaking in the sense
that the matter spectrum is supersymmetric above this scale and non-supersymmetric
below it.  Such issues, while very important, are clearly model-dependent
and must be studied case by case.

However, even in this simple
toy model, we see that RG flow has great potential to lead
to a number of novel phase transitions and boundary crossings
in the landscape --- effects which are
highly sensitive to our initial landscape location $(\xi_{1,0},\xi_{2,0})$. 
Of course, as already emphased, we stress that the possible vacuum-structure 
phase transitions that we have discussed in this section must be understood in 
terms of temperature phase transitions in the early universe.

%======================================================================
\subsection{Adding soft masses for the chiral fields}

Finally, before concluding this section, let us briefly comment on the effects
which are induced in our toy model
when non-zero masses are introduced for our chiral scalar fields.
Towards this end, let us consider adding the mass term
\beq
            V_{\rm soft} ~=~ \tilde m^2 \, \sum_{i=1}^3 |\phi_i|^2 ~
\eeq
to the scalar potential given in Eq.~(\ref{potential}). 

\begin{figure}[ht]
\centerline{ 
   \epsfxsize 4.8  truein \epsfbox {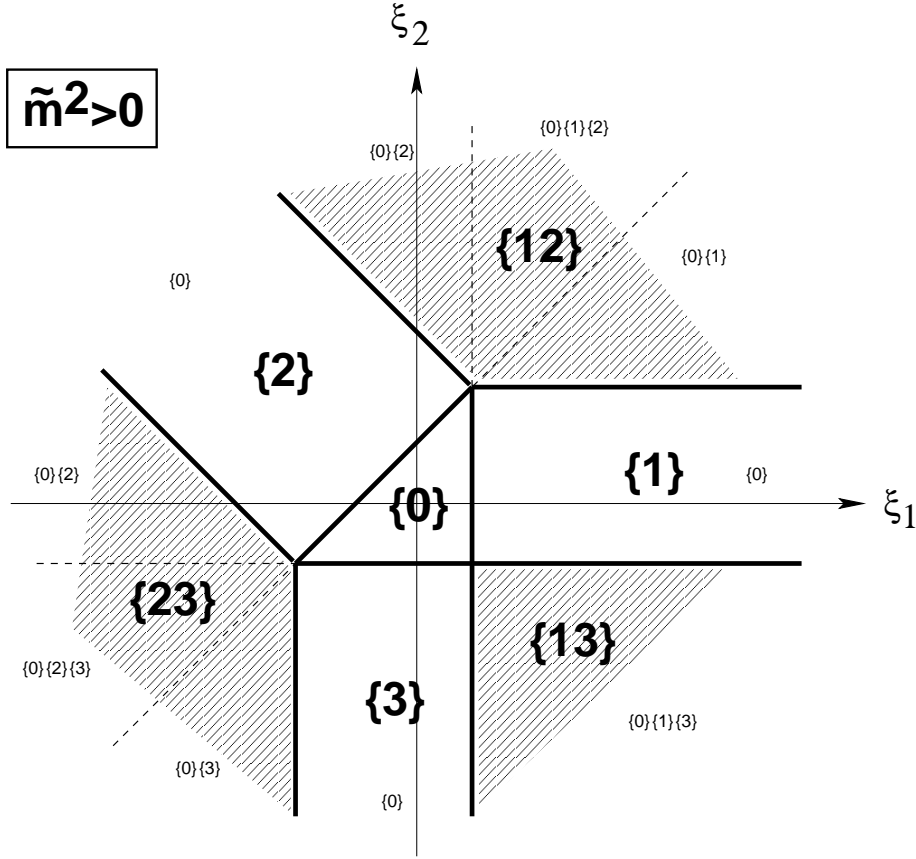}
    }
\caption{Same $\lambda=0$ landscape as shown in Fig.~\protect\ref{fig2}(b),
     but now with non-zero soft masses $\tilde m^2>0$ for the chiral superfields.  
     The  non-zero masses open up a central triangular region of stable vacua 
     in Class~\vac{$\emptyset$}, 
     and likewise introduce three semi-infinite rectangular ``strips''
     of stable vacua in Classes~\vac{1}, \vac{2}, and \vac{3}.      
     Note that very far from the origin,
     the relative effects of the
     mass terms generally vanish.  In this limit, only those regions with 
     non-zero opening angles survive, and the diagram reduces back to 
     that shown in Fig.~\protect\ref{fig2}(b).}
\label{fig7}
\end{figure}

It is straightforward to repeat the above vacuum-structure analysis 
for this case as well.  Let us first focus on the case with $\lambda=0$.
We then find that Class~\vac{$\emptyset$} solutions, which were formerly unstable 
everywhere, now become stable in a small, central, triangular region 
given by
\beq
              \xi_1 <  \half \tilde m^2~,~~~
              \xi_2 > -\half \tilde m^2~,~~~
             \xi_2 - \xi_1  < \half \tilde m^2~.
\label{Aregion}
\eeq
Likewise, the solutions in Classes~\vac{1}, \vac{2}, and \vac{3}, which were also
unstable everywhere for $\lambda= 0$, now become stable within
rectangular semi-infinite ``strips'' emanating outward from this central triangle.
The final landscape is shown in Fig.~\ref{fig7}.
Note that as $(\xi_1,\xi_2)\to \pm \infty$ (\ie, as we move increasingly
far from the origin or either axis), this diagram reduces back to
that shown in Fig.~\protect\ref{fig2}(b).  This is in accordance with
our expectation that the effects of the mass terms should vanish
if these masses are small compared with the Fayet-Iliopoulos (FI) coefficients.
Of course, these masses remain important if either FI coefficient
$\xi_i$ is small, or if their sum $\xi_1+\xi_2$ is small with $\xi_1<0$,
$\xi_2>0$.

%================== FIGURE ============================================
%================== FIGURE ============================================

For $\lambda>0$, the situation generalizes as we might expect.
First, we observe that stable \vac{$\emptyset$} vacua continue to
exist within the triangular region 
in Eq.~(\ref{Aregion}) for all $\lambda$, since
this result is $\lambda$-independent.
In general, for any $\lambda>0$, the one-vev stability regions then emanate 
outwards from the three {\it sides}\/ of this central triangle, with the same opening 
angles as they have for $\tilde m^2=0$.
Similarly, the stability regions that emanate from the {\it vertices}\/ of this central triangle
are the same as they would have been for $\tilde m^2=0$, corresponding
either to two-vev vacua or to overlapping one-vev vacua depending on the value of 
$\lambda$.  In all cases, we see that these landscapes with $\tilde m^2>0$
reduce back to the previous landscapes with $\tilde m^2=0$ when we look
far from the origin and therefore retain only those stability regions with 
non-zero opening angles.

Finally, let us briefly consider the case with $\tilde m^2<0$.  In this case, 
and with $\lambda=0$, 
it is easy to verify that no stable vacua exist anywhere in the landscape, \ie, for
any values of $(\xi_1,\xi_2)$.
This makes sense, since we already saw that the $\lambda=0$ landscape
comprises a single flat direction when $\tilde m^2=0$, and this flat
direction is necessarily destabilized for $\tilde m^2<0$.
However, for non-zero $\lambda$, this expectation can change since we can 
balance the stabilizing $F$-terms against the 
destabilizing soft masses.
In general, we find that as $\lambda$ increases from zero,
stable pie-slice regions of stable vacua begin to exist at the outer edges
of the landscape.  
These vacua have the same opening angles and properties as functions
of $\lambda$ as 
they have in the $\tilde m^2=0$ case,
but their origination points are
displaced away from the origin, leaving a hollow, unstable center.
Thus, this is our first example of a {\it hole in the landscape}\/ --- \ie, a bounded
finite region of the landscape in which there are no stable vacua of any sort.

As $\lambda$ increases, each of these pie-slice regions begins to 
originate from a point closer and closer to the origin.
Together, they eventually fill the hole completely when $\lambda$
reaches a critical value related to $|\tilde m|$.  
However,
as $\lambda$ continues to increase {\it beyond}\/ this critical value, 
each of these origination points
continues to move {\it past}\/ the origin, 
thereby creating a crowded new region of {\it three}\/ overlapping single-vev
vacua in the vicinity of the origin.  This is thus our first example of more than 
two simultaneously overlapping vacua in the same landscape region.

%=====================================================================================
\setcounter{footnote}{0}
\section{Adding more U(1) gauge groups}

In this section we shall consider generalizing the model in Sect.~3
by adding more $U(1)$ gauge groups.  This will significantly increase
the number of vacua and the complexity of the corresponding landscape.
More importantly, since the Wilson-line superpotential can in principle contain
more fields, we see from dimensional analysis that the $F$-terms will generically 
be suppressed.  Thus, $R$-symmetry will be tend to be broken  
only at very low energies and only for relatively few vacua. 

There are, of course, many different ways in which we can 
introduce further $U(1)$ gauge groups.
The simplest example containing a large number of vacua is provided by
considering $n$ copies of the Fayet-Iliopoulos model analyzed in Sect.~2.
Such a setup is easily realized in string models containing  
large numbers of Abelian factors, corresponding to branes with significant
geometric separations
in the six-dimensional internal space. 
Such a model would clearly have $3^n$ classes of vacua, 
and requires the various Abelian
gauge group factors to be anomalous in the string-theory sense.  
While this is certainly possible, 
one must bear in mind that in
perturbative string constructions, the number of Abelian factors is limited by
the total rank of the gauge group, as determined by Ramond-Ramond (RR) tadpole 
cancellation conditions.  In
$2^n$ of these classes of vacua, the Abelian gauge group is completely
broken.\footnote{Models with several Abelian factors 
     have been considered phenomenologically in various papers;  see, \eg, 
     Ref.~\cite{kn} and references therein.}

\begin{table}[ht]
\centerline{
   \begin{tabular}{||c||ccccc||}
   \hline
   \hline
    $~$ &  $U(1)_1$  & $U(1)_2 $  &  $U(1)_3$ &  ... & $U(1)_n$ \\
   \hline
   \hline
    $\Phi_1$  &  $-1$ & ~  &  ~ & ~ & ~ \\
    $\Phi_2$  &  $1$ & $-1$ & ~ & ~ & ~  \\
    $\Phi_3$  &  ~ & $1$  & $-1$ & ~ & ~  \\ 
    $\vdots$  &  ~ & ~  &  $\ddots$  & $\ddots$ & ~  \\
    $\Phi_n$  &  ~ & ~  &  ~ & $1$ & $-1$  \\
    $\Phi_{n+1}$  &  ~ & ~   & ~ & ~ & $1$  \\
   \hline
   \hline
   \end{tabular}
 }
\caption{$U(1)$ charge assignments for chiral superfields, inspired by
      ``deconstruction'' models of extra dimensions.}
\label{table2}
\end{table}

Things become less straightforward if the various Abelian factors are
not decoupled from one another.  For this purpose, let us therefore consider a
generalization of the two-$U(1)$ model of Sect.~3 to the case of $n$ different
$U(1)$ gauge group factors, with $n+1$ chiral superfields.
Inspired by deconstruction models of extra dimensions~\cite{deconstruction},
we shall take our charge
assignments to follow the pattern indicated in Table~\ref{table2}.
Thus, as evident from Table~\ref{table2}, we 
are only considering bi-fundamental ``nearest-neighbor'' charges;
other configurations will be briefly discussed in Sect.~6.
Likewise, we shall assume for simplicity that only the boundary $U(1)$
gauge-group factors, \ie, $U(1)_1$ and $U(1)_n$, have Fayet-Iliopoulos
coefficients $\xi_1$ and $\xi_n$ respectively.
Given these charges, we can in general write down a Wilson-line
superpotential of the
form\footnote{
   This implies that the $\Phi_i$ superfields 
   have $R$-charge $2/(n+1)$, giving 
   the $F$-terms $R$-charge $-2n/(n+1)$.  Thus,
   as claimed earlier, $F$-term breaking will correspond
   to $R$-symmetry breaking in this model.}
\beq
           W ~=~ \lambda \prod_{i=1}^{n+1} \Phi_i~. \label{wilson}
\eeq
Note that unlike the previous case with $n=2$, this superpotential is not
renormalizable for $n>2$.  
However, we can continue to study the landscape 
of this model as a function of the Fayet-Iliopoulos coefficients 
$(\xi_1,\xi_n)$ for arbitrary values of the Yukawa parameter $\lambda$. 
We can also consider the addition of soft scalar masses by adding
a term
\beq
             V_{\rm soft}~=~ \tilde m^2 \,\sum_{i=1}^{n+1} |\phi_i|^2~
\eeq
to the scalar potential, as in Sect.~3.

%==========================================================================
\subsection{Special case $n=3$}

Let us first consider the case with $n=3$. 
Since the superpotential $W$ is nonrenormalizable,
its contribution to the overall potential is negligible at
energy scales below the fundamental physics scale.
We shall therefore take $\lambda=0$ in our analysis.
However, we shall continue to leave the common scalar soft mass $\tilde m^2$ 
arbitrary.

Since there are four chiral superfields when $n=3$, 
there are now sixteen possible classes of vacua which can be characterized
by the vacuum expectation values (vev's) $v_i$ of the scalar fields $\phi_i$.
We shall continue to employ a notation in which \vac{ijk...} denotes
the class of extrema in which $v_i$, $v_j$, $v_k$, $...$, are non-zero.
Likewise, the notation \vac{$\emptyset$} will  denote the class of extrema
in which all vev's vanish.
For example, \vac{14} denotes extrema in which $v_2=v_3=0$ while
$v_1\not=0$, $v_4\not=0$.
In general, just as in the $n=2$ case,
we can calculate a $(8\times 8)$-dimensional mass matrix as in Eq.~(\ref{massmatrix}).
There will be a zero eigenvalue for each spontaneously
broken $U(1)$;  these eigenvalues correspond to the resulting Nambu-Goldstone bosons.
Our extrema 
then represent
stable (or metastable) vacua only if each of the remaining eigenvalues 
is positive.

We find the following results.
In general, of the sixteen potential classes of extrema, we find 
that only eleven classes of extrema actually exist;  
there are no solutions for extrema in Classes
\vac{12}, \vac{34}, \vac{124}, \vac{134}, or \vac{1234}.

%================== FIGURE ============================================
\begin{figure}[ht]
\centerline{ 
   \epsfxsize 4.3  truein \epsfbox {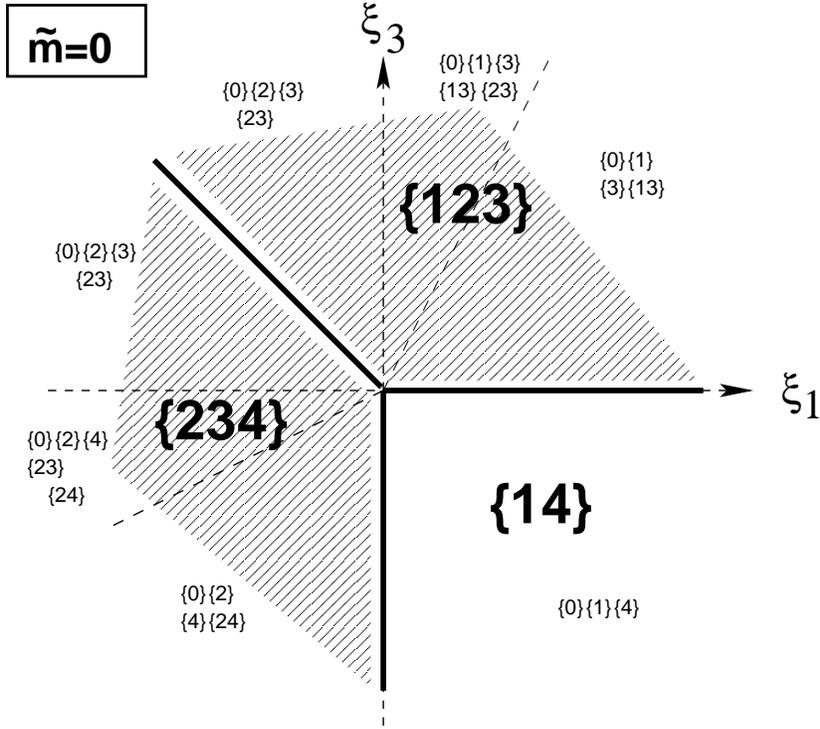}
    }
\caption{The landscape of our $n=3$ model for $\tilde m=0$.
  Stable vacua are indicated in large bold type, whereas unstable
  extrema are indicated in smaller non-bold type.  Vacua breaking
  $R$-symmetry are shaded;
   note that the \vac{14} region is
  supersymmetric and hence $R$-symmetry preserving.
  Compared with the analogous $n=2$ case in 
  Fig.~\protect\ref{fig2}(b),
  we see that not all vacua break $R$-symmetry for $n=3$.
  There are also slightly more unstable extrema for each
  stable vacuum.
}
\label{fig11}
\end{figure}
%========================================================================

Of the remaining eleven classes of extrema, we find
that only three of these give rise to stable (non-overlapping) regions
when $\tilde m=0$.  This case is sketched in Fig.~\ref{fig11},
which should be compared with the analogous $n=2$ case
Fig.~\ref{fig2}(b).
Compared with the $n=2$ case, we see that the stable
regions have a virtually identical shape, with
the \vac{12} vacua for $n=2$ 
corresponding to the \vac{123} vacua for $n=3$;
the \vac{23} vacua for $n=2$ 
corresponding to the \vac{234} vacua for $n=3$;
and the \vac{13} vacua for $n=2$ 
corresponding to the \vac{14} vacua for $n=3$.
There are some differences, however.
For example, in the $n=2$ case, the entire landscape
was dominated by vacua that break $R$-symmetry.
By contrast, if we classify our $n=3$ vacua according
to their general properties for arbitrary $\lambda$,
we see that the \vac{14} vacua are actually supersymmetric
and consequently preserve
$R$-symmetry.
This is therefore our first example
of a landscape containing both supersymmetric and non-supersymmetric
regions of the same dimensionality [as opposed to lines of
supersymmetric vacua within 
two- (or higher-)dimensional non-supersymmetric regions].
We also notice that the $n=3$ case gives rise to
slightly more unstable extrema for each
stable vacuum than we found in the $n=2$ case.
Finally, we note that the dashed lines in Fig.~\ref{fig11} have slopes
$2$ and $1/2$, as opposed to the single slope $1$ in Fig.~\ref{fig2}(b).

%================== FIGURE ============================================
\begin{figure}[ht]
\centerline{ 
   \epsfxsize 5.1  truein \epsfbox {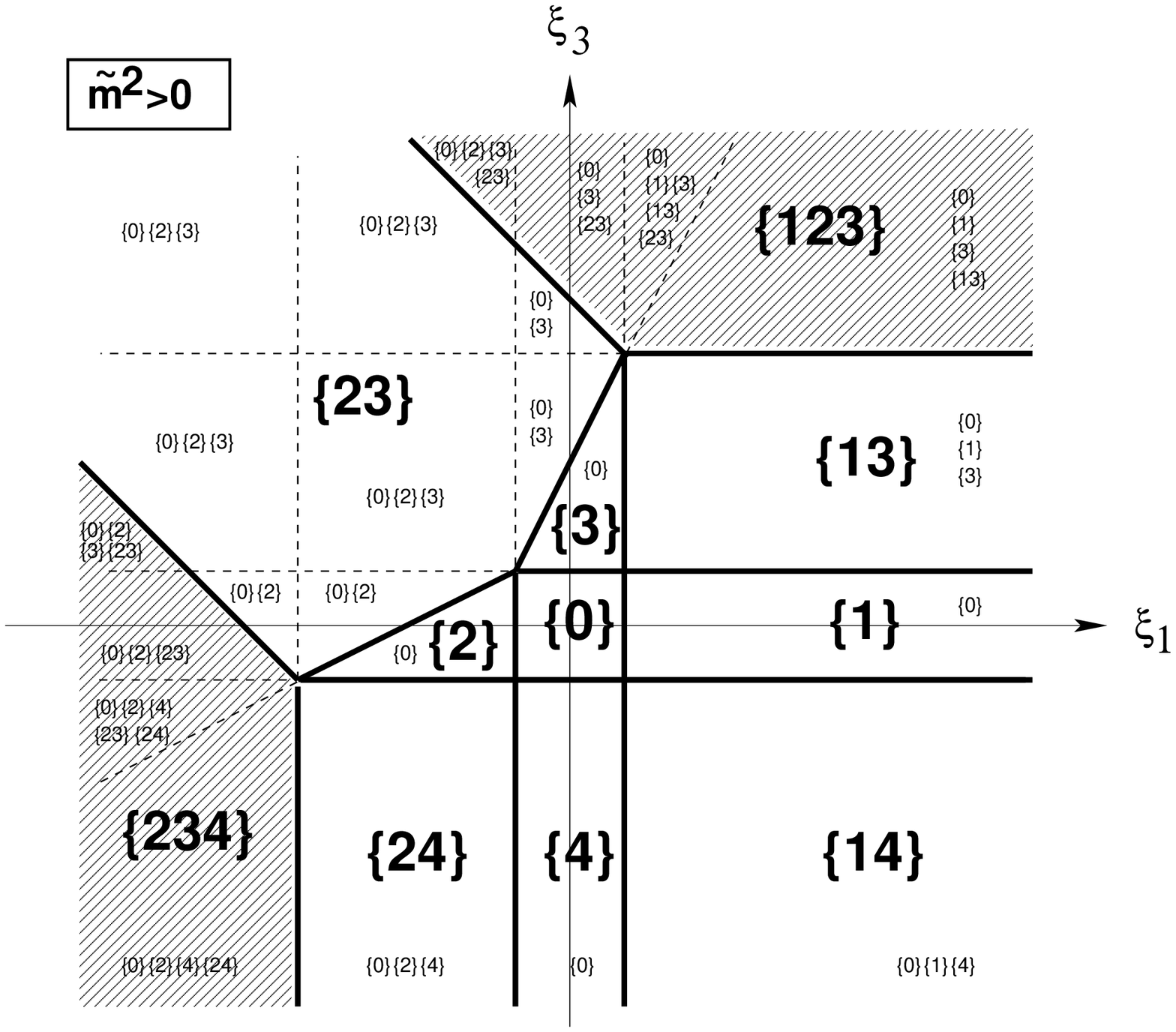}
    }
\caption{The landscape of our $n=3$ model, now drawn 
  for $\tilde m^2>0$.
  Stable vacua are indicated in large bold type, whereas unstable
  extrema are indicated in smaller non-bold type.  Vacua breaking
  $R$-symmetry are shaded.
  This landscape is to be compared with its $n=2$ equivalent
  in Fig.~\protect\ref{fig7}.
  Once again, we observe that in the $\tilde m\to 0$ limit
  (\ie, as we move radially outwards from the origin),
  only those regions with non-zero opening angles survive;
  this diagram then reduces back to that in Fig.~\protect\ref{fig11}.
}
\label{fig10}
\end{figure}
%========================================================================

We stress that Fig.~\ref{fig11} must be interpreted in 
the same way as Fig.~\ref{fig2}(b), namely as the landscape
that emerges for infinitesimal $\lambda$ approaching zero.
When $\lambda=0$, the entire landscape in Fig.~\ref{fig11} actually
becomes supersymmetric and $R$-symmetry preserving, completely dominated by 
the flat directions implicit in the $\lambda=0$ solution $\vac{1234}$.

We can also consider the effects of turning on the
soft masses $\tilde m$.  As expected from the $n=2$ case, 
we find that the
\vac{$\emptyset$} solutions, which were formerly unstable everywhere,
now become stable in a small, central, square region given by
\beq
        |\xi_1| < \tilde m^2~,~~~
        |\xi_3| < \tilde m^2~.
\eeq
Likewise, of the seven remaining unstable classes of extrema,
two of these (Classes \vac{2} and \vac{3}) become stable within
small triangular regions attached to this central square,
while the remaining five classes of vacua
now become stable within rectangular semi-infinite strips
emanating outward from the central square and triangles.
The resulting landscape is shown in Fig.~\ref{fig10}.
Detailed solutions for each class of extrema in this model are
given  in the Appendix.

%==========================================================================
\subsection{Arbitrary $n$ and large-$n$ limit}

We now turn our attention to the general-$n$ case, with particular interest
in the large-$n$ limit.
In the general case, the $D$- and $F$-terms are now given by 
\beq
            D_{a} ~=~ \sum_{i=1}^{n+1}
          \, q_i^{(a)} |\phi_i|^2 ~+~ \xi_1 \delta_{a1}
            ~+~ \xi_n \delta_{an}~,~~~~~~~
            F_i~=~ {\partial W\over \partial \phi_i}~,
\eeq
where $a=1,\dots,n$. This leads to the scalar potential
\beq
     V ~=~ \half \sum_{a=1}^n D_a^2 ~+~ \sum_{i=1}^{n+1} |F_i|^2~.
\label{genpotential}
\eeq
As we already remarked for the $n=3$ case, the $F$-term contributions coming from the Wilson line
superpotential (\ref{wilson}) in the large-$n$ case have a negligible effect on the vacuum
structure for FI terms smaller than the Planck (or string) scale.
The extrema of the scalar potential are then obtained as solutions to the field equations
\beq
     \phi_i\left(q_i^{(1)}\xi_1 + q_i^{(n)}\xi_n + \sum_{a=1}^n
     q_i^{(a)}\sum_{j=1}^{n+1} q_j^{(a)} |\phi_j|^2 
      \\+|\lambda|^2\sum_{\stackrel{j=1}{j\neq i}}^{n+1} |\phi_1\dots
     \hat\phi_i\dots\hat\phi_j\dots\phi_{n+1}|^2~\right)=0~,
\eeq
where the hatted fields refer to fields that are omitted from the product.
These equations each have 
two solutions:  $v_k \equiv \langle\phi_k\rangle=0$, and a solution obtained
by setting the term in parentheses to zero. This gives a large number $\sim{\cal O}(2^{n+1})$
of different extrema which were explicitly
analyzed in Sect.~3 for $n=2$ and  Sect.~4.1 for $n=3$. 

Certain features of the general-$n$ solution can easily be 
perceived directly by generalizing from the explicit $n=2$ and $n=3$ solutions.
For example, it is straightforward to show 
that the $\tilde m=0$, $\lambda=0$ landscape always has three non-overlapping regions 
with the same shapes as in Fig.~\ref{fig4}(b) and Fig.~\ref{fig11},
and with the dashed lines (separating regions
with different unstable extrema) having slopes $n-1$ and $1/(n-1)$. 
Moreover, it is easy to see that the three regions of stability always
correspond to vacua in the classes
    \vac{1,n+1}, \vac{1,2,3,..,n}, and \vac{2,3,...,n+1}.

Similar general statements can also be made for the 
$\tilde m\not=0$ case.  
When $\tilde m^2>0$, the presence of the non-zero soft mass always opens up a central
region of stability for the \vac{$\emptyset$}
extrema.  In general, this region is an $(n+1)$-gon,
with regions of stability for otherwise unstable
    extrema either emanating outwards in semi-infinite strips, or 
    attached to the $(n+1)$-gon 
    with bounded regions shaped like $n$-gons.
    Note that the appearance of a central $(n+1)$-gon can be 
    understood as the result of having $n+1$ independent eigenvalue
    constraints defining the borders of the stability region.

While all of these properties hold true for general $n$, it is also
instructive to examine certain specific vacuum solutions.
Instead of attempting a
general classification, we will concentrate on presenting some general features
such as the absolute minima of the theory.

In the following, we shall focus on the case in which $\lambda/M_P^{n-2} \to 0$ and
$\tilde m=0$.
Since $\lambda/M_P^{n-2}\to 0$,
we expect that the stable vacua are dominated by the three classes
\vac{1,n+1}, \vac{1,...,n}, and \vac{2,...,n+1}.
Specifically, we find that the \vac{1,n+1} solution exists for
$\xi_1>0$, $\xi_n<0$, and is given by
\beq
   \vac{1,n+1}:~~~~~  |v_1|^2 = \xi_1~, ~~~ |v_{n+1}|^2 = - \xi_n~, ~~~
           v_2 = ... =v_n =0~.  
\label{solut1}
\eeq
Similarly, the \vac{1,...,n} solution exists for 
$\xi_n > 0$,  $\xi_1 + \xi_n > 0$, and is given by
\beqn
&& \vac{1,...,n}:~~~~~  |v_1|^2 = (\xi_1 + \xi_n)(1+ {\cal O}(\epsilon_n^2))~\nonumber\\
&& \phantom{\vac{1,...,n}:~~~~~}
  |v_2|^2, ...,  |v_n|^2 = \xi_n (1+{\cal O}(\epsilon_n^2))~,~~~
  v_{n+1} = 0~
\label{solut2}
\eeqn
where $\epsilon_n^2 \sim (\lambda/M_P^{n-2})^2 (\xi_n/M_P^2)^{n-2}$, while
the \vac{2,...,n+1} solution exists for
$\xi_1<0$, $\xi_1+\xi_n<0$, and is given by
\beqn
  && \vac{2,...,n+1}:~~~~~  v_{1} = 0~,~~~ 
      |v_{n+1}|^2 = - (\xi_1 + \xi_n)(1+ {\cal O}(\epsilon_1^2))~\nonumber\\
  && \phantom{\vac{2,...,n+1}:~~~~~}
   |v_2|^2, ..., |v_n|^2 = - \xi_1 (1 + {\cal O}(\epsilon_1^2))~
\label{solut3}
\eeqn
where $\epsilon_1^2 \sim (\lambda/M_P^{n-2})^2 (\xi_1/M_P^2)^{n-2}$.
Note that $\epsilon_1$ and $\epsilon_n$ are very small numbers 
for $n \gg1$ and $\xi_i \ll M_P^2$. 
The vacuum in Eq.~(\ref{solut2}) was discussed in 
detail in Ref.~\cite{dfp}.  Each of these classes of vacua
occupy non-overlapping regions in the two-dimensional 
$(\xi_,\xi_n)$ parameter space.
Note that the \vac{1,2,...,n} and \vac{2,3,...,n+1} vacua
completely break all of the $U(1)$ gauge factors,  
while the \vac{1,n+1} vacuum is supersymmetric 
and $\lambda$-independent 
for all $n\geq 3$.

We shall also be interested in several other explicit solutions for general $n$.
All of the following solutions are $\lambda$-independent.
For example, the \vac{2,3,...,n} solution is 
given by
\beqn
&&  \vac{2,3,...,n}:~~~~~ v_1 = v_{n+1} = 0~, \nonumber\\
&&  \phantom{\vac{2,3,...,n}:~~~~~}
    |v_k|^2 = {1 \over n} \left[(\xi_1+\xi_n) (k-1) - n \xi_1 \right] 
   ~~{\rm for}~ k = 2,...,n~.~~~~~
\label{solut4}
\eeqn
This solution has an unbroken $U(1)$ generator $Q_1 + ...+ Q_n$, 
where $Q_i$ are the generators of $U(1)_i$, and gives rise to the $D$-terms
\beq
   \langle D_1 \rangle ~=~... ~ =~\langle D_n \rangle ~=~ {\xi_1+\xi_n
   \over n}~. \label{solut04}
\eeq
Note that this solution has a linear ``profile'', in the sense
that the sequence of non-zero vacuum expectation values $|v_k|^2$ 
in Eq.~(\ref{solut4}) grows linearly with $k$.

Another solution of interest is the \vac{2,3,...,n-1,n+1} solution, given by
\beqn
 &&  \vac{2,3,...,n-1,n+1}: ~~~~~ v_1 = v_{n} = 0~,~~~  |v_{n+1}|^2 = -\xi_n~\nonumber\\
 &&  \phantom{\vac{2,3,...,n-1,n+1}: ~~~~~}
 |v_k|^2 =  {k-n \over n-1}  \,\xi_1 ~~~{\rm for}~k=2,...,n-1~.~~~~~
\label{solut5}
\eeqn
This solution has the unbroken generator $Q_1 +Q_2 + \cdots + Q_{n-1}$,
and gives rise to the $D$-terms
\beq
 \langle D_1 \rangle ~=~ ... ~=~\langle D_{n-1} \rangle ~=~ 
              {\xi_1 \over n-1}~,~~~~    \langle D_n \rangle ~=~ 0~.
\eeq
This solution also exhibits a linear profile.

Another solution of interest is the \vac{1,3,4,...,n} solution, given by
\beqn
&&  \vac{1,3,4,...,n}: ~~~~~ v_2 = v_{n+1} = 0~,~~~ |v_{1}|^2 = \xi_1\nonumber\\ 
&&  \phantom{\vac{1,3,4,...,n}: ~~~~~}
 |v_k|^2 = {k-2 \over n-1}\,  \xi_n ~~~{\rm for}~k=3,4,...,n~.
\label{solut6}
\eeqn
This solution has an unbroken $U(1)$ generator $Q_2+...+Q_n$, and gives rise
to the $D$-terms 
\beq
 \langle D_1 \rangle ~=~ 0 ~,~~~ 
 \langle D_2 \rangle~=~ \cdots 
    ~=~\langle D_{n} \rangle ~=~ {\xi_n \over n-1} \ . 
\eeq
Clearly, this solution also exhibits a linear profile.

Finally, the last solutions we shall consider are the 
\vac{2,3,...,j-1,j+1,...,n}
solutions, given by
\beqn
 &&  \vac{2,3,...,j-1,j+1,...,n}:~~~~~ v_1= v_j=v_{n+1}~=~0\nonumber\\
 &&  ~~~~~~~~~~~~~
    |v_k|^{2} =\cases{
             {\displaystyle k-j \over \displaystyle j-1} \,\xi_1 & for $k=2,...,j-1$\cr
             {\displaystyle k-j \over \displaystyle n-j+1} \, \xi_n  & for $k = j+1,...,n$~.\cr}~~~~
\label{solut7}
\eeqn
These solutions leave unbroken the two generators
$Q_1 +Q_2 +...+ Q_{j-1}$ and 
$Q_j + Q_{j+1} + ...+ Q_n$, and gives rise to the $D$-terms
\beqn
\langle D_1 \rangle ~=~ \cdots ~=~ \langle D_{j-1} \rangle ~=~& \xi_1 /(j-1)
&~\equiv~ d_1 \nonumber\\
 \langle D_j \rangle ~=~ \cdots ~=~ \langle D_n \rangle
~=~& \xi_n /(n-j+1) &~\equiv~ d_2 ~. \label{twofluxes}
\eeqn 
Unlike the previous solutions, each of these solutions
has {\it two}\/ independent linear profiles.

Needless to say,  there are numerous other solutions which can be
generated for general $n$.  However, the above solutions will be sufficient
for our purposes.

Note that these vacua, as well as all vacua
with smaller numbers of non-zero vev's, 
are unstable at the level of our discussion
(in which we are taking $\lambda/M_P^{n-2} \to 0$ and $\tilde m=0$).
Moreover, one of the obvious properties of such solutions
is the fact that, at first sight, they unfortunately appear to 
give rise to a large number of unbroken Abelian gauge symmetries 
which would survive down to low energies. 
However, as we shall  see in Sect.~4.4,
this is not the case due to 
the presence of mixed gauge anomalies.

%==========================================================================
\subsection{Higher-dimensional flux interpretations}

We shall now demonstrate that many of the above general-$n$ solutions
have natural interpretations in terms of higher-dimensional flux compactifications.

First, let us consider the 
\vac{1,2,...,n} and \vac{2,3,...,n+1} vacua.
These clearly can be interpreted as emerging from
a five-dimensional supersymmetric $U(1)$ gauge theory compactified on
the orbifold $S^1/\IZ_2$, with compactification radii
\beq
          R ~\sim~ n/\sqrt{\xi_n}~~  {\rm and } 
           ~\sim~ n/\sqrt{|\xi_1|}~, 
\eeq
respectively.  In each case, the  four-dimensional zero-mode gauge
field receives a mass from the four-dimensional FI term $\xi_1+\xi_n$. 
The supersymmetry-breaking scale in these two cases is controlled by the 
Wilson-line superpotential, and all soft masses are $\sim{\cal O}(\epsilon_n)$ in the first 
case and $\sim{\cal O}(\epsilon_1)$ in the second case. 
By contrast, the third vacuum 
$\vac{1,n+1}$ is supersymmetric and probably does not have an 
extra-dimensional interpretation.  

We now turn to the 
\vac{2,3,...,n} solution in Eq.~(\ref{solut4}).
We shall now argue 
that this vacuum can be given the higher-dimensional interpretation of 
having magnetic flux on a torus in a six-dimensional Abelian gauge theory. 
Indeed, as we will see, the smoking gun for such a magnetic flux 
interpretation is the presence of a linear profile in the associated
vacuum expectation values.

 From a five-dimensional $\IR^4 \times S^1/\IZ_2$ perspective, where
the fifth dimension is the interval $0< y < \pi R$, a supersymmetric 
Abelian gauge theory contains a gauge field and a $\IZ_2$-odd real scalar 
$\Sigma$. The $D$-term from the four-dimensional point of view in the 
continuous limit of the deconstruction setup discussed above is
given by
\beq
   D ~=~ \partial_5 \Sigma ~+~ (-|\phi_1|^2 + \xi_1) \, \delta(y) ~+~ 
    (|\phi_{n+1}|^2 + \xi_n) \,\delta(y-\pi R)~. 
\label{d2}
\eeq
The standard profile for the scalar $\Sigma$, largely discussed in the
literature, is of the form $\langle\Sigma \rangle= \epsilon (y) \xi_n/2$, 
which in the case $\xi_1 + \xi_n =0$ is the needed profile for preserving
supersymmetry and the gauge symmetry~\cite{pr}. Notice, however, that
the field equations 
\beq
  \delta \Sigma \, \biggl\{ \partial_4^2 \Sigma + \partial_5 
\left[ \partial_5 \Sigma + (\xi_1-|v_1|^2) \delta (y) +  (\xi_n + |v_{n+1}|^2) \delta 
(y-\pi R) \right] \biggr\} ~=~ 0 
\label{d3}
\eeq 
have another solution on the orbifold $S_1/\IZ_2$, namely
\beq
\langle \Sigma \rangle ~=~ {\xi_1 + \xi_n \over 2 \pi R} \, y ~-~ {\xi_1 \over 2} 
    \, \epsilon (y) ~,~~~~ \langle D \rangle ~=~ {\xi_1 + \xi_n \over 2 \pi R}~.  
\label{d4} 
\eeq   
This solution does not describe the absolute minimum of the theory,
since it has a large positive vacuum energy, but it is an extremum
of the theory. By using $R \sim n$, it is clear
that Eq.~(\ref{d4}) matches the deconstructed result Eq.~(\ref{solut4}). 
On the other hand, from a six-dimensional perspective, $\Sigma$ 
corresponds to the sixth-component $A_6$ of the gauge field. 
Then the flux in the two-dimensional compact space is given by
\beq
     F_{56} ~\equiv~ \langle \partial_5 \Sigma \rangle ~=~  
              {\xi_1 + \xi_n \over 2 \pi R} 
         ~-~ \xi_1 \, \delta (y) ~-~ \xi_n \, \delta (y-\pi R) ~.
\label{d5}
\eeq  
The first term is the magnetic flux we were searching for,
whereas the localized terms, already discussed in the literature,
have the interpretation of fluxes 
localized at the orbifold fixed points. 
Note that the integrated flux in the compact space is actually zero,
\beq
       \int_{-\pi R}^{\pi R}  dy~\langle \partial_5 \Sigma \rangle ~=~0~,
\eeq
the magnetic flux cancelling
the localized contributions at the fixed points.     

The other extrema listed in Sect.~4.2 also have a flux-compactification
interpretation in extra dimensions. 
For example, let us consider the
\vac{2,3,...,n-1,n+1} solution given in Eq.~(\ref{solut5}).
In the extra-dimensional interpretation, the bulk field $\Sigma$ has a 
different profile: 
\beq
      \langle \Sigma \rangle ~= ~ {\xi_1 \over 2 \pi R} \, y ~-~ {\xi_1 \over 2}\, \epsilon(y)
    ~,~~~~~~
     \langle D \rangle ~=~ \cases{
             \xi_1 /( 2 \pi R)  &  for $0 \leq y < \pi R $\cr
             0  &  for $y = \pi R$~.  \cr} 
\label{d11}
\eeq
This gives rise to a flux which again integrates to zero:
\beq
     F_{56} ~\equiv ~ \langle \partial_5 \Sigma \rangle ~=~  
    {\xi_1 \over 2 \pi R} - \xi_1 \delta (y) ~, ~~~~~~~~ \int_{-\pi R}^{\pi R}  
           dy~\langle \partial_5 \Sigma \rangle ~=~0 ~ . 
\eeq  
It may seem strange that the 
discontinuity at $y=\pi R$ in the auxiliary field $D$ can 
be consistent with the field 
equations~(\ref{d3}).  However, in Eq.~(\ref{d11}) we have 
$\langle \Sigma (\pi R) \rangle =0$, 
and by choosing the consistent boundary condition $\delta \Sigma (\pi R) =0$, 
the field equations~(\ref{d3}) are indeed satisfied. 

This will be the 
general rule for the flux interpretation of the solutions with lower numbers
of non-zero vev's:
discontinuities in the auxiliary $D$-terms (and therefore the flux) 
correspond to points where the wave function of $\Sigma$ has Dirichlet boundary
conditions. A solution with $\ell$ zero vev's for the link fields therefore 
corresponds to a wave function of $\Sigma$ having $\ell$ nodes (and 
corresponding Dirichlet boundary conditions) on the corresponding points. 
The solution in Eq.~(\ref{solut5}) is also unstable, having a tachyon at the position 
($\phi_n$) of the link field having a zero vev, but by adding the Wilson-line 
superpotential contributions, this solution becomes a stable vacuum 
close to the vertical axis $\xi_1=0$.      

The final example with $n-1$ non-zero vev's
is the \vac{1,3,4,...,n} solution given in Eq.~(\ref{solut6}).
In the extra-dimensional interpretation, the bulk field $\Sigma$ has the profile 
\beq
 \langle \Sigma \rangle ~=~ {\xi_n \over 2 \pi R} \, y  ~,~~~~~~~
\langle D \rangle ~=~ \cases{
          \xi_n /(2 \pi R)  &  for $ 0 < y \leq \pi R$\cr
               0  &  for  $y = 0$~,\cr}
\label{d13}
\eeq
leading to a flux
\beq
         F_{56} ~=~   {\xi_n \over 2 \pi R} 
    ~-~ \xi_n \, \delta (y-\pi R) ~ ,~~~~~~
    \int_{-\pi R}^{\pi R}  dy~\langle \partial_5 \Sigma \rangle ~=~0 ~ . 
\eeq  
Moreover, consistent with our previous explanations, 
$\langle \Sigma \rangle $ vanishes at the 
position $y=0$ where the auxiliary field is discontinuous. 
Just as with the two previous flux solutions, this solution is unstable. 
However, by adding the Wilson-line superpotential, 
this solution becomes stable close to the horizontal axis $\xi_n=0$.

The last examples we shall discuss are the 
\vac{2,3,...,j-1,j+1,...,n} solutions.
In these cases, the profile for the bulk 
scalar $\Sigma$ satisfies
\beqn
    && \delta \Sigma \, \biggl\{\langle \partial_5 \Sigma \rangle + \xi_1 \delta (y) +  \xi_n \delta 
    (y-\pi R)\biggr\} \nonumber\\ 
   && ~~~~~~~~~~~~~~~~ = ~\delta \Sigma \, \biggl\{ d_2 + \half (d_2-d_1)[\epsilon (y-y_j)- \epsilon
            (y+y_j)] \biggr\} , 
\label{d7}
\eeqn 
where $y_j$ is the location at which the auxiliary field $\langle D \rangle $ 
jumps from
$\langle D \rangle = d_1$ for $0<|y|<y_j$ to 
$\langle D \rangle = d_2$ for $y_j<|y|<\pi R$.  
The profile has the form 
\beq
   \langle \Sigma \rangle ~=~ \cases{
       d_1\, y -{\xi_1 \over 2} \, \epsilon (y) & for  $0 < |y| < y_j $ \cr
       d_2\, y + \left(\half \, \xi_n  - \pi R d_2\right) \,\epsilon (y) & 
              for $y_j < |y| < \pi R $~,\cr}
\label{d8}
\eeq
where the constraint
\beq
          d_1 y_j ~+~ d_2  (\pi R-y_j) ~=~  \half (\xi_1 + \xi_n ) 
\eeq
serves as the continuity condition for the profile at $y=y_j$. 
By matching the 
deconstructed version to the continuous version via $y_j \sim j-1$, 
we find the expected result that $\langle \Sigma (y_j) \rangle=0$.  
The flux in this case is
\beq
    F_{56} ~=~  d_2 + \half\, (d_2-d_1)\,
    \left \lbrack \epsilon (y-y_j)- \epsilon (y+y_j)\right\rbrack  - 
         \xi_1 \, \delta (y) -  
           \xi_n \, \delta (y-\pi R) ~,
\label{d9} 
\eeq
which again integrates to zero.

%==========================================================================
\subsection{Anomalies in the deconstructed $U(1)^n$ theory}

As we remarked earlier,
the solutions with $n-1$ non-zero vev's --- and indeed all of the
solutions with fewer non-zero vev's ---
are unstable at the level of our discussion
(in which we are taking $\lambda/M_P^{n-2} \to 0$ and $\tilde m=0$).
However, at first sight it may also seem that
these vacua have a further unpleasant feature, namely
a large number of unbroken Abelian gauge symmetries 
which survive down to low energies.  However,
as we shall now demonstrate, this is not the case
thanks to the presence of mixed gauge anomalies in such theories.

It is easy to see why subtleties related to mixed anomalies
will be present in any model containing multiple Abelian gauge fields 
related to higher-dimensional compactifications and/or string theory. 
The basic reason is that bi-fundamental fields, which have been the building 
blocks of our toy landscape models, create mixed anomalies which typically
require
axionic fields and four-dimensional Green-Schwarz mechanisms for
their cancellation. 
The axionic fields will come in complex
(moduli) superfields due to supersymmetry, which will also fix the values of
the FI terms of the model. 
The dynamics and the stabilization of these 
moduli is, of course, very important in the landscape picture.  Indeed, we
have thus far been assuming that these moduli are 
already stabilized by the dynamics
(through, \eg,  NS-NS and RR fluxes, gaugino condensation, {\it etc}\/.).
 
In the $U(1)^n$ example, the mixed anomalies are described by the 
$(n \times n)$-dimensional anomaly matrix 
\beq
       C_{ab} ~=~ { 1\over 4 \pi^2} \,  {\rm Tr} \, (Q_a Q_b^2)~~~~~~
         (a,b = 1,...,n)~,
\eeq
where $Q_a$ is the generator of the $U(1)_a$ gauge group. 
The explicit matrix elements of $C_{ab}$ are given by~\cite{dfp}
\beq
    C_{ab} ~=~ {1 \over 4 \pi^2} \, \left\lbrack
     \delta_{a,b-1}~-~\delta_{a,b+1} ~+~ 
      {\rm Tr} \, Q_1^3\, \delta_{a,1}\delta_{b,1} ~+~ 
      {\rm Tr} \, Q_n^3 \, \delta_{a,n}\delta_{b,n}\right\rbrack ~,
\label{d011}  
\eeq
where for generality we have allowed for additional particles (\eg, MSSM particles)
at the fixed points.  Otherwise, 
in our minimal example discussed earlier, 
we would have
${\rm Tr} \, Q_1^3={\rm Tr} \, Q_n^3=0$. 
Whereas at the four-dimensional
level the anomaly-cancelling mechanism is not uniquely fixed (and we shall
shortly provide a trivial illustrative example of this),
the requirement of
realizing a five-dimensional Lagrangian in the continuous limit
is much more constraining. 
This was consistently done in Ref.~\cite{dfp}, where it was shown that the
axionic-type fields taking care of the mixed anomalies in Eq.~(\ref{d011}) 
are themselves bi-fundamental.
More precisely, there are axionic couplings of
the type $- \ln (\phi_i \phi_j) \delta_{ia} \delta_{j,a+1}
 W^{\alpha}_a  W_{\alpha, a}$, in superfield notation. 
There are also two boundary axionic fields
$S_{L,R}$ which transform only under the first and the last gauge group.\footnote{
    Further details and a discussion of
    the five-dimensional limit are given in Ref.~\cite{dfp}.}

The presence of the non-minimal kinetic
(and K\"ahler) functions clearly modifies the minimization procedure. 
It is therefore important to verify that our previous results remain intact.
We have explicitly verified that the absolute
minima of the model, as well as the first $(n-1)$-vev example with a clean 
magnetic flux interpretation, are affected in the full theory
only at subleading order in $\xi_i/M_P^2$.
However, the vacua with fewer than $(n-1)$ non-zero vev's 
do not remain extrema
in the full theory, since a vanishing vev for a bi-fundamental 
represents a singular point of the theory. 
The natural solution in the case where a zero vev occurs in a link
field $\phi_k$ is, in the extra-dimensional interpretation, to add a
brane at the corresponding point in the compact space.
We should also  introduce a
superfield $S_k$ with the gauge transformation 
$\delta S_k \sim \Lambda_{k-1} - \Lambda_{k}$, and replace 
$\ln \phi_k \rightarrow S_k$ in the gauge kinetic functions. 
This new field would then have
the interpretation of coupling to the $U(1)$ gauge field of the brane.
This automatically cancels the anomalies, explains why a new massless
gauge field appears in the spectrum, and gives a rationale for the
Dirichlet boundary condition which we found in Sect.~4.3
at the position of the vanishing vev.

This new axionic coupling can, however, render
the new gauge field massive.  Indeed, more and more vanishing vev's require
the addition of more and more $U(1)$ branes and corresponding new axionic 
couplings, which
eventually render most of the new brane $U(1)$'s massive. 
   
If we completely abandon the five-dimensional interpretation, 
there is certainly more freedom
in anomaly cancellation.  For example, we can introduce $n$ moduli fields
$S_a$ which have $U(1)_a$  gauge transformations 
$\delta S_1 \sim - \Lambda_2 $, 
$\delta S_a \sim \Lambda_{a-1} - \Lambda_{a+1}$ for $a=2, ...,  n-1$, 
and $\delta S_n \sim  \Lambda_{n-1} $ 
where $\Lambda_a$ is
the superfield describing $U(1)_a$ gauge transformations.
Coupling these moduli fields  to the gauge fields through a 
term of the form
$\sum_a S_a  W^{\alpha}_a W_{\alpha, a}$
will indeed cancel the mixed anomalies. 
In this case, assuming an {\it a priori}\/ 
stabilization of moduli, 
all the results of our toy model follow accordingly. 
Generically however, this
generates FI terms at all sites (\ie, for all $U(1)$'s), 
unless the Lagrangian of the moduli
fields and their vev's are such that FI terms vanish 
at all sites but the first and last.
Indeed, such examples are relatively easy to find. 
However, the price to pay in such a construction is the loss of
a five-dimensional interpretation in the limit of large $n$, and 
the unsolved remaining issue of moduli stabilization fixing the 
FI terms.

The upshot of the present discussion, then, is that in order to
cancel gauge anomalies, 
we must introduce new
axionic (super)fields with appropriate couplings under the various 
gauge groups.  In this case, 
most of the $U(1)$ gauge-group factors 
become massive, a welcome feature for the phenomenological viability of 
these models.

%========================================================================
\setcounter{footnote}{0}
\section{Extension to supergravity}

Among the main motivations for seriously considering the landscape picture
of string theory are its ramifications for vacuum stability and 
the cosmological constant problem.
 From the perspective of both of these problems,
the inclusion of supergravity interactions
is of crucial importance. 
We shall now briefly discuss each of these issues in turn.

\subsection{Vacuum stability, soft masses, and  split supersymmetry}

If we consider only the Abelian gauge fields which are relevant for our purposes,
the scalar supergravity potential is given in terms of the 
K\"ahler potential $K$ and the superpotential $W$ by~\cite{cfgp}
\beqn
    && V (\phi_i) ~=~ e^{K} \, ( K^{i \bar j} D_i W D_{\bar j} {\bar W} \, -\,  3 |W|^2)\nonumber\\ 
   && ~~~~~~~~~~~~~~~~~~~+ ~\half \, \sum_{a,b} \,f_{ab} \, 
          (q_{i}^{(a)} \phi^i K_i + \xi_a)  \, (q_{j}^{(b)} 
\phi^j K_j + \xi_b)~   
\label{s1}
\eeqn  
where we have temporarily set $M_P=1$ and where
we have employed the usual notation $K_i \equiv \partial_i K$,  
 $K_{i \bar j} \equiv \partial_i \partial_{\bar j}K$, 
and $F_i \equiv D_i W = \partial_i W + K_i W$.
Here $f_{ab}$ 
are the gauge kinetic functions of the Abelian gauge groups, which depend 
on the moduli fields which we are implicitly assuming to be stabilized.
Of course, after moduli stabilization, we should 
be considering gauged $R$-symmetries 
at the supergravity level~\cite{freedman}.  However, since our main focus
is on the scalar potential, this subtlety does not affect the
present discussion  and it is enough to consider the expectation value   
$\langle f_{ab}\rangle  = \delta_{ab} /g_a^2$. 
Note that the gravitino mass is $m_{3/2}^2 = |W|^2 \exp (K)$. 

As before, the stability of a given extremum in the supergravity
framework depends on
the scalar mass matrix:  an extremum represents  a stable vacuum 
or flat direction
only if the mass matrix has no negative eigenvalues. 
Of a particular importance for our purposes are the $D$-term contributions  
to the scalar masses, which after moduli stabilization are given by
\beq
\cases{  
   \partial_i \partial_{\bar j} V_D ~=~ g_a^2 q_i^{(a)} (K_{\bar j}^i + {\bar
      \phi}^{\bar l} K_{{\bar j} {\bar l}}^i) D_a + Q_{ik} \phi^k  {\bar
      \phi}^{\bar l} K_{\bar l}^i  K_{\bar j}^k  & \cr 
   \partial_i \partial_j V_D ~=~ g_a^2 q_i^{(a)} {\bar
      \phi}^{\bar l} K_{j {\bar l}}^i D_a + Q_{ij} {\bar
      \phi}^{\bar l} K_{\bar l}^i  K^j + Q_{ik} \phi^k  {\bar
      \phi}^{\bar l} K_{\bar l}^i  K_{\bar j}^k  & \cr 
  }
\label{s01}  
\eeq
where the charge matrix $Q_{ij}$ is defined as
\beq
   Q_{ij} ~\equiv~ \sum_{a=1}^n \, q_i^{(a)}  q_j^{(a)} ~. 
\label{s02} 
\eeq

The supergravity effects on the $U(1)^n$ deconstruction model described in the previous section can be
easily described in the limit in which the supergravity moduli fields are heavy and can be integrated out.
This leads to a low-energy theory with soft breaking terms. 
It is clear from gauge invariance that the resulting
soft breaking terms are of the form 
\beq
         V_{\rm soft} ~=~ \sum_{i=1}^{n+1} {\tilde m}_i^2 |\phi_i|^2 ~+~ 
         \lbrace  A \, \lambda\,  \prod_{i=1}^{n+1} \phi_i  \, +\,  {\rm h.c.}\rbrace ~ . 
\label{s03} 
\eeq    
The $A$-term induced by the Wilson-line superpotential has 
a negligible effect on the dynamics for large $n$, and can
safely be neglected in the following.
However, the effects of the soft terms ${\tilde m}_i^2$ on the vacua described in 
previous sections are quite clear. 
Specifically, just as we saw in previous sections,
the large number of flux vacua which were previously unstable 
become now stable in a certain region of the parameter space $(\xi_1,\xi_n)$.

Clearly, this method of increasing the number of stable vacua 
only works for large
soft masses, of the order of the FI terms. 
 From this point of view, a large supersymmetry breaking scale
stabilizing most of the landscape vacua fits well with the ``split supersymmetry'' 
proposal in Ref.~\cite{split}, 
in which a large supersymmetry-breaking scale was 
assumed to go together with the presence of a large number of vacua.

%===============================================================================
\subsection{A supergravity toy model and the cosmological constant problem}

In order to quantify 
the issue of the cancellation of the cosmological constant
in a simpler setting, 
we shall now propose
a simple supergravity model, inspired by the toy models discussed in the 
previous sections and by the flux compactifications of string theory~\cite{fluxes}. 
Since the deconstruction model of Sect.~3 is difficult to analyze
in this context, we will instead consider the supergravity extension
of $n$ copies of the original FI model.  
Thus, our model will be based on the gauge group $U(1)^n$ and chiral
superfields $\Phi_i^{(+)}, \Phi_i^{(-)}$ with charges $\pm 1$ under the gauge 
group factor $U(1)_i$.  These fields  are therefore coupled to each other only
through supergravity interactions.  
The model has a single K\"ahler modulus 
$T$ describing
the volume of a Calabi-Yau or orbifold compactification,
and has supergravity potentials described by
\beqn
     K &=& -3  \ln \left (T+{\bar T} - \sum_{i=1}^n |\Phi_i^{(+)}|^2 -
        \sum_{i=1}^n |\Phi_i^{(-)}|^2\right) \nonumber \\
    W &=& \sum_{i=1}^n \, m_i \, \Phi_i^{(+)} \Phi_i^{(-)} + W_0 + h \,e^{-b_0 T}~,
\label{s2}  
\eeqn   
where $W_0$ is determined by the fluxes according to Ref.~\cite{fluxes} 
and the nonperturbative (exponential) term in the 
superpotential, responsible for the stabilization of the K\"ahler modulus $T$, 
could come from
D7-branes, gaugino condensation, or Euclidean D3-branes~\cite{kklt}.  
All other moduli are assumed to be stabilized by other dynamical
effects, \eg, by adding appropriate fluxes. 

If we first set the FI terms to zero, \ie,  $\xi_i =0$, there is a supersymmetric
solution giving an AdS space, obtained by solving $F_T = 0$.
However, in the presence of the FI terms, the vev's of the moduli fields
will be changed. If the FI terms and the vev's of the fields 
$\phi_i^{(\pm)}$ are sufficiently small compared to the Planck scale, 
we can effectively integrate out the $T$-field.  This will give rise
to  explicit soft terms in the Lagrangian~\cite{bfs}. 
The part of the scalar potential which is 
relevant for discussing the cosmological constant in the various vacua of the
theory is then given by
\beqn
   V &=& \sum_{i=1}^n \, \biggl[  (m_i^2+  {\tilde m}_i^2) \, (|\phi_i^{(+)}|^2 +
   |\phi_i^{(-)}|^2)
   ~+~  B_i \, (\phi_i^{(+)}\phi_i^{(-)} + {\bar \phi_i}^{(+)} {\bar \phi_i}^{(-)})
     \biggr]  \nonumber \\
   && ~~~+~\sum_{i=1}^n \, {g_i^2 \over 2}
    ( |\phi_i^{(+)}|^2 - |\phi_i^{(-)}|^2+ \xi_i)^2 ~ - ~ 3 \, m_{3/2}^2 M_P^2~,   
\label{s4}
\eeqn
where ${\tilde m}_i^2$ and $B_i$ describe the soft terms and where
we have assumed equal soft masses for the fields
$\phi_i^{(+)}$ and $\phi_i^{(-)}$  for simplicity. 

A few words of caution are in order here.
Note that the potential~(\ref{s4}) has been written
to the lowest order in the matter fields $\phi_i^{(\pm)}$.  Thus,
for example, the gravitino mass, which is a function of all the fields,
becomes a constant. 
As a result of these approximations, 
the vacua derived from Eq.~(\ref{s4}) are certainly only
approximations to the true vacua. 
Note, in addition, that we have also neglected perturbative loop effects 
as well as non-perturbative effects. 
Moreover, integrating out the field $T$ in Eq.~(\ref{s2})
is a good approximation only if the field $T$ is heavier than 
the charged fields $\phi_i$.   This is 
not always a good approximation to the dynamics of the model. 
Despite these issues,
we believe that 
our discussion concerning the possible cancellation of the
cosmological constant will still be of  qualitative value,
and that the qualitative features of such a cancellation --- which relies
on balancing the positive contributions to the vacuum energy
coming from the supersymmetry-breaking against the negative contributions
coming from the supergravity interactions --- capture the qualitative 
features of the full quantum problem.
 
Minimizing the scalar potential in Eq.~(\ref{s4}),
we find $3^n$ possible vacua,
determined as simultaneous solutions of the coupled equations
\beqn
&& (m_i^2 + {\tilde m}_i^2 + g_i^2 D_i) \, v_i^{(+)} ~+~ 
             B_i {\bar v}_i^{(-)} ~=~ 0  \nonumber \\
&& (m_i^2 +  {\tilde m}_i^2 - g_i^2 D_i) \, v_i^{(-)} ~+~ 
             B_i {\bar v}_i^{(+)} ~=~ 0  
\label{s5}
\eeqn
where  $v_i^{(\pm)} \equiv \langle \phi_i^{(\pm)} \rangle $. 
Indeed, the field equations~(\ref{s5})
have three solutions for each $i$. 
By performing $U(1)$ rotations,
the solutions can always be made real. 
 
First, it is useful to define
\beq
     m_{S,i}^2 ~\equiv~  m_i^2 + {\tilde m}_i^2 ~,~~~~~~ 
        \Delta_i^2 ~\equiv~ \sqrt{ m_{S,i}^4-B_i^2}
\label{s05}
\eeq
for notational simplicity. 
We then find that the first class of solutions are given by
\beq
|v_i^{(+)}|^2 = - { (\Delta_i^2 + m_{S,i}^2)(\Delta_i^2+ g_i^2 \xi_i) 
    \over 2 g_i^2 \Delta_i^2}~,~~~~ 
   |v_i^{(-)}|^2 = { (\Delta_i^2 - m_{S,i}^2)(\Delta_i^2+ g_i^2 \xi_i) 
    \over 2 g_i^2 \Delta_i^2} ~,
\eeq
which leads to $D$-terms and vacuum energy $\Lambda_0$ of the form
\beq
    g_i^2 \langle D_i \rangle ~=~ - \Delta_i^2 ~,~~~~~~
   \Lambda_0 ~=~ \sum_i {\Delta_i^4 \over g_i^2} \left(-{1 \over 2} - {g_i^2
  \xi_i \over \Delta_i^2} \right) - 3 \ m_{3/2}^2 M_P^2 ~.
\label{s6}
\eeq  
The second class of solutions, by contrast, are given by
\beq
        |v_i^{(+)}|^2  =  { (\Delta_i^2-m_{S,i}^2)(\Delta_i^2- g_i^2 \xi_i) 
       \over 2 g_i^2 \Delta_i^2}~,~~~~ 
      |v_i^{(-)}|^2  = - { (\Delta_i^2 + m_{S,i}^2)(\Delta_i^2- g_i^2 \xi_i) 
          \over 2 g_i^2 \Delta_i^2} ~,
\eeq
which lead to 
\beq
        g_i^2 \langle D_i \rangle ~=~  \Delta_i^2 ~,~~~~~~
         \Lambda_0 ~=~  \sum_i {\Delta_i^4 \over g_i^2} \left(-{1 \over 2} + {g_i^2
  \xi_i \over \Delta_i^2} \right) - 3  m_{3/2}^2 M_P^2 ~.
\label{s7}
\eeq
Clearly these solutions exist only when $|v_i^{(\pm)}|^2 >0$, 
which means that $g_i^2 |\xi_i| > \Delta^2$ and $m_{S,i}^2 >\Delta^2$ 
where $\xi_i < 0$ for Eq.~(\ref{s6}) and $\xi_i > 0$ for Eq.~(\ref{s7}). 
One can also verify that these solutions are stable.
Finally, the third class of solutions is trivially
\beq
            v_i^{(+)} ~ = ~ v_i^{(-)} ~ = ~ 0 ~,
\label{s8}
\eeq
which is easily verified to be stable if
\beq
            m_{S,i}^2 \pm \sqrt{B_i^2 + g_i^4 \xi_i^2} ~\geq~  0 ~ . 
\label{s8a}
\eeq  
Of course, in the limit of vanishing soft terms, \ie, ${\tilde m_i}^2 , B_i \to 0$,
all three classes of solutions reduce 
to the standard global minima of the Fayet-Iliopoulos model discussed in Sect.~2. 

From a string-theory perspective, 
the limit of interest corresponds to taking
$|\xi_i| \gg m_i^2,  {\tilde m_i}^2, B_i$. 
In this limit, the cancellation of the cosmological constant then 
takes the approximate form
\beq
     \sum_{i=1}^n |\xi_i| \sqrt{( m_i^2 + {\tilde m}_i^2)^2-B_i^2} ~\simeq~ 
     3 m_{3/2}^2 M_P^2~,
\eeq 
for the $2^n$ vacua in the classes in Eqs.~(\ref{s6}) and (\ref{s7}).
In gravity-mediated supersymmetry breaking scenarios such as we are considering,
we expect ${\tilde m_i}^2, B_i \sim m_{3/2}^2$. 
Since typical values for the Fayet-Iliopoulos coefficients $\xi_i$ are 
smaller by one or two orders of magnitude compared to the Planck scale, 
we see that there are therefore basically
only two possible ways in which these $2^n $ vacua
can yield a small cosmological constant:
\begin{itemize}
\item   We can have $m_i^2 \gg m_{3/2}^2$.  However, in this case,
    partial cancellation can occur only for a 
     small number of gauge factors, which means that a reasonable 
     accommodation for a small cosmological constant is difficult to achieve.
\item  Alternatively, we can have $m_i^2 \leq  m_{3/2}^2$.
       In this case, the cosmological constant can be
      cancelled by combining the two different types of vacua in Eqs.~(\ref{s6}) and
       (\ref{s7}).   For example, we would require $n \sim 10-100$ 
             for generic values $\xi_i \sim (10^{-1}-10^{-2}) M_P^2$ . 
     Indeed, in such cases, there will be a large
     number of landscape vacua with small vacuum energies. 
     Of course, any vacuum solutions which also  contain the third class
     of solutions in Eq.~(\ref{s8}) will add contributions 
      $\half g_j^2 \xi_j^2$ to the vacuum energy $\Lambda_0$. 
     These contributions are very large, and 
     will upset the cancellation of the predicted cosmological constant unless 
         $|\xi_j| \ll m_{3/2} M_P$.     
\end{itemize}
By contrast, in the opposite limit of small FI terms, cancellation occurs for
typical values of soft terms ${\tilde m}^2 \sim \sqrt{3} m_{3/2} M_P/n$. 

The outcome of this discussion, then, is that it is the second possibility itemized
above which can best accommodate a small cosmological constant
in our toy model. 
Barring additional fine-tunings, this clearly points towards    
requiring $m_i^2 \ll {\tilde m_i}^2$, from which we conclude that 
even a qualitative supergravity analysis of the vacuum structure
of any landscape model is crucial for addressing the smallness
of the cosmological constant.\footnote{
     Of course, in addition to the stability of the vacua and the fine-tuning of the
     cosmological constant (both of which seem to favor large soft masses),
     a second crucial fine-tuning in this case is the light Higgs mass.
     This can only be addressed within the context 
     of a specific model in  which Standard Model or MSSM fields are coupled to the
     $U(1)$ fields we are investigating, and is thus beyond the scope of the current 
     paper.}

High-scale supersymmetry breaking therefore seems
to have two positive features in our $D$-term landscape models. 
First, they render most of the vacua stable by adding large diagonal soft masses.
Second, they allow a possible cancellation of the cosmological constant.

%=============================================================================
\setcounter{footnote}{0}
\section{Discussion}

In this paper, we proposed a field-theoretic framework giving rise to 
models containing large numbers of vacuum solutions. 
The field-theoretic nature of these models therefore allowed us to explicitly calculate 
such quantities of interest as
the ratio of stable versus total numbers of vacua, the number of
$R$-symmetry preserving vacua, and the supersymmetry-breaking scale.
While at the field-theory level many models with large
numbers of vacua can certainly be given, our examples have the
advantage of describing large classes of string compactifications
with Abelian gauge groups and FI terms. 
Moreover, within this large class, 
we presented specific examples involving discretized versions of magnetic
fluxes in the internal space, as obtained by deconstructing (supersymmetric)
models with $U(1)$ gauge fields on the orbifold $S^1/\IZ_2$. 
By examining the extrema involving vanishing vev's for bi-fundamental 
fields, we found that these solutions correspond to 
profiles which are linear in $x_5$ for the odd-scalar 
$\Sigma = A_6$ in the five-dimensional vector multiplet.  These solutions 
therefore correspond
to constant magnetic fluxes $\partial_5 \Sigma = F_{56}$ from a six-dimensional
perspective.  The number of deconstructed magnetic fluxes is large, and
although they all have Nielsen-Olesen instabilities, 
they can be stabilized by $F$-term contributions coming from supergravity
interactions.  This therefore generates a field-theory landscape which is similar
in spirit and closely related to the landscapes currently under discussion 
in string-theory contexts~\cite{landscape}. 
We also showed that (super)gravity effects turn out to be consistent with 
(and actually needed for) a qualitative accommodation of the
smallness of the cosmological constant. 
 
One of the interesting results of the landscape picture emerging in
the class of models we considered in this paper is the possibility
of passing from one vacuum to another by renormalization group
flow.  This possibility arises because our fundamental defining parameters,
such as $\xi_i$ and $\lambda$,  can change with the energy scale.
We showed in an explicit toy model that this renormalization group flow
can indeed induce boundary crossings.   
A general feature of the boundary
separating two different vacua is the presence of a massless scalar in
the spectrum, which suggests an interpretation in terms of phase transitions.
This type of boundary crossing could therefore be of potential interest for 
inflationary models. Moreover, the gauge symmetry-breaking pattern
can also change as a result of boundary crossings.
The corresponding phase transitions
could therefore also be relevant for electroweak symmetry-breaking 
in the early Universe and for baryogenesis.
We hope to return to some
of these interesting phenomena in the future. 

The example(s) that we analyzed in some detail in this paper
were motivated by
providing a purely four-dimensional description 
of higher-dimensional fluxes, in the spirit of deconstruction.
There are clearly several possible generalizations of this
setup.  One possibility would be to add flavors to the models discussed in Sect.~3, 
again in the spirit of deconstruction. 
Note that the net effect of adding such flavors is
to add new fields. 
For example, if we add flavors in the bulk [\ie, at all $U(1)$ sites 
except the first and last], the total number of fields for the $U(1)^n$ 
model would grow to $3(n-1)$. 
The naive number of extrema in this case 
would then be $\sim {\cal O}(2^{3n-4})$ but may actually be smaller in practice. 
There are also new $F$-term contributions to the scalar
potential which render some of these vacua stable. 

The field-theoretic framework proposed in this paper also allowed us
to explicitly calculate the ratio of stable vacua to 
the total numbers of extrema.
We found that this ratio is much larger
than a na\"\i ve estimate
based on random sign assignments for mass eigenvalues would suggest. 
We also found in Sect.~3 that these sorts of results --- and  indeed the entire vacuum
structure of the models we considered ---  are very sensitive not only
to dimensionful mass parameters, such as FI terms and
various type of scalar masses (all of which are natural landscape
coordinates), but also to dimensionless parameters such as the Yukawa
and gauge couplings.
 
There are clearly more general examples that can arise in string
models with D-branes.  However, these will depend on the particular 
classes of string models under consideration.
For example, in heterotic string constructions, 
there is at most one anomalous $U(1)$ factor, with the
universal axion mixing \`a la St\"uckelberg~\cite{dsw}.  
By contrast,
in a Type~I string model (or equivalently a Type~II orientifold),
there can be a large number of anomalous $U(1)$ factors, each equipped with
its own FI term~\cite{sagnotti}. 

One natural framework to consider consists of taking a large  number  of
$U(1)$'s, each with its FI term,  and all possible charged bi-fundamental fields.
If there are $n$ different $U(1)$ gauge factors, this would imply 
$n(n+1)/2$ bi-fundamental fields.  Such a setup would emerge naturally 
in Type I or Type II string models
containing large numbers of $U(1)$ D-branes which either sit at the same point in
the compact space or intersect each other.  In such a geometry,
the bi-fundamental fields would emerge
as open-string excitations stretching from one $U(1)$ brane to another. 
This setup would not only yield a large number of
vacua, but would also naturally give rise to various new types of superpotential terms.
These include cubic, quartic, and higher-order interactions,
each of which would produce additional
$F$-term contributions to the scalar potential. 
It is easy to imagine that in
some region of the parameter space,   
these additional contributions could easily stabilize most of the huge numbers of vacua 
that such models would produce.

Regardless of the particular field-theoretic model under discussion,
it is apparent that string models of all sorts generically 
give rise to multiple $U(1)$ gauge factors.
We therefore believe that even though the models we presented in this paper
are field-theoretic in nature, they 
will emerge naturally in realistic string contexts and should be viewed,
quite literally, as at least one component of the full string landscape.
As such, we believe that these types of models should be analyzed further,
as they undoubtedly play a direct role in affecting the statistical
properties of string vacua.

%=============================================================================
\setcounter{footnote}{0}
\section*{Acknowledgments}

KRD is supported in part by the US National Science Foundation
under Grant~PHY/0301998, 
by the US Department of Energy under Grant~DE-FG02-04ER-41298,
and by a Research Innovation Award from 
Research Corporation.  ED is supported in part by the CNRS PICS \#2530, 
INTAS grant 03-51-6346, the RTN grants MRTN-CT-2004-503369 and 
MRTN-CT-2004-005104, and by a European Union Excellence Grant, 
MEXT-CT-2003-509661.
TG is supported in part by a US Department of Energy grant~DE-FG02-94ER-40823, 
a grant from the Office of the Dean of the Graduate School of the University 
of Minnesota, and a Research Innovation Award from Research 
Corporation.  KRD and ED would like to acknowledge the hospitality of the 
William~I. Fine Theoretical Physics Institute at the University of Minnesota,
and KRD also wishes to acknowledge the hospitality of
both CERN and the CPhT-Ecole Polytechnique.
We also wish 
to acknowledge the hospitality of the Aspen Center for Physics
where this work was initiated.

%  \vfill
%  \eject

%===============================================================================
\setcounter{footnote}{0}
\section*{Appendix}

In this Appendix we present the details behind the $n=3$ landscape 
discussed in Sect.~4.1 and sketched in Fig.~\ref{fig10}. 
As discussed in Sect.~4.1, we shall set $\lambda=0$ and take $\tilde m^2\geq 0$.
The surviving extrema are then in the following classes:

\bigskip
\noindent \vac{$\emptyset$}:
These are extrema with all $v_i\equiv \langle \phi_i\rangle = 0$. 
The eigenvalues are:
\begin{equation}
   \tilde m^2 \pm \xi_1~~(2), \quad \tilde m^2 \pm \xi_3~~(2)~,
\end{equation}
where the number in parentheses corresponds to the multiplicity of the 
eigenvalues. Thus, provided that
\begin{equation}
   |\xi_1| < \tilde m^2\quad {\rm and}\quad |\xi_3| < \tilde m^2~,
\end{equation}
this vacuum is stable. Clearly, when $\tilde m =0$ this vacuum is unstable.
The $D$-terms are given by
\begin{equation}
    \langle {\vec D}\rangle =\{\xi_1,0,\xi_3\}~.
\end{equation}

\noindent
\vac{1}:  $|v_1|^2= \xi_1 - \tilde m^2, v_2=v_3=v_4=0$, with 
eigenvalues
\begin{equation}
   0,\quad 2(\xi_1-\tilde m^2),~\quad 2\tilde m^2~~(2),\quad
   \tilde m^2 \pm \xi_3~~(2)~.
\end{equation}
The stability region is:
\begin{equation}
   \xi_1 > \tilde m^2\quad {\rm and}\quad |\xi_3| < \tilde m^2~.
\end{equation}
When $\tilde m=0$ this vacuum is no longer stable.
The $D$-terms are given by
\begin{equation}
    \langle {\vec D} \rangle =\{\tilde m^2,0,\xi_3\}~.
\end{equation}
%  \\

\noindent
\vac{2}:  $|v_2|^2= -\frac{1}{2}(\xi_1 + \tilde m^2), 
v_1=v_3=v_4=0$, with eigenvalues
\begin{equation}
   0,~~-2(\xi_1+\tilde m^2),
   ~~\frac{1}{2}(3\tilde m^2-\xi_1)~~(2),
   ~~\frac{1}{2}(3\tilde m^2+\xi_1-2\xi_3)~~(2),
   ~~\tilde m^2 + \xi_3~~(2)~.
\end{equation}
The stability region is:
\begin{equation}
   \xi_1 < -\tilde m^2,~\quad \xi_3 > -\tilde m^2,\quad
    {\rm and}~\quad 2\xi_3-\xi_1-3\tilde m^2 <0~.
\end{equation}
When $\tilde m=0$ this vacuum is no longer stable.
The $D$-terms are given by
\begin{equation}
    \langle {\vec D} \rangle =\{\frac{1}{2}(\xi_1-\tilde m^2),\frac{1}{2}(\xi_1+\tilde m^2),
     \xi_3\}~.
\end{equation}
%  \\

\noindent
\vac{3}: $|v_3|^2= \frac{1}{2}(\xi_3 - \tilde m^2), 
v_1=v_2=v_4=0$, with eigenvalues
\begin{equation}
   0,~~2(\xi_3-\tilde m^2),
   ~~\frac{1}{2}(3\tilde m^2+\xi_3)~~(2),
   ~~\frac{1}{2}(3\tilde m^2+2\xi_1-\xi_3)~~(2),
   ~~\tilde m^2 - \xi_1~~(2)~.
\end{equation}
The stability region is:
\begin{equation}
   \xi_1 < \tilde m^2,~\quad \xi_3 > \tilde m^2,\quad
    {\rm and}~\quad \xi_3-2\xi_1-3\tilde m^2 <0~.
\end{equation}
When $\tilde m=0$ this vacuum is no longer stable.
The $D$-terms are given by
\begin{equation}
    \langle {\vec D} \rangle =\{\xi_1,\frac{1}{2}(\xi_3-\tilde m^2),
     \frac{1}{2}(\xi_3+\tilde m^2)\}~.
\end{equation}
%  \\

\noindent
\vac{4}:  $|v_4|^2= -\xi_3 - \tilde m^2, v_1=v_2=v_3=0$, 
with eigenvalues
\begin{equation}
   0,\quad -2(\xi_3+\tilde m^2),~\quad 2\tilde m^2~~(2),\quad
   \tilde m^2 \pm \xi_1~~(2)~.
\end{equation}
The stability region is:
\begin{equation}
   \xi_3 < -\tilde m^2\quad {\rm and}\quad |\xi_1| < \tilde m^2~.
\end{equation}
When $\tilde m=0$ this vacuum is no longer stable.
The $D$-terms are given by
\begin{equation}
    \langle {\vec D} \rangle =\{\xi_1,0,-\tilde m^2\}~.
\end{equation}

\noindent
\vac{12}:  There are no solutions in this class.
%  \\

\noindent
\vac{13}: $|v_1|^2= \xi_1 - \tilde m^2, 
|v_3|^2=\frac{1}{2}(\xi_3-\tilde m^2), v_2=v_4=0$, with eigenvalues
\begin{equation}
   0~~(2),\quad 2(\xi_1-\tilde m^2),\quad 2(\xi_3-\tilde m^2),
    \quad \frac{1}{2}(5\tilde m^2-\xi_3)~~(2),\quad
    \frac{1}{2}(3\tilde m^2+\xi_3)~~(2)~.
\end{equation}
The stability region is
\begin{equation}
   \xi_1> \tilde m^2,\quad {\rm and}\quad\tilde m^2 < \xi_3 < 5\tilde m^2~.
\end{equation}
When $\tilde m=0$ this vacuum is no longer stable.
The $D$-terms are given by
\begin{equation}
   \langle {\vec D} \rangle =\{\tilde m^2,\frac{1}{2}(\xi_3-\tilde m^2),
\frac{1}{2}(\xi_3+\tilde m^2)\}~.
\end{equation}
%  \\

\noindent
\vac{14}:  $|v_1|^2= \xi_1 - \tilde m^2, |v_4|^2=-\xi_3-\tilde m^2, 
v_2=v_3=0$, with eigenvalues
\begin{equation}
   0~~(2),\quad 2(\xi_1-\tilde m^2),\quad -2(\xi_3+\tilde m^2),
    \quad 2\tilde m^2~~(4)~.
\end{equation}
The stability region is
\begin{equation}
   \xi_1> \tilde m^2\quad {\rm and}\quad \xi_3 < -\tilde m^2~.
\end{equation}
When $\tilde m=0$ this vacuum becomes supersymmetric. 
The $D$-terms are given by
\begin{equation}
   \langle {\vec D} \rangle =\{\tilde m^2,0,-\tilde m^2\}~.
\end{equation}
%  \\

\noindent
\vac{23}:  $|v_2|^2= \frac{1}{3}(-3\tilde m^2-2\xi_1+\xi_3), 
|v_3|^2= \frac{1}{3}(-3\tilde m^2-\xi_1+2\xi_3), 
v_1=v_4=0$, with eigenvalues
\beqn
   && 0~~(2),\quad \frac{1}{3}(6\tilde m^2\pm\xi_1\pm\xi_3)~~(2), \nonumber\\
   && -4\tilde m^2-2\xi_1+2\xi_3\pm \frac{2}{\sqrt{3}}
   \sqrt{3\tilde m^4+3\tilde m^2(\xi_1-\xi_3)+\xi_1^2-\xi_1\xi_3+\xi_3^2}~.
\eeqn
The stability region is
\begin{equation}
   \xi_1 <2\xi_3-3\tilde m^2,\quad
   \xi_3 > 2\xi_1+3\tilde m^2, \quad{\rm and}\quad
     |\xi_1 +\xi_3| < 6\tilde m^2~.
\end{equation}
When $\tilde m=0$ this vacuum is no longer stable.
The $D$-terms are given by
\begin{equation}
    \langle {\vec D} \rangle =\{-\tilde m^2+\frac{1}{3}(\xi_1+\xi_3),\frac{1}{3}(\xi_1+\xi_3),
      \tilde m^2+\frac{1}{3}(\xi_1+\xi_3)\}~.
\end{equation}
%  \\

\noindent
\vac{24}:  $ |v_2|^2=\frac{1}{2}(-\xi_1-\tilde m^2), 
|v_4|^2= -\xi_3 - \tilde m^2, v_1=v_3=0$, with eigenvalues
\begin{equation}
   0~~(2),\quad -2(\xi_1+\tilde m^2),\quad -2(\xi_3+\tilde m^2),
    \quad \frac{1}{2}(5\tilde m^2+\xi_1)~~(2),
    \quad \frac{1}{2}(3\tilde m^2-\xi_1)~~(2)~.
\end{equation}
The stability region is
\begin{equation}
   -5\tilde m^2 < \xi_1 < -\tilde m^2\quad {\rm and}\quad
     \xi_3 < -\tilde m^2~.
\end{equation}
When $\tilde m=0$ this vacuum is no longer stable.
The $D$-terms are given by
\begin{equation}
  \langle  {\vec D} \rangle =\{\frac{1}{2}(\xi_1-\tilde m^2),\frac{1}{2}(\xi_1+\tilde m^2),
     -\tilde m^2\}~.
\end{equation}
%  \\

\noindent
\vac{34}:  There are no solutions in this class.

\noindent
\vac{123}:   $|v_1|^2= \xi_1 +\xi_3 - 6\tilde m^2, 
|v_2|^2=\xi_3-5\tilde m^2, |v_3|^2=\xi_3-3\tilde m^2, v_4=0$. Some of the 
eigenvalue expressions are quite complicated but one
can verify that the stability region is:
\begin{equation}
   \xi_1 +\xi_3 > 6 \tilde m^2,\quad {\rm and}\quad \xi_3 > 5\tilde m^2~.
\end{equation}
When $\tilde m=0$ this vacuum remains stable.
The $D$-terms are given by
\begin{equation}
 \langle {\vec D} \rangle =\{\tilde m^2,2\tilde m^2,3\tilde m^2\}~.
\end{equation}
%  \\

\noindent
\vac{124}:  There are no solutions in this class.

\noindent
\vac{134}: There are no solutions in this class.

\noindent
\vac{234}:  $|v_2|^2=-\xi_1-3\tilde m^2, |v_3|^2=-\xi_1-5\tilde m^2, 
|v_4|^2= -\xi_1 -\xi_3 - 6\tilde m^2, v_1=0$. Some of the eigenvalue expressions
are quite complicated but one can verify that the stability 
region is:
\begin{equation}
   \xi_1 +\xi_3 < -6 \tilde m^2\quad {\rm and}\quad \xi_1 < -5\tilde m^2~.
\end{equation}
When $\tilde m=0$ this vacuum remains stable.
The $D$-terms are given by
\begin{equation}
  \langle {\vec D} \rangle =\{-3\tilde m^2,-2\tilde m^2,-\tilde m^2\}~.
\end{equation}

\noindent
\vac{1234}: There are no solutions in this class unless $\tilde m=0$.
For $\tilde m=0$, we have a line of solutions given by
\beq
           |v_1|^2 ~=~ |v_3|^2 + \xi_1~,~~~~
           |v_4|^2 ~=~ |v_3|^2 - \xi_3~,~~~~
           |v_2|^2 ~=~ |v_3|^2~.
\eeq
These solutions are necessarily supersymmetric, with all $D$-terms vanishing.

\bigskip
\bigskip
\noindent For $\tilde m^2 <0$, we obtain solutions for the remaining cases:
\bigskip

\noindent
\vac{12}:  $|v_1|^2= \xi_1 - 3\tilde m^2, |v_2|^2=-2\tilde m^2, 
v_3=v_4=0$, with eigenvalues
\begin{equation}
   0~~(2),\quad \tilde m^2+\xi_3~~(2),\quad 3\tilde m^2-\xi_3~~(2),
   \quad -7\tilde m^2+\xi_1\pm\sqrt{25 \tilde m^4-6\xi_1\tilde m^2+\xi_1^2}~.
\end{equation}
This vacuum has no stable regions, and the $D$-terms are given by
\begin{equation}
  \langle {\vec D} \rangle =\{\tilde m^2,2\tilde m^2,\xi_3\}~.
\end{equation}

\noindent
\vac{34}:
$|v_3|^2=-2\tilde m^2, |v_4|^2= -\xi_3-3\tilde m^2, 
v_1=v_2=0$, with eigenvalues
\begin{equation}
   0~~(2),\quad \tilde m^2-\xi_1~~(2),\quad 3\tilde m^2+\xi_1~~(2),
   \quad -7\tilde m^2-\xi_3\pm\sqrt{25 \tilde m^4 +6\tilde m^2\xi_3+\xi_3^2}~.
\end{equation}
This vacuum has no stable regions, and the $D$-terms are given by
\begin{equation}
\langle {\vec D} \rangle =\{\xi_1,-2\tilde m^2,-\tilde m^2\}~.
\end{equation}

\noindent
\vac{124}:
$|v_1|^2= -\xi_3 - \tilde m^2, 
|v_2|^2=\xi_1-3\tilde m^2, |v_4|^2= -2\tilde m^2, v_3=0$. This vacuum
has no stable regions, and the $D$-terms are given by
\begin{equation}
 \langle {\vec D} \rangle =\{\tilde m^2,2\tilde m^2,-\tilde m^2\}~.
\end{equation}

\noindent
\vac{134}:
$|v_1|^2= \xi_1 - \tilde m^2, |v_3|^2= -2\tilde m^2, 
|v_4|^2=-\xi_3-3\tilde m^2, v_2=0$. This vacuum has no stable regions,
and the $D$-terms are given by
\begin{equation}
  \langle {\vec D} \rangle =\{\tilde m^2,-2\tilde m^2,-\tilde m^2\}~.
\end{equation}

%  \vfill\eject

%=============================================================================
\bibliographystyle{unsrt}

\begin{thebibliography}{99}
\bibitem{bp}  
R.~Bousso and J.~Polchinski,
%``Quantization of four-form fluxes and dynamical neutralization of the
%cosmological constant,''
JHEP {\bf 0006}, 006 (2000)
[arXiv:hep-th/0004134]; 
J.~L.~Feng, J.~March-Russell, S.~Sethi and F.~Wilczek,
%``Saltatory relaxation of the cosmological constant,''
Nucl.\ Phys.\ B {\bf 602}, 307 (2001)
[arXiv:hep-th/0005276];
%%CITATION = HEP-TH 0005276;%%
for a general review, see
%%CITATION = HEP-TH 0004134;%%
R.~Bousso and J.~Polchinski,
%``The string theory landscape,''
Sci.\ Am.\  {\bf 291}, 60 (2004).
%%CITATION = SCAMA,291,60;%%

\bibitem{landscape}
L.~Susskind,
%``The anthropic landscape of string theory,''
arXiv:hep-th/0302219;
%%CITATION = HEP-TH 0302219;%%
M.~R.~Douglas,
%``The statistics of string / M theory vacua,''
JHEP {\bf 0305}, 046 (2003)
[arXiv:hep-th/0303194];
%%CITATION = HEP-TH 0303194;%%
M.~Dine,
%``Supersymmetry, naturalness and the landscape,''
arXiv:hep-th/0410201.
%%CITATION = HEP-TH 0410201;%%

\bibitem{statistics}
S.~Ashok and M.~R.~Douglas,
%``Counting flux vacua,''
JHEP {\bf 0401}, 060 (2004)
[arXiv:hep-th/0307049];
M.~R.~Douglas, B.~Shiffman and S.~Zelditch,
%``Critical points and supersymmetric vacua,''
Commun.\ Math.\ Phys.\  {\bf 252}, 325 (2004)
[arXiv:math.cv/0402326].
%%CITATION = MATH-CV 0402326;%%
%%CITATION = HEP-TH 0307049;%%
F.~Denef and M.~R.~Douglas,
%``Distributions of flux vacua,''
JHEP {\bf 0405}, 072 (2004)
[arXiv:hep-th/0404116];
%%CITATION = HEP-TH 0404116;%%
A.~Giryavets, S.~Kachru and P.~K.~Tripathy,
%``On the taxonomy of flux vacua,''
JHEP {\bf 0408}, 002 (2004)
[arXiv:hep-th/0404243];
M.~R.~Douglas, B.~Shiffman and S.~Zelditch,
%``Critical points and supersymmetric vacua, II: Asymptotics and extremal
%metrics,''
arXiv:math.cv/0406089;
%%CITATION = MATH-CV 0406089;%%
%%CITATION = HEP-TH 0404243;%%
O.~DeWolfe, A.~Giryavets, S.~Kachru and W.~Taylor,
%``Enumerating flux vacua with enhanced symmetries,''
arXiv:hep-th/0411061;
%%CITATION = HEP-TH 0411061;%%
J.~P.~Conlon and F.~Quevedo,
%``On the explicit construction and statistics of Calabi-Yau flux vacua,''
JHEP {\bf 0410}, 039 (2004)
[arXiv:hep-th/0409215];
%%CITATION = HEP-TH 0409215;%%
F.~Denef and M.~R.~Douglas,
%``Distributions of nonsupersymmetric flux vacua,''
arXiv:hep-th/0411183;
%%CITATION = HEP-TH 0411183;%%
R.~Blumenhagen, F.~Gmeiner, G.~Honecker, D.~Lust and T.~Weigand,
%``The statistics of supersymmetric D-brane models,''
arXiv:hep-th/0411173.
%%CITATION = HEP-TH 0411173;%%


\bibitem{dgt} 
T.~Banks, M.~Dine and E.~Gorbatov,
%``Is there a string theory landscape?,''
JHEP {\bf 0408}, 058 (2004)
[arXiv:hep-th/0309170];
%%CITATION = HEP-TH 0309170;%%
M.~R.~Douglas,
%``Statistical analysis of the supersymmetry breaking scale,''
arXiv:hep-th/0405279;
%%CITATION = HEP-TH 0405279;%%
D.~Robbins and S.~Sethi,
%``A barren landscape,''
arXiv:hep-th/0405011;
%%CITATION = HEP-TH 0405011;%%
H.~Firouzjahi, S.~Sarangi and S.~H.~H.~Tye,
%``Spontaneous creation of inflationary universes and the cosmic landscape,''
JHEP {\bf 0409}, 060 (2004)
[arXiv:hep-th/0406107];
%%CITATION = HEP-TH 0406107;%%
X.~Calmet,
%``Minimal grand unification model in an anthropic landscape,''
arXiv:hep-ph/0406314;
%%CITATION = HEP-PH 0406314;%%
E.~Silverstein,
%``Counter-intuition and scalar masses,''
arXiv:hep-th/0407202;
%%CITATION = HEP-TH 0407202;%%
B.~Freivogel and L.~Susskind,
%``A framework for the landscape,''
arXiv:hep-th/0408133;
%%CITATION = HEP-TH 0408133;%%
J.~Kumar and J.~D.~Wells,
%``Landscape cartography: A coarse survey of gauge group rank and stabilization
%of the proton,''
arXiv:hep-th/0409218;
%%CITATION = HEP-TH 0409218;%%
G.~Dvali,
%``Large hierarchies from attractor vacua,''
arXiv:hep-th/0410286;
%%CITATION = HEP-TH 0410286;%%
J.~D.~Wells,
%``PeV-scale supersymmetry,''
arXiv:hep-ph/0411041;
%%CITATION = HEP-PH 0411041;%%
R.~Kallosh and A.~Linde,
%``Landscape, the scale of SUSY breaking, and inflation,''
arXiv:hep-th/0411011.
%%CITATION = HEP-TH 0411011;%%


\bibitem{anthropic}
S.~Weinberg,
%``Anthropic Bound On The Cosmological Constant,''
Phys.\ Rev.\ Lett.\  {\bf 59}, 2607 (1987);
%%CITATION = PRLTA,59,2607;%%
A.~Vilenkin,
%``Predictions from quantum cosmology,''
Phys.\ Rev.\ Lett.\  {\bf 74}, 846 (1995)
[arXiv:gr-qc/9406010];
%%CITATION = GR-QC 9406010;%%
V.~Agrawal, S.~M.~Barr, J.~F.~Donoghue and D.~Seckel,
%``Anthropic considerations in multiple-domain theories and the scale of
%electroweak symmetry breaking,''
Phys.\ Rev.\ Lett.\  {\bf 80}, 1822 (1998)
[arXiv:hep-ph/9801253];
%%CITATION = HEP-PH 9801253;%%
M.~L.~Graesser, S.~D.~H.~Hsu, A.~Jenkins and M.~B.~Wise,
%``Anthropic distribution for cosmological constant and primordial density
%perturbations,''
Phys.\ Lett.\ B {\bf 600}, 15 (2004)
[arXiv:hep-th/0407174];
%%CITATION = HEP-TH 0407174;%%
A.~Vilenkin,
%``Anthropic predictions: The case of the cosmological constant,''
arXiv:astro-ph/0407586.
%%CITATION = ASTRO-PH 0407586;%%



\bibitem{fluxes}
S.~B.~Giddings, S.~Kachru and J.~Polchinski,
%``Hierarchies from fluxes in string compactifications,''
Phys.\ Rev.\ D {\bf 66}, 106006 (2002) [arXiv:hep-th/0105097];
%%CITATION = HEP-TH 0105097;%%
K.~Becker and M.~Becker,
%``Supersymmetry breaking, M-theory and fluxes,''
JHEP {\bf 0107}, 038 (2001)
[arXiv:hep-th/0107044];
%%CITATION = HEP-TH 0107044;%%
K.~Becker, M.~Becker, M.~Haack and J.~Louis,
%``Supersymmetry breaking and alpha'-corrections to flux induced  potentials,''
JHEP {\bf 0206}, 060 (2002)
[arXiv:hep-th/0204254];
%%CITATION = HEP-TH 0204254;%%
S.~Kachru, M.~B.~Schulz and S.~Trivedi,
%``Moduli stabilization from fluxes in a simple IIB orientifold,''
JHEP {\bf 0310}, 007 (2003)  [arXiv:hep-th/0201028];
%%CITATION = HEP-TH 0201028;%%
S.~Kachru, R.~Kallosh, A.~Linde and S.~P.~Trivedi,
%``De Sitter vacua in string theory,''
Phys.\ Rev.\ D {\bf 68}, 046005 (2003)  [arXiv:hep-th/0301240];
%%CITATION = HEP-TH 0301240;%%
R.~Blumenhagen, D.~Lust and T.~R.~Taylor,
%``Moduli stabilization in chiral type IIB orientifold models with
%  fluxes,'' 
Nucl.\ Phys.\ B {\bf 663}, 319 (2003)
[arXiv:hep-th/0303016];
%%CITATION = HEP-TH 0303016;%%
J.~F.~G.~Cascales and A.~M.~Uranga,
%``Chiral 4d N = 1 string vacua with D-branes and NSNS and RR fluxes,''
JHEP {\bf 0305}, 011 (2003) [arXiv:hep-th/0303024];
%%CITATION = HEP-TH 0303024;%%
P.~G.~Camara, L.~E.~Ibanez and A.~M.~Uranga,
%``Flux-induced SUSY-breaking soft terms on D7-D3 brane systems,''
arXiv:hep-th/0408036;
%%CITATION = HEP-TH 0408036;%%
M.~Grana, T.~W.~Grimm, H.~Jockers and J.~Louis,
%``Soft supersymmetry breaking in Calabi-Yau orientifolds with D-branes and
%fluxes,''
Nucl.\ Phys.\ B {\bf 690}, 21 (2004) [arXiv:hep-th/0312232].
%%CITATION = HEP-TH 0312232;%%
F.~Marchesano and G.~Shiu,
%``MSSM vacua from flux compactifications,''
arXiv:hep-th/0408059;
%%CITATION = HEP-TH 0408059;%%
%  F.~Marchesano and G.~Shiu,
%``Building MSSM flux vacua,''
JHEP {\bf 0411}, 041 (2004)
[arXiv:hep-th/0409132];
%%CITATION = HEP-TH 0409132;%%
A.~Font,
%``Z(N) orientifolds with flux,''
JHEP {\bf 0411}, 077 (2004)
[arXiv:hep-th/0410206];
%%CITATION = HEP-TH 0410206;%%
F.~Marchesano, G.~Shiu and L.~T.~Wang,
%``Model building and phenomenology of flux-induced supersymmetry breaking on
%D3-branes,''
arXiv:hep-th/0411080.
%%CITATION = HEP-TH 0411080;%%



\bibitem{deconstruction}
N.~Arkani-Hamed, A.~G.~Cohen and H.~Georgi,
%``(De)constructing dimensions,''
Phys.\ Rev.\ Lett.\  {\bf 86}, 4757 (2001)
[arXiv:hep-th/0104005];
%%CITATION = HEP-TH 0104005;%%
C.~T.~Hill, S.~Pokorski and J.~Wang,
%``Gauge invariant effective Lagrangian for Kaluza-Klein modes,''
Phys.\ Rev.\ D {\bf 64}, 105005 (2001)
[arXiv:hep-th/0104035].
%%CITATION = HEP-TH 0104035;%%




\bibitem{intersecting} 
C.~Bachas,
%``A Way to break supersymmetry,''
arXiv:hep-th/9503030;
%%CITATION = HEP-TH 9503030;%%
M.~Berkooz, M.~R.~Douglas and R.~G.~Leigh,
%``Branes intersecting at angles,''
Nucl.\ Phys.\ B {\bf 480}, 265 (1996) [arXiv:hep-th/9606139];
%%CITATION = HEP-TH 9606139;%%
N.~Ohta and P.~K.~Townsend,
%``Supersymmetry of M-branes at angles,''
Phys.\ Lett.\ B {\bf 418}, 77 (1998)  [arXiv:hep-th/9710129];
%%CITATION = HEP-TH 9710129;%%
R.~Blumenhagen, L.~Gorlich and B.~Kors,
%``Supersymmetric orientifolds in 6D with D-branes at angles,''
Nucl.\ Phys.\ B {\bf 569}, 209 (2000)  [arXiv:hep-th/9908130],
%%CITATION = HEP-TH 9908130;%%
%``Supersymmetric 4D orientifolds of type IIA with D6-branes at angles,''
JHEP {\bf 0001}, 040 (2000)  [arXiv:hep-th/9912204];
%%CITATION = HEP-TH 9912204;%%
G.~Pradisi,
%``Type I vacua from diagonal Z(3)-orbifolds,''
Nucl.\ Phys.\ B {\bf 575}, 134 (2000)  [arXiv:hep-th/9912218];
%%CITATION = HEP-TH 9912218;%%
R.~Blumenhagen, L.~Goerlich, B.~Kors and D.~Lust,
%``Noncommutative compactifications of type I strings on tori with  magnetic
%background flux,''
JHEP {\bf 0010},006 (2000)  [arXiv:hep-th/0007024];
%%CITATION = HEP-TH 0007024;%%
C.~Angelantonj, I.~Antoniadis, E.~Dudas and A.~Sagnotti,
%``Type-I strings on magnetised orbifolds and brane transmutation,''
Phys.\ Lett.\ B {\bf 489}, 223 (2000)  [arXiv:hep-th/0007090];
%%CITATION = HEP-TH 0007090;%%
C.~Angelantonj and A.~Sagnotti,
%``Type-I vacua and brane transmutation,''
arXiv:hep-th/0010279;
%%CITATION = HEP-TH 0010279;%%
G.~Aldazabal, S.~Franco, L.~E.~Ibanez, R.~Rabadan and
A.~M.~Uranga,
%``Intersecting brane worlds,''
JHEP {\bf 0102}, 047 (2001)  [arXiv:hep-ph/0011132];
%%CITATION = HEP-PH 0011132;%%
M.~Cvetic, G.~Shiu and A.~M.~Uranga,
%``Chiral four-dimensional N = 1 supersymmetric type IIA orientifolds from
%intersecting D6-branes,''
Nucl.\ Phys.\ B {\bf 615}, 3 (2001) [arXiv:hep-th/0107166];
%%CITATION = HEP-TH 0107166;%%
C.~Kokorelis,
%``New standard model vacua from intersecting branes,''
JHEP {\bf 0209}, 029 (2002) [arXiv:hep-th/0205147];
%%CITATION = HEP-TH 0205147;%%
D.~Bailin, G.~V.~Kraniotis and A.~Love,
%``Standard-like models from intersecting D5-branes,''
Phys.\ Lett.\ B {\bf 553}, 79 (2003)  [arXiv:hep-th/0210219];
%%CITATION = HEP-TH 0210219;%%
G.~Honecker,
%``Chiral supersymmetric models on an orientifold of Z(4) x Z(2) with
%intersecting D6-branes,''
Nucl.\ Phys.\ B {\bf 666}, 175 (2003) [arXiv:hep-th/0303015];
%%CITATION = HEP-TH 0303015;%%
M.~Larosa and G.~Pradisi,
%``Magnetized four-dimensional Z(2) x Z(2) orientifolds,''
Nucl.\ Phys.\ B {\bf 667}, 261 (2003) [arXiv:hep-th/0305224];
%%CITATION = HEP-TH 0305224;%%
G.~L.~Kane, P.~Kumar, J.~D.~Lykken and T.~T.~Wang,
%``Some phenomenology of intersecting D-brane models,''
arXiv:hep-ph/0411125.
%%CITATION = HEP-PH 0411125;%%


\bibitem{fi}
P.~Fayet and J.~Iliopoulos,
%``Spontaneously Broken Supergauge Symmetries And Goldstone Spinors,''
Phys.\ Lett.\ B {\bf 51}, 461 (1974);
%%CITATION = PHLTA,B51,461;%%
P.~Fayet,
%``Spontaneously Broken Supersymmetric Theories Of Weak, Electromagnetic And
%Strong Interactions,''
Phys.\ Lett.\ B {\bf 69}, 489 (1977).
%%CITATION = PHLTA,B69,489;%%



\bibitem{ibanez}
L.~E.~Ibanez,
%``The fluxed MSSM,''
arXiv:hep-ph/0408064;
%%CITATION = HEP-PH 0408064;%%
D.~Lust, S.~Reffert and S.~Stieberger,
%``MSSM with soft SUSY breaking terms from D7-branes with fluxes,''
arXiv:hep-th/0410074.
%%CITATION = HEP-TH 0410074;%%



\bibitem{kn}
K.~R.~Dienes, A.~E.~Faraggi and J.~March-Russell,
%``String Unification, Higher Level Gauge Symmetries, and Exotic Hypercharge
%Normalizations,''
Nucl.\ Phys.\ B {\bf 467}, 44 (1996)
[arXiv:hep-th/9510223];
%%CITATION = HEP-TH 9510223;%%
M.~Cvetic and P.~Langacker,
%``Implications of Abelian Extended Gauge Structures From String Models,''
Phys.\ Rev.\ D {\bf 54}, 3570 (1996)
[arXiv:hep-ph/9511378];
%%CITATION = HEP-PH 9511378;%%
K.~R.~Dienes, C.~F.~Kolda and J.~March-Russell,
%``Kinetic mixing and the supersymmetric gauge hierarchy,''
Nucl.\ Phys.\ B {\bf 492}, 104 (1997)
[arXiv:hep-ph/9610479];
%%CITATION = HEP-PH 9610479;%%
J.~K.~Elwood, N.~Irges and P.~Ramond,
%``A model of Yukawa hierarchies,''
Phys.\ Lett.\ B {\bf 413}, 322 (1997)
[arXiv:hep-ph/9705270];
%%CITATION = HEP-PH 9705270;%%
B.~Kors and P.~Nath,
%``Hierarchically split supersymmetry with Fayet-Iliopoulos D-terms in string
%theory,''
arXiv:hep-th/0411201.
%%CITATION = HEP-TH 0411201;%%


\bibitem{dfp}
E.~Dudas, A.~Falkowski and S.~Pokorski,
%``Deconstructed U(1) and supersymmetry breaking,''
Phys.\ Lett.\ B {\bf 568}, 281 (2003)
[arXiv:hep-th/0303155];
%%CITATION = HEP-TH 0303155;%%
A.~Falkowski, H.~P.~Nilles, M.~Olechowski and S.~Pokorski,
%``Deconstructing 5D supersymmetric U(1) gauge theories on orbifolds,''
Phys.\ Lett.\ B {\bf 566}, 248 (2003)
[arXiv:hep-th/0212206]. 
%%CITATION = HEP-TH 0212206;%%


\bibitem{pr}
C.~A.~Scrucca, M.~Serone, L.~Silvestrini and F.~Zwirner,
%``Anomalies in orbifold field theories,''
Phys.\ Lett.\ B {\bf 525}, 169 (2002)
[arXiv:hep-th/0110073];
%%CITATION = HEP-TH 0110073;%%
L. Pillo and  A.~Riotto,
%``On anomalies in orbifold theories,''
Phys.\ Lett.\ B {\bf 546}, 135 (2002)
[arXiv:hep-th/0202144];
%%CITATION = HEP-TH 0202144;%%
R.~Barbieri, R.~Contino, P.~Creminelli, R.~Rattazzi and C.~A.~Scrucca,
%``Anomalies, Fayet-Iliopoulos terms and the consistency of orbifold field
%theories,''
Phys.\ Rev.\ D {\bf 66}, 024025 (2002)
[arXiv:hep-th/0203039].
%%CITATION = HEP-TH 0203039;%%



\bibitem{cfgp}
E.~Cremmer, S.~Ferrara, L.~Girardello and A.~Van Proeyen,
%``Yang-Mills Theories With Local Supersymmetry: Lagrangian, Transformation
%Laws And Superhiggs Effect,''
Nucl.\ Phys.\ B {\bf 212}, 413 (1983).
%%CITATION = NUPHA,B212,413;%%




\bibitem{freedman} 
D.~Z.~Freedman,
%``Supergravity With Axial Gauge Invariance,''
Phys.\ Rev.\ D {\bf 15}, 1173 (1977); for a general formalism, see   
%%CITATION = PHRVA,D15,1173;%%
P.~Binetruy, G.~Dvali, R.~Kallosh and A.~Van Proeyen,
%``Fayet-Iliopoulos terms in supergravity and cosmology,''
Class.\ Quant.\ Grav.\  {\bf 21}, 3137 (2004)
[arXiv:hep-th/0402046].
%%CITATION = HEP-TH 0402046;%%



\bibitem{split}
N.~Arkani-Hamed and S.~Dimopoulos,
%``Supersymmetric unification without low energy supersymmetry and signatures
%for fine-tuning at the LHC,''
arXiv:hep-th/0405159;
G.~F.~Giudice and A.~Romanino,
%``Split supersymmetry,''
Nucl.\ Phys.\ B {\bf 699}, 65 (2004)
[arXiv:hep-ph/0406088];
%%CITATION = HEP-PH 0406088;%%
%%CITATION = HEP-TH 0405159;%%
A.~Arvanitaki, C.~Davis, P.~W.~Graham and J.~G.~Wacker,
%``One loop predictions of the finely tuned SSM,''
arXiv:hep-ph/0406034;
%%CITATION = HEP-PH 0406034;%%
A.~Pierce,
%``Dark matter in the finely tuned minimal supersymmetric standard model,''
Phys.\ Rev.\ D {\bf 70}, 075006 (2004)
[arXiv:hep-ph/0406144];
%%CITATION = HEP-PH 0406144;%%
C.~Kokorelis,
%``Standard models and split supersymmetry from intersecting brane orbifolds,''
arXiv:hep-th/0406258;
%%CITATION = HEP-TH 0406258;%%
S.~h.~Zhu,
%``Chargino pair production at linear collider and split supersymmetry,''
Phys.\ Lett.\ B {\bf 604}, 207 (2004)
[arXiv:hep-ph/0407072];
%%CITATION = HEP-PH 0407072;%%
B.~Mukhopadhyaya and S.~SenGupta,
%``Sparticle spectrum and phenomenology in split supersymmetry: Some
%possibilities,''
arXiv:hep-th/0407225;
%%CITATION = HEP-TH 0407225;%%
W.~Kilian, T.~Plehn, P.~Richardson and E.~Schmidt,
%``Split supersymmetry at colliders,''
arXiv:hep-ph/0408088;
%%CITATION = HEP-PH 0408088;%%
R.~Mahbubani,
%``Bounds on the Higgs mass in variations of Split Supersymmetry,''
arXiv:hep-ph/0408096;
%%CITATION = HEP-PH 0408096;%%
M.~Binger,
%``The Higgs boson mass at 2 loops in the finely tuned split supersymmetric
%standard model,''
arXiv:hep-ph/0408240;
%%CITATION = HEP-PH 0408240;%%
J.~L.~Hewett, B.~Lillie, M.~Masip and T.~G.~Rizzo,
%``Signatures of long-lived gluinos in split supersymmetry,''
JHEP {\bf 0409}, 070 (2004)
[arXiv:hep-ph/0408248];
%%CITATION = HEP-PH 0408248;%%
L.~Anchordoqui, H.~Goldberg and C.~Nunez,
%``Probing split supersymmetry with cosmic rays,''
arXiv:hep-ph/0408284;
%%CITATION = HEP-PH 0408284;%%
K.~Cheung and W.~Y.~Keung,
%``Split supersymmetry, stable gluino, and gluinonium,''
arXiv:hep-ph/0408335;
%%CITATION = HEP-PH 0408335;%%
N.~Arkani-Hamed, S.~Dimopoulos, G.~F.~Giudice and A.~Romanino,
%``Aspects of split supersymmetry,''
arXiv:hep-ph/0409232; 
%%CITATION = HEP-PH 0409232;%%
V.~Barger, C.~W.~Chiang, J.~Jiang and T.~Li,
%``Axion models with high-scale supersymmetry breaking,''
arXiv:hep-ph/0410252;
%%CITATION = HEP-PH 0410252;%%
I.~Antoniadis and S.~Dimopoulos,
%``Splitting supersymmetry in string theory,''
arXiv:hep-th/0411032;
%%CITATION = HEP-TH 0411032;%%
A.~Masiero, S.~Profumo and P.~Ullio,
%``Neutralino Dark Matter Detection in Split Supersymmetry Scenarios,''
arXiv:hep-ph/0412058.
%%CITATION = HEP-PH 0412058;%%



\bibitem{kklt}
S.~Kachru, R.~Kallosh, A.~Linde and S.~P.~Trivedi,
%``De Sitter vacua in string theory,''
Phys.\ Rev.\ D {\bf 68}, 046005 (2003)
[arXiv:hep-th/0301240]; 
%%CITATION = HEP-TH 0301240;%%
C.~P.~Burgess, R.~Kallosh and F.~Quevedo,
%``de Sitter string vacua from supersymmetric D-terms,''
JHEP {\bf 0310}, 056 (2003)
[arXiv:hep-th/0309187].
%%CITATION = HEP-TH 0309187;%%


\bibitem{bfs}
R.~Barbieri, S.~Ferrara and C.~A.~Savoy,
%``Gauge Models With Spontaneously Broken Local Supersymmetry,''
Phys.\ Lett.\ B {\bf 119}, 343 (1982);
%%CITATION = PHLTA,B119,343;%%
A.~H.~Chamseddine, R.~Arnowitt and P.~Nath,
%``Locally Supersymmetric Grand Unification,''
Phys.\ Rev.\ Lett.\  {\bf 49}, 970 (1982);
%%CITATION = PRLTA,49,970;%%
L.~J.~Hall, J.~Lykken and S.~Weinberg,
%``Supergravity As The Messenger Of Supersymmetry Breaking,''
Phys.\ Rev.\ D {\bf 27}, 2359 (1983);
%%CITATION = PHRVA,D27,2359;%%
E.~Cremmer, P.~Fayet and L.~Girardello,
%``Gravity Induced Supersymmetry Breaking And Low-Energy Mass Spectrum,''
Phys.\ Lett.\ B {\bf 122}, 41 (1983);
%%CITATION = PHLTA,B122,41;%%
N.~Ohta,
%``Grand Unified Theories Based On Local Supersymmetry,''
Prog.\ Theor.\ Phys.\  {\bf 70}, 542 (1983).
%%CITATION = PTPKA,70,542;%%




\bibitem{dsw}
M.~Dine, N.~Seiberg and E.~Witten,
%``Fayet-Iliopoulos Terms In String Theory,''
Nucl.\ Phys.\ B {\bf 289}, 589 (1987);
%%CITATION = NUPHA,B289,589;%%
J.~J.~Atick, L.~J.~Dixon and A.~Sen,
%``String Calculation Of Fayet-Iliopoulos D Terms In Arbitrary Supersymmetric
%Compactifications,''
Nucl.\ Phys.\ B {\bf 292}, 109 (1987).
%%CITATION = NUPHA,B292,109;%%



\bibitem{sagnotti}
A.~Sagnotti,
%``A Note on the Green-Schwarz mechanism in open string theories,''
Phys.\ Lett.\ B {\bf 294}, 196 (1992)
[arXiv:hep-th/9210127];
%%CITATION = HEP-TH 9210127;%%
E.~Poppitz,
%``On the one loop Fayet-Iliopoulos term in chiral four dimensional type I
%orbifolds,''
Nucl.\ Phys.\ B {\bf 542}, 31 (1999)
[arXiv:hep-th/9810010];
%%CITATION = HEP-TH 9810010;%%
L.~E.~Ibanez, R.~Rabadan and A.~M.~Uranga,
%``Anomalous U(1)'s in type I and type IIB D = 4, N = 1 string vacua,''
Nucl.\ Phys.\ B {\bf 542}, 112 (1999)
[arXiv:hep-th/9808139];
%%CITATION = HEP-TH 9808139;%%
Z.~Lalak, S.~Lavignac and H.~P.~Nilles,
%``String dualities in the presence of anomalous U(1) symmetries,''
Nucl.\ Phys.\ B {\bf 559}, 48 (1999)
[arXiv:hep-th/9903160];
%%CITATION = HEP-TH 9903160;%%
I.~Antoniadis, C.~Bachas and E.~Dudas,
%``Gauge couplings in four-dimensional type I string orbifolds,''
Nucl.\ Phys.\ B {\bf 560}, 93 (1999)
[arXiv:hep-th/9906039].
%%CITATION = HEP-TH 9906039;%%






\end{thebibliography}

\end{document}